# A Survey of Big Data Machine Learning Applications Optimization in Cloud Data Centers and Networks

Sanaa Hamid Mohamed, *Student Member, IEEE*, Taisir E.H. El-Gorashi, *Member, IEEE*, and Jaafar M.H. Elmirghani, *Senior Member, IEEE*

*Abstract*— **This survey article reviews the challenges associated with deploying and optimizing big data applications and machine learning algorithms in cloud data centers and networks. The MapReduce programming model and its widely-used open-source platform; Hadoop, are enabling the development of a large number of cloud-based services and big data applications. MapReduce and Hadoop thus introduce innovative, efficient, and accelerated intensive computations and analytics. These services usually utilize commodity clusters within geographically-distributed data centers and provide cost-effective and elastic solutions. However, the increasing traffic between and within the data centers that migrate, store, and process big data, is becoming a bottleneck that calls for enhanced infrastructures capable of reducing the congestion and power consumption. Moreover, enterprises with multiple tenants requesting various big data services are challenged by the need to optimize leasing their resources at reduced running costs and power consumption while avoiding under or over utilization. In this survey, we present a summary of the characteristics of various big data programming models and applications and provide a review of cloud computing infrastructures, and related technologies such as virtualization, and software-defined networking that increasingly support big data systems. Moreover, we provide a brief review of data centers topologies, routing protocols, and traffic characteristics, and emphasize the implications of big data on such cloud data centers and their supporting networks. Wide ranging efforts were devoted to optimize systems that handle big data in terms of various applications performance metrics and/or infrastructure energy efficiency. This survey aims to summarize some of these studies which are classified according to their focus into applications-level, networking-level, or data centers-level optimizations. Finally, some insights and future research directions are provided.**

## I Introduction

THE evolving paradigm of big data is essential for critical advancements in data processing models and the underlying acquisition, transmission, and storage infrastructures [1]. Big data differs from traditional data in being potentially unstructured, rapidly generated, continuously changing, and massively produced by a large number of distributed users or devices. Typically, big data workloads are transferred into powerful data centers containing sufficient storage and processing units for real-time or batch computations and analysis. A widely used characterization for big data is the "5V" notion which describes big data through its unique attributes of Volume, Velocity, Variety, Veracity, and Value [2]. In this notation, the volume refers to the vast amount of data produced which is usually measured in Exabytes (i.e. $2^{60}$ or $10^{18}$ bytes) or Zettabytes (i.e. $2^{70}$ or $10^{21}$ bytes), while the velocity reflects the high speed or rate of data generation and hence potentially the short lived useful lifetime of data. Variety indicates that big data can be composed of different types of data which can be categorized into structured and unstructured. An example of structured data is bank transactions which can fit into relational database systems, and an example of the unstructured data is social media content that could be a mix of text, photos, animated Graphics Interchange Format (GIF), audio files, and videos contained in the same element (e.g. a tweet, or a post). The veracity measures the trustworthiness of the data as some generated portions could be erroneous or inaccurate, while the value measures the ability of the user or owner of the data to extract useful information from the data.

In 2020, the global data volume is predicted to be around 40,000 Exabytes which represents a 300 times growth factor compared to the global data volume in 2005 [3]. An estimate of the global data volume in 2010 is about 640 Exabytes [4], and in 2015 is about 2,700 Exabytes [5]. This huge growth in data volumes is the result of continuous developments in various applications that generate massive and rich content related to a wide range of human activities. For example, online business transactions are expected to have a rate of 450 Billion transactions per day by 2020 [4]. Social media such as Facebook, LinkedIn, and Twitter, which have between 300

Million and 2 Billion subscribers who access these social media platforms through web browsers in personal computers (PCs), or through applications installed in tablets and smart phones are enriching the content of the Internet with content in the range of several Terabytes ($2^{40}$ bytes) per day [5]. Analyzing the thematic connections between the subscribers, for example by grouping people with similar interests, is opening remarkable opportunities for targeted marketing and e-commerce. Moreover, the subscriber's behaviours and preferences tracked by their activities, clickstreams, requests, and collected web log files can be analyzed with big data mining tools for profound psychological, economical, business-oriented, and product improvement studies [6], [7]. To accelerate the delay-sensitive operations of web searching and indexing , distributed programming models for big data such as MapReduce were developed [8]. MapReduce is a powerful, reliable, and cost-effective programming model that performs parallel processing for large distributed datasets. These features have enabled the development of different distributed programming big data solutions and cloud computing applications.

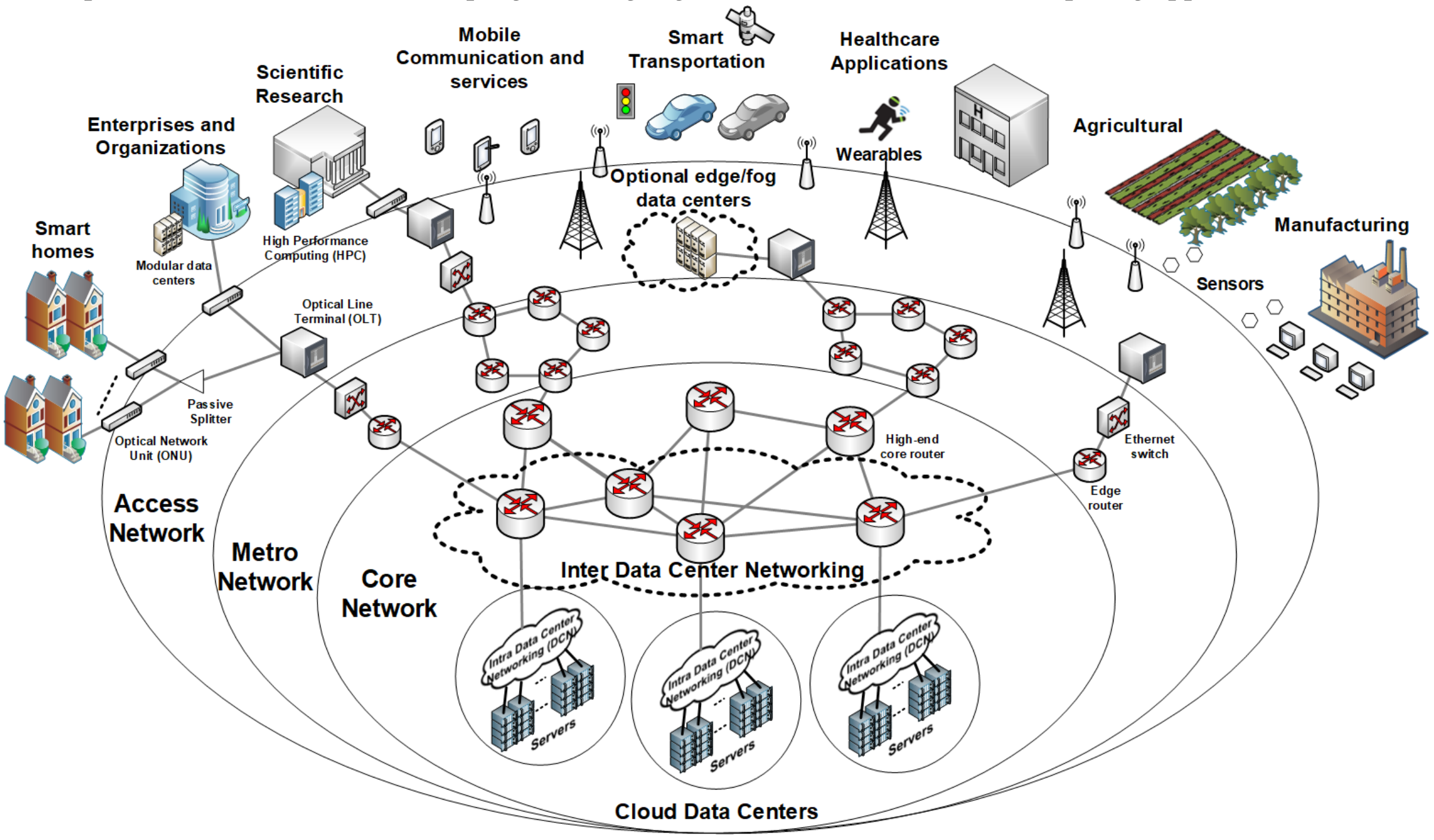

Fig. 1. Big data communication, networking, and processing infrastructure, and examples of big data applications.

A wide range of applications are considered big data applications from data-intensive scientific applications that require extensive computations to massive datasets that require manipulation such as in earth sciences, astronomy, nanotechnology, genomics, and bioinformatics [9]. Typically, the computations, simulations, and modelling in such applications are carried out in High Performance Computing (HPC) clusters with the aid of distributed and grid computing. However, the datasets growth beyond these systems capacities in addition to the desire to share datasets for scientific research collaborations in some disciplines are encouraging the utilization of big data applications in cloud computing infrastructures with commodity devices for scientific computations despite the resultant performance and cost tradeoffs [10].

With the prevalence of mobile applications and services that have extensive computational and storage demands exceeding the capabilities of the current smart phones, emerging technologies such as Mobile Cloud Computing (MCC) were developed [11]. In MCC, the computational and storage demands of applications are outsourced to remote (or close as in mobile edge computing (MEC)) powerful servers over the Internet. As a result, on-demand rich services such as video streaming, interactive video, and online gaming can be effectively delivered to the capacity and battery limited devices. Video content accounted for 51% of the total mobile data traffic in 2012 [11], and is predicted to account for 78% of an expected total volume of 49 Exabytes by 2021 [12]. Due to these huge demands, in addition to the large sizes of video files, big video data platforms are fronting several challenges related to video streaming, storage, and replication management, while needing to meet strict quality-of-experience (QoE) requirements [13].

In addition to mobile devices, the wide range of everyday physical objects that are increasingly interconnected for automated operations has formed what is known as the Internet-of-Things (IoT). In IoT systems, the underlying communication and networking infrastructure are typically integrated with big data computing systems for data collection, analysis, and decision-making. Several technologies such as RFID, low power communication technologies, Machine-to-Machine (M2M) communications, and wireless sensor networking (WSN) have been suggested for improved IoT communications and networking infrastructure [14]. To process the big data generated by IoT devices, different solutions such as cloud and fog computing were proposed [15]-[31]. Existing cloud computing infrastructures could be utilized by aggregating and processing big data in powerful central data centers. Alternatively, data could be processed at the edge where fog computing units, typically with limited processing capacities compared to cloud, are utilized [32]. Edge computing reduces both; the traffic in core networks, and the latency by being closer to end devices. The connected devices could be sensors gathering different real-time measurements, or actuators performing automated control operations in industrial, agricultural, or smart building applications. IoT can support vehicle communication to realize smart transportation systems. IoT can also support medical applications such as wearables and telecare applications for remote treatment, diagnosis and monitoring [33]. With this variety in IoT devices, the number of Internet-connected things is expected to exceed 50 Billion by 2020, and the services provided by IoT are expected to add $15 Trillion to the global Gross Domestic Product (GDP) in the next 20 years [14]. Figure 1 provides generalized big data communication, networking, and processing infrastructure and examples of applications that can utilize it.

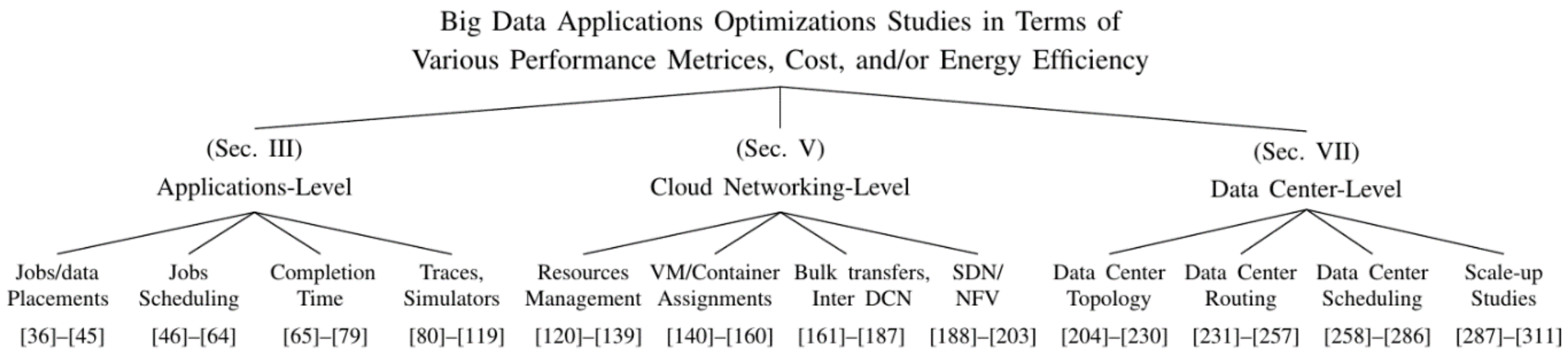

Fig.2. Classification of big data applications optimization studies.

Achieving the full potential of big data requires a multidisciplinary collaboration between computer scientists, engineers, data scientists, as well as statisticians and other stakeholders [4]. It also calls for huge investments and developments by enterprises and other organizations to improve big data processing, management, and analytics infrastructures to enhance decision making and services offerings. Moreover, there are urgent needs for integrating new big data applications with existing Application Program Interfaces (API) such as Structured Query Language (SQL), and R language for statistical computing. More than $15 billion have already been invested in big data systems by several leading Information Technology (IT) companies such as IBM, Oracle, Microsoft, SAP, and HP [34]. One of the challenges of big data in enterprise and cloud infrastructures is the existence of various workloads and tenants with different Service Level Agreement (SLA) requirements that need to be hosted on the same set of clusters. An early solution to this challenge at the application level is to utilize a distributed file system to control the access and sharing of data within the clusters [35]. On the infrastructure level, solutions such as Virtual Machines (VMs) or Linux containers dedicated to each application or tenant were utilized to support the isolation between their assigned resources [1], [34]. Big data systems are also challenged by security, privacy, and governance related concerns. Furthermore, as the increasing computational demands of the increasing data volumes are exceeding the capabilities of existing commodity infrastructures, future enhanced and energy efficient processing and networking infrastructure for big data have to be investigated and optimized.

This survey paper aims to summarize a wide range of studies that use different state-of-art and emerging networking and computing technologies to optimize and enhance big data applications and systems in terms of various performance metrics such as completion time, data locality, load balancing, fairness, reliability, and resources utilization, and/or their energy efficiency. Due to the popularity and wide applicability of the MapReduce programming model, its related optimization studies will be the main focus in this survey. Moreover, optimization studies for big data management and streaming, in addition to generic cloud computing applications and bulk data transfer are also considered. The optimization studies in this survey are classified according to their focus into applications-level, cloud networking-level, and data center-level studies as summarized in Figure 2. The first category at the application level targets the studies that extend or optimize existing framework parameters and mechanisms such as optimizing jobs and data placements, and scheduling, in addition to developing benchmarks, traces, and simulators [36]-[119]. As the use of big data applications is evolving from clusters with controlled environments into cloud environments with geo-distributed data centers, several additional challenges are encountered. The second category at the networking level focuses on optimizing cloud networking infrastructures for big data applications such as inter-data centers networking, virtual machine assignments, and bulk data transfer optimization studies [120]-[203]. The increasing data volumes and processing demands are also challenging the data centers that store and process big data. The third category at the data center level targets optimizing the topologies, routing, and scheduling in data centers for big data applications, in addition to the studies that utilize, demonstrate, and suggest scaled-up computing and networking infrastructures to replace commodity hardware in the future [204]-[311]. For the performance evaluations in the aforementioned studies, either realistic traces, or deployments in experimental testbed clusters are utilized.

Although several big data related surveys and tutorials are available, to the best of our knowledge, none has extensively addressed optimizing big data applications while considering the technological aspects of their hosting cloud data centers and networking infrastructures. The tutorial in [1] and the survey in [312] considered mapping the role of cloud computing, IoT, data centers, and applications to the acquisition, storage and processing of big data. The authors in [313]-[316] extensively surveyed the advances in big data processing frameworks and compared their components, usage, and performance. A review of benchmarking big data systems is provided in [317]. The surveys in [318], [319] focused on optimizing jobs scheduling at the application level, while the survey in [320] additionally tackled extensions, tuning, hardware acceleration, security, and energy efficiency for MapReduce. The environmental impacts of big data and its usage to green applications and systems were discussed in [321]. Security and privacy concerns of MapReduce in cloud environments were discussed in [322], while the challenges and requirements of geo-distributed batch and streaming big data frameworks were outlined in [323]. The surveys in [324]-[326] addressed the use of big data analytics to optimize wired and wireless networks, while the survey in [327] overviewed big data mathematical representations for networking optimizations. The scheduling of flows in data centers for big data is addressed in [328] and the impact of data centers frameworks on the scheduling and resource allocation is surveyed in [329] for three big data applications.

This survey paper is structured as follows: For the convenience of the reader, brief overviews for the state-of-art and advances in big data programming models and frameworks, cloud computing and its related technologies, and cloud data centers are provided before the corresponding optimization studies. Section II reviews the characteristics of big data programming models and existing batch, streaming processing and storage management applications while Section III summarizes the applications-focused optimization studies. Section IV discusses the prominence of cloud computing for big data applications, and the implications of big data applications on cloud networks. It also reviews some related technologies that support big data and cloud computing systems such as machine and network virtualization, and Software-Defined Networking (SDN). Section V summarizes the cloud networking-focused optimization studies. Section VI briefly reviews data center topologies, traffic characteristics and routing protocols, while Section VII summarizes the data center-focused optimization studies. Finally, Section VIII provides future research directions, and Section IX concludes the survey. Key Acronyms are provided below.

APPENDIX A LIST OF KEY ACRONYMS

| | |
|---|---|
| AM | Application Master. |
| ACID | Atomicity, Consistency, Isolation, and Durability. |
| BASE | Basically Available Soft-state Eventual consistency. |
| CAPEX | Capital Expenditure. |
| CPU | Central Processing Unit. |
| CSP | Cloud Service Provider. |
| DAG | Directed Acyclic Graph. |
| DCN | Data Center Networking. |
| DFS | Distributed File System. |
| DVFS | Dynamic Voltage and Frequency Scaling. |
| EC2 | Elastic Compute Cloud. |
| EON | Elastic Optical Network. |
| ETSI | European Telecom Standards Institute. |
| HDFS | Hadoop Distributed File System. |
| HFS | Hadoop Fair Scheduler. |
| HPC | High Performance Computing. |
| I/O | Input/Output. |
| ILP | Integer Linear Programming. |
| IP | Internet Protocol. |
| ISP | Internet Service Provider. |
| JT | Job Tracker. |
| JVM | Java Virtual Machine. |
| MILP | Mixed Integer Linear Programming. |
| NFV | Network Function Virtualization. |
| NM | Node Manager. |
| NoSQL | Not only SQL. |
| OF | Open Flow. |
| O-OFDM | Optical Orthogonal Frequency Division Multiplexing. |
| OPEX | Operational Expenses. |
| OVS | Open v Switch. |
| P2P | Peer-to-Peer. |
| PON | Passive Optical Network. |
| QoE | Quality of Experience. |
| QoS | Quality of Service. |
| RAM | Random Access Memory. |
| RDBMS | Relational Data Base Management System. |
| RDD | Resilient Distributed Data sets. |
| RM | Resource Manager. |
| ROADM | Reconfigurable Optical Add Drop Multiplexer. |
| SDN | Software Defined Networking. |
| SDON | Software Defined Optical Networks. |
| SLA | Service Level Agreement. |
| SQL | Structured Query Language. |
| TE | Traffic Engineering. |
| ToR | Top-of-Rack. |
| TT | Task Tracker. |
| VM | Virtual Machine. |
| VNE | Virtual Network Embedding. |
| VNF | Virtual Network Function. |
| WAN | Wide Area Network. |
| WC | WordCount. |
| WDM | Wavelength Division Multiplexing. |
| WSS | Wavelength Selective Switch. |
| YARN | Yet Another Resource Negotiator. |

## II Programming Models, Platforms, and Applications for Big Data Analytics:

This Section reviews some of the programming models developed to provide parallel computation for big data. These programming models include MapReduce [8], Dryad [330], and Cloud Dataflow [331]. The input data in these models can be generally categorized into bounded and unbounded data depending on the required computational speed. In applications that have no strict processing speed requirements, the input data can be aggregated, bounded, and then processed in batch mode. In applications that require instantaneous analysis of continuous data flows (i.e. unbounded data), streaming mode is utilized. The MapReduce programming model, with the support of open source platforms such as Apache Hadoop, has been widely used for efficient, scalable, and reliable computations at reduced costs especially for batch processing. The maturity of such platforms have supported the development of several scalable commercial or open-source big data applications and services for processing and data storage management. The rest of this Section is organized as follows: Subsection II-A illustrates the aforementioned programming models, while Subsection II-B briefly describes the characteristics and components of Apache Hadoop and its related applications. Subsection II-C focuses on big data storage management applications, while Subsection II-D summarizes some of the in-memory big data applications. Finally, Subsections II-E, and II-F elaborate on big data stream processing applications, and the Lambda hybrid architecture, respectively.

### *A. Programming Models:*

The MapReduce programming model was introduced by Google in 2003 as a cost-effective solution for processing massive data sets. MapReduce utilizes distributed computations in commodity clusters that run in two phases; *map* and *reduce* which are adopted from the Lisp functional programming language [8]. The MapReduce user defines the required functions of each phase by using a programming language such as C++, Java, or Python, and then submits the code as a single MapReduce job to process the data. The user also defines a set of parameters to configure the job. Each MapReduce job consists of a number of map and reduce tasks depending on the input data size and the configurations, respectively. Each map task is assigned to process a unique portion of the input data set, preferably available locally, and hence can run independently from other map tasks. The processing starts by transforming the input data into the key-value schema and applies it to the map function to compute another key-value pair also known as the intermediate results. These results are then shuffled to reduce tasks according to their keys where each reduce tasks is assigned to process intermediate results with a unique set of keys. Finally, each reduce task generates the final outputs. The internal operational details of MapReduce such as assigning the nodes within the cluster to map or reduce tasks, partitioning the input data, tasks scheduling, fault tolerance, and inter-machine communications are typically performed by the run-time system and are hidden from the users. The input and output data files are typically managed by a Distributed File System (DFS) that provides a unified view of the files and their details, and allows various modifications such as replication, read, and write operations. An example of DFSs is the Google File System (GFS) [332] which is a fault-tolerant, reliable, and scalable chunk-based distributed file system designed to support MapReduce in Google's commodity servers. The typical chunk size in GFS is 64 MB and each chunk is replicated in different nodes with a default value of 3 to support fault-tolerance.

The components of a typical MapReduce cluster and the implementation details of the MapReduce programming model are illustrated in Figure 3. One of the nodes in the cluster is set to be a *Master*, while the others are set to be *Workers* and are assigned to either a map or a reduce task by the master. Beside task assignments, the master is also responsible for monitoring the performance of the running tasks and for checking the statuses of the nodes within the cluster. The master also manages the information about the location of the running jobs, data replicas, and intermediate results. The detailed steps of implementing a MapReduce job are as follows [8]:

1) The MapReduce code is copied to all cluster nodes. In this code, the user defines the map and reduce functions and provides additional parameters such as input and output data types, names of the output files, and the number of reduce workers.
2) The master assigns map and reduce tasks to available workers where typically, map workers are more than reduce workers and are assigned to several map tasks.
3) The input data, in the form of key-value pairs, is split into smaller partitions. The splits (S) and their replicas ($S_R$) are distributed in the map workers' local disks as illustrated in Figure 3. The splits are then processed concurrently in their assigned map workers according to their scheduling. Each map function produces intermediate results (IR) consisting of the intermediate key-value pairs. These results are then materialized (i.e. saved persistently) in the local disks of the map workers.
4) The intermediate results are divided into (R) parts to be processed by R reduce workers. Partitioning can be done through hash functions (e.g. hash(*key*) mod R) to ensure that each key is assigned to only one reduce worker. The locations of the hashed intermediate results and their file sizes are sent to the master node.

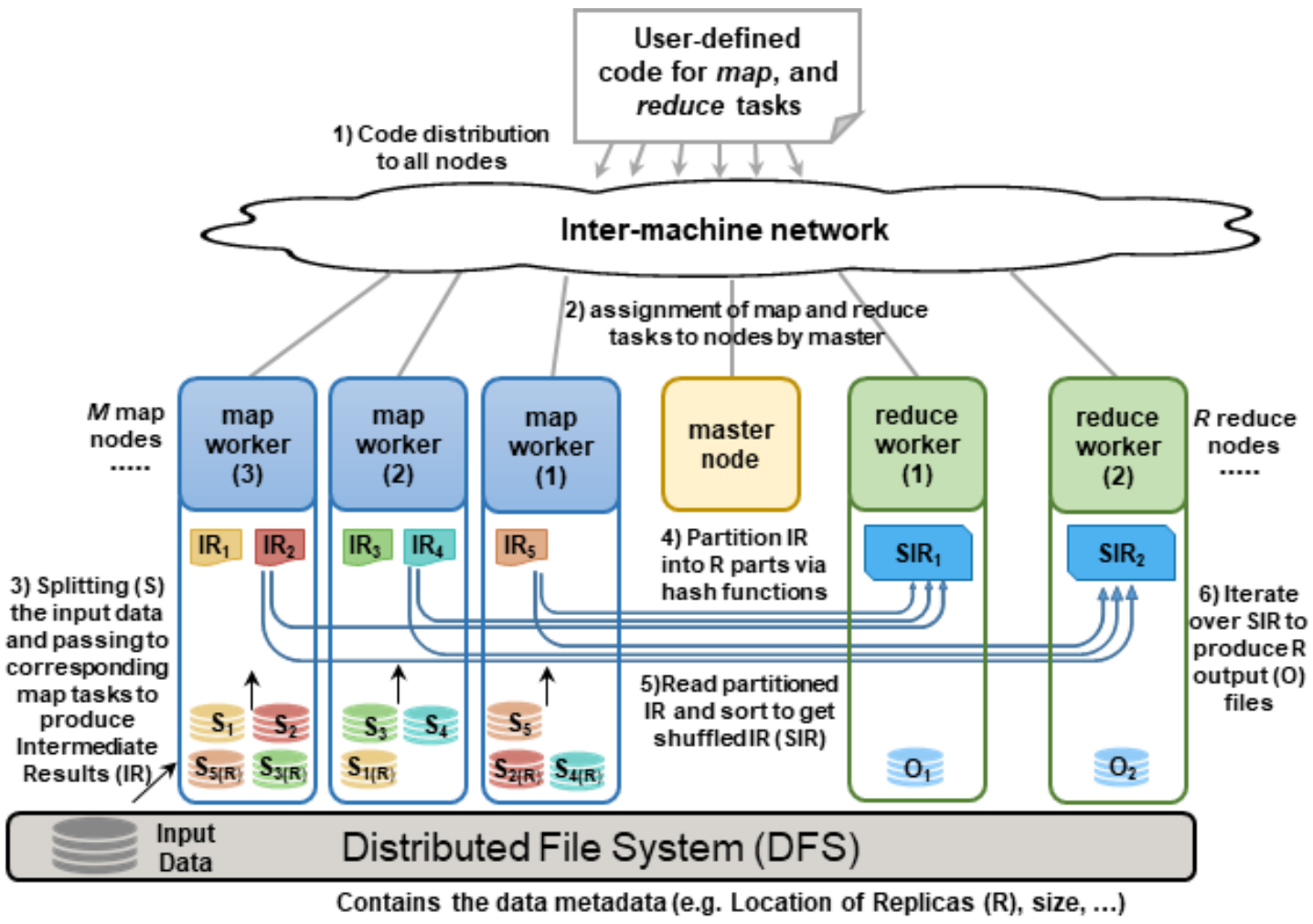

Fig. 3. Google's MapReduce cluster components and the programming model implementation.

5) A reduce task is composed of shuffle, sort, and reduce phases. The shuffle phase can start when 5% of the map results are generated, however the last reduction phase cannot start unless all map tasks are completed. For shuffling, each reduce worker obtains the locations of the intermediate pairs with the keys assigned to it and fetches the corresponding results from the map workers' local disks typically via the HyperText Transfer Protocol (HTTP).
6) Each reduce worker then sorts its intermediate results by the keys. The sorting is performed in the Random Access Memory (RAM) if the intermediate results can fit, otherwise, external sort-merge algorithms are used. The sorting groups all the occurrences of the same key and forms the shuffled intermediate results (SIR).
7) Each reduce worker applies the assigned user-defined reduce function on the shuffled data to generate the final key-value pairs output (O), the final output files are then saved in the distributed file system.

In MapReduce, fault-tolerance is achieved by re-executing the failed tasks. Failures can occur due to hardware causes such as disk failures, out of disk, out of memory, and socket time out. Each map or reduce task can be in one of three statuses which are idle, in-progress, and completed. If an in-progress map task fails, the master changes its status to idle to allow it to be re-scheduled on an available map node containing a replica of the data. If a map worker fails while having some completed map tasks, all contained map tasks must be re-scheduled as the intermediate results, which are only saved on local disks, are no longer accessible. In this case, all reduce workers must be re-scheduled to obtain the correct and complete set of intermediate results. If a reduce worker fails, only in-progress tasks are re-scheduled as the results of completed reduce jobs are saved in the distributed file system. To improve MapReduce performance, speculative execution can be activated, where backup tasks are created to speed up the lacking in-progress tasks known as stragglers [8].

The MapReduce programming model is capable of solving several common programming problems such as *words count and sort in addition to implementing complex graph processing, data mining, and machine learning applications*. However, the speed requirements for some computations might not be satisfied by MapReduce due to several limitations [333]. Moreover, developing efficient MapReduce applications requires advanced programming skills to fit the computations into the map and reduce pipeline, and deep knowledge of underlying infrastructures to properly configure and optimize a wide range of parameters [334]-[337]. One of MapReduce limitations is that transferring non-local input data to map workers and shuffling intermediate results to reduce workers typically require intensive networking bandwidth and disk I/O operations. Early efforts to minimize the effects of these bottlenecks included maximizing data locality, where the computations are carried closer to data [8]. Another limitation is due to the fault-tolerance mechanism that requires materializing the entire output of MapReduce jobs in the disks managed by the DFS before being accessible for further computations. Hence, MapReduce is generally less suitable for interactive and iterative computations that require repetitive access to results. An implementation variation known as MapReduce Online [338], supports shuffling by utilizing RAM resources to pipeline intermediate results between map and reduce stages before the materialization.

Several other programming models were developed as variants to MapReduce [314]. One of these variants is Dryad which is a high-performance general-purpose distributed programming model for parallel applications with coarse-grain data [330]. In Dryad, a Directed Acyclic Graph (DAG) is used to describe the jobs by representing the computations as the graph *vertices* and the data communication patterns as the graph *edges*. The Job Manager within Dryad, schedules the vertices, which contain sequential programs, to run concurrently in a set of machines that are available at the run time. These machines can be different cores within the same multi-core PC or can be thousands of machines within a large data center. Dryad provides fault-tolerance and efficient resource utilization by allowing graph modification during the computations. Unlike MapReduce, which restricts the programmer to provide a single input file and produces a single output file, Dryad allows for an arbitrary number of input and output files.

As a successor to MapReduce, Google has recently introduced "Cloud Dataflow" which is a unified programming model with enhanced processing capabilities for bounded and unbounded data. It provides a balance between correctness, latency, and cost when processing massive, unbounded, and out-of-order data [331]. In Cloud Dataflow, the data is represented as tuples containing the key, value, event-time, and the required time window for the processing. This supports sophisticated user requirements such as event-time ordering of results by utilizing data windowing that divide the data streams into finite chunks to be processed in groups. Cloud Data flow utilizes the features of both FlumeJava which is a batch engine and MillWheel which is a streaming engine [331]. The core primitives of Cloud Dataflow, are *ParDo*, which is an element-wise generic parallel processing function, and *GroupByKey* and *GroupByKeyandWindow* which aggregate data with the same key according to the user requests.

*B. Apache Hadoop Architecture and Related Software:*

Hadoop, which is currently under the auspices of the Apache Software Foundation, is an open source software framework written in Java for reliable and scalable distributed computing [339]. This framework was initiated by Doug Cutting who utilized the MapReduce programming model for indexing web crawls and was open-sourced by Yahoo in 2005. Beside Apache, several organizations have developed and customized Hadoop distributions tailored for their infrastructures such as HortonWorks, Cloudera, Amazon Web Services (AWS), Pivotal, and MAPR technologies. The Hadoop ecosystem allows other programs to run on the same infrastructure with MapReduce which made it a natural choice for enterprise big data platforms [35].

The basic components of the first versions of Hadoop; Hadoop 1.x are depicted in Figure 4. These versions contain a layer for the Hadoop Distributed File System (HDFS), a layer for the MapReduce 1.0 engine which resembles Google's MapReduce, and can have other applications on the top layer. The MapReduce 1.0 layer follows the master-slave architecture. The master is a single node containing a Job Tracker (JT), while each slave node contains a Task Tracker (TT). The JT handles jobs assignment and scheduling and maintains the data and metadata of jobs, in addition to resources information. It also monitors the liveness of TTs and the availability of their resources by sending periodic heartbeat messages typically each 3 seconds. Each TT contains a predefined set of slots. Once it accepts a map or a reduce task, it launches a Java Virtual Machine (JVM) in one of its slots to perform the task, and periodically updates the JT with the task status [339].

The HDFS layer consists of a name node in the master and several data nodes in each slave node. The name node stores the details of the data nodes and the addresses of the data blocks and their replicas. It also checks the data nodes via heartbeat messages and manages load balancing. For reliability, a secondary name node is typically assigned to save snapshots of the primary name node. As in GFS, the default file size in HDFS is 64 MB and three replicas are maintained for each file for fault-tolerance, performance improvements, and load balancing. Beside GFS and HDFS, several distributed files systems were developed such as Amazon's simple Storage Service (S3), Moose File System (MFS), Kosmos distributed file system (KFS), and Colossus [314], [340].

Default tasks scheduling mechanisms in Hadoop are First-In First-Out (FIFO), capacity scheduler, and Hadoop Fair Scheduler (HFS). FIFO schedules the jobs according to their arrival time which leads to undesirable delays in environments with a mix of long batch jobs and small interactive jobs [319]. The Capacity scheduler developed at Yahoo reserves a pool containing minimum resources guarantees for each user, and hence suits systems with multiple users [319]. FIFO scheduling is then used for the jobs of the same user. The Fair scheduler developed at Facebook dynamically allocates the resources equally between jobs. It thus improves the response time of small jobs [46].

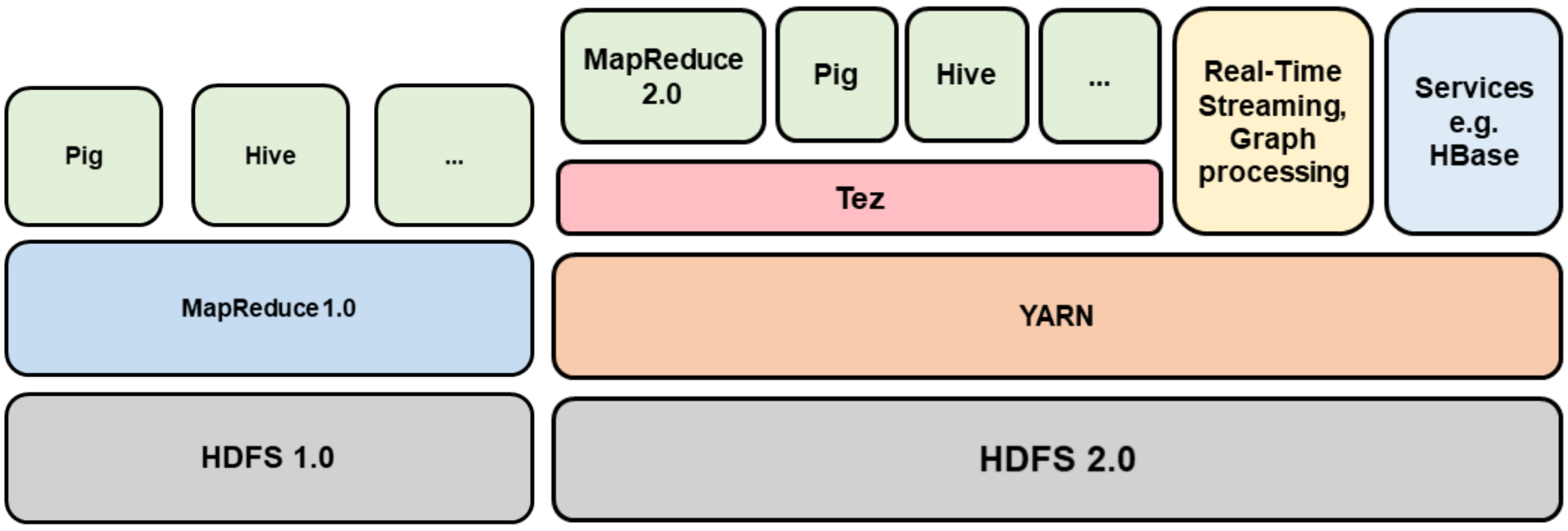

(a) (b)

Fig. 4. Framework components in Hadoop (a) Hadoop 1.x, and (b) Hadoop 2.x.

Hadoop 2.x, which is also depicted in Figure 4, introduced a resource management platform named YARN; Yet Another Resource Negotiator [341]. YARN decouples the resource management infrastructure from the processing components and enables the coexistence of different processing frameworks beside MapReduce which increases the flexibility in big data clusters. In YARN, the JT and TT are replaced with three components which are the Resource Manager (RM), the Node Manager (NM), and the Application Master (AM). The RM is a per-cluster global resources manager which runs as daemon on a dedicated node. It contains a scheduler that dynamically leases the available cluster resources in the form of containers (further explained in Subsection IV-B4), which are considered as logical bundles (e.g. 2 GB RAM, 1 Central Processing Unit (CPU) core), among competing MapReduce jobs and other applications according to their demands and scheduling priorities. A NM is a per-server daemon that is responsible for monitoring the health of its physical node, tracking its containers assignments, and managing the containers lifecycle (i.e. starting and killing). An AM is a per-application container that manages the resources consumption, the jobs execution flow, and also handles the fault-tolerance tasks. The AM, which typically needs to harness resources from several nodes to finish its job, issues a resource request to the RM indicating the required number of containers, the required resources per container, and the locality preferences.

Figure 5 illustrates the differences between Hadoop 1.x, and Hadoop 2.x with YARN. A detailed study of various releases of Hadoop is presented in [342]. The authors also provided a brief comparison of these releases in terms of their energy efficiency and performance.

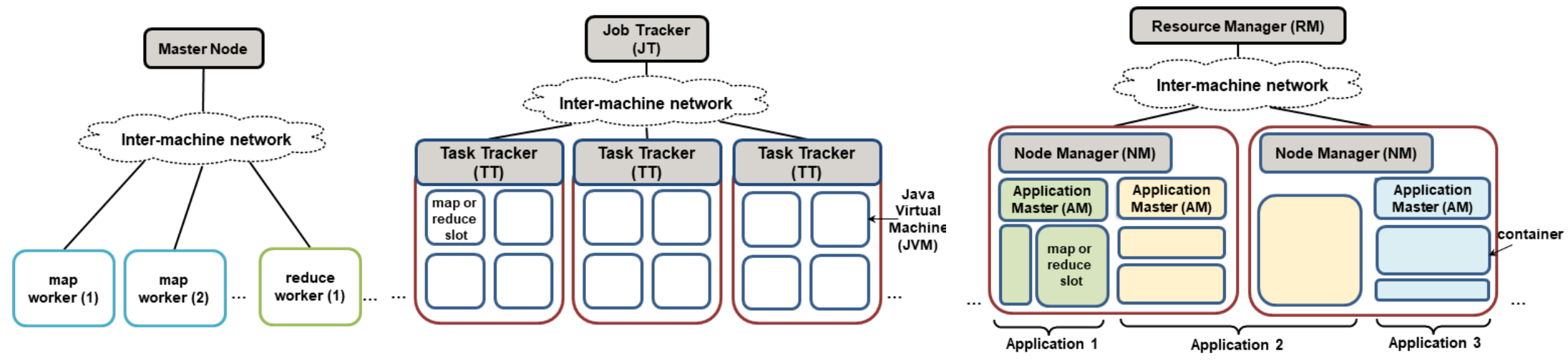

(a) (b) (c)

Fig. 5. Comparison of clusters with (a) Google's MapReduce, (b) Hadoop 1.x, and (b) Hadoop 2.x with YARN.

YARN increases the resource allocation flexibility in MapReduce 2 as it utilizes flexible containers for resources allocations and hence eliminates the use of fixed resources slots as in MapReduce 1. However, this advantage comes at the expenses of added system complexity and a slight increase in the power consumption when compared to MapReduce 1 [342]. Another difference is that the intermediate results shuffling operations in MapReduce 2 are performed via auxiliary services that preserve the output of a container's results before killing it. The communications between the AMs, NMs, and the RM are heartbeat-based. If a node fails, the NM sends an indicating heartbeat to the RM which in turn informs the affected AMs. If an in-progress job fails, the NM marks it as idle and re-executes it. If an AM fails, the RM restarts it and synchronizes its tasks. If an RM fails, an old checkpoint is used and a secondary RM is activated [341].

A wide range of applications and programming frameworks can run natively in Hadoop as depicted in Figure 4. The difference in their implementation between the Hadoop versions is that in the 1.x versions, they are forced to follow the MapReduce framework while in the 2.x versions they are no longer restricted to it. Examples of these applications and frameworks are Pig [343], Tez [344], Hive [345], HBase [314], Storm [346], Giraph for graph processing, and Mahout for machine learning [315], [340]. Due to the lack of built-in declarative languages in Hadoop, Pig, Tez, and Hive were introduced to support querying and to replace ad-hoc users-written programs which were hard to maintain and reuse. Pig is composed of an execution engine and a declarative scripting language named Pig Latin that compiles SQL-like queries to an equivalent set of sequenced MapReduce jobs. Pig hides the complexity of MapReduce and directly provides advanced operations such as filtering, joining, and ordering [343]. Tez is a flexible input-processor output-runtime model that transforms the queries into abstracted DAG where the vertices represent the parallel tasks, and the edges represent the data movement between different map and reduce stages [344]. It supports in-memory operations which makes it more suitable for interactive processing than MapReduce. Hive [345] is a data warehouse software developed at Facebook. It contains a declarative language; HiveQL that automatically generates MapReduce jobs from SQL-like user queries.

### *C. Big Data Storage Management Applications:*

As with the programming models, the increasing data volume is challenging legacy storage management systems. Using relational centralized database systems with big data will typically lead to several inefficiencies. This has encouraged the use of large-scale distributed cloud-based data storage management systems known as "key-value stores" or, "Not only SQL (NoSQL)". It is well known that no single computing system is capable of providing effective processing to all types of workloads. To design a generic platform that suits several types of requests and workloads, some trade-offs have be considered. A set of measures for these trade-offs for data storage management systems is defined and follows the Consistency, Availability, and Partition-tolerance (CAP) theorem which states that any distributed system can only satisfy two of these three properties. Consistency reflects the fact that all replicas of an entry must have the same value at all times, and that reading operations should return the latest values of that entry. Availability implies that the requested operations should be allowed always and performed promptly, while partition-tolerance indicates that the system can function if some parts of it are disconnected [313]. Most of the traditional Relational Data Base Management Systems (RDBMS), which utilize SQL for querying, are classified as "CA" which stands for consistency and availability. As partition-tolerance is a key requirement in cloud distributed infrastructures, NoSQL systems are classified as either "CP" where the availability is relaxed or "AP" where the consistency is relaxed and replaced by eventual, timeline, or session consistency.

For any transaction to be processed concurrently, RDBMS are required to provide the Atomicity, Consistency, Isolation, and Durability (ACID) guarantees. Most of RDBMS rely on expensive shared-memory or shared-disks hardware to ensure high performance [314]. On the other hand, NoSQL systems utilize commodity share-nothing hardware while ensuring scalability, fault-tolerance, and cost-effectiveness, traded for consistency. Thus, the ACID guarantees are typically relaxed or replaced by the Basically Available Soft-state Eventual consistency (BASE) guarantees. While RDBMS are considered mature and well-established, NoSQL systems still lack experienced programmers and most of these systems are still in pre-production phases [313].

Several big data storage management applications were developed for commercial internal production workloads operations such as BigTable [347], PNUTS [348], and DynamoDB [349]. Big Table is a Column oriented storage system developed at Google for managing structured large-scale data sets in petabyte scale. It only supports single row transactions and performs an atomic read-modify-write sequence that relies on a distributed locking system called Chubby. PNUTS is a massive-scale data base system designed at Yahoo to support their web-based applications, while DynamoDB is a highly scalable and available distributed key-value data store developed at Amazon to support their cloud-based applications [340]. Examples of open-sourced big data management systems include HBase, HadoopDB, and Cassandra [314]. HBase is a key-value column-oriented data base management system while HadoopDB is a hybrid system that combines the scalability of MapReduce with the performance guarantees of parallel databases. In HadoopDB, the queries are expressed in SQL but are executed in parallel through the MapReduce framework [313], [350]. Cassandra is a highly scalable eventually consistent, distributed structured key-value store developed at Facebook [351]. Examples of other systems in efforts to integrate indexing capabilities with MapReduce are Hadoop++ [352], and Hadoop

Aggressive Indexing Library (HAIL). Hadoop++ adds indexing and joining capabilities to Hadoop without changing its framework, while HAIL creates different clustered indexes for each data replica and supports multi-attribute querying [315].

*D. Distributed In-memory Processing:*

Distributed in-memory processing systems are widely preferred for iterative, interactive, and real-time big data applications due to the bottleneck of the relatively slow data materialization processes in existing disk-based systems [316]. Running these applications in legacy MapReduce systems requires repetitive materialization as map and reduce tasks have to be re-launched iteratively. This in turn leads to excessive un-utilized disk Input/Output (I/O), CPU, and network bandwidth resources [315]. In-memory systems use fast memory units such as Dynamic Random Access Memory (DRAM) or cache units to provide rapid access to data during the run-time to avoid the slow disk I/O operations. However, as memory units are volatile, in-memory systems are required to adopt advanced mechanisms to guarantee fault-tolerance and data durability. Examples of in-memory NoSQL databases are RAMCloud [353] and HANA [354]. RAMCloud aggregates DRAM resources from thousands of commodity servers to provide large-scale database system with low latency. Distributed cache systems can also enhance the performance of large-scale web applications. Examples of these systems are Redis, which is an in-memory data structure store, and Memcached which is a light-weight in-memory key-value object caching system with strict Least Recently Used (LRU) eviction mechanism [316].

To support in-memory interactive and iterative data processing, Spark which is a scalable data analytics platform written in Scala, was introduced [355]. In Spark, the data is stored as Resilient Distributed Data sets (RDD) [356] which are general purpose and fault-tolerant abstraction for data sharing in distributed computations. RDDs are created in the memory through course-grained deterministic transformations to datasets such as map, flatmap, filter, join, and GroupByKey. In cases of insufficient RAM, Spark performs lazy materialization of RDDs. RDDs are immutable and by applying further transformations, new RDDs are created. This provides fault-tolerance without the need for replication as the information about the transformations that created each RDD can be retrieved and reapplied to obtain the lost portions.

*E. Distributed Graph Processing:*

The partitioning and processing of graphs (i.e. a description of entities by vertices and their relationships by connecting edges) is considered a key class of big data applications especially for social networks that contain graphs with up to Billions of entities and edges [357]. Most of big data graph processing applications utilize in-memory systems due to the iterative nature of their algorithms. Pregel [358], developed at Google, targets the processing of massive graphs on distributed clusters of commodity machines by using the Bulk Synchronous Parallel (BSP)-based programming model. Giraph, which is the open-source implementation of Pregel, uses online hashing or range-based partitioning to dispatch sub-graphs to workers [39]. Trinity is a distributed graph engine that optimizes distributed memory usage and communication cost under the assumption that the whole graph is partitioned across a cloud memory [359]. Other examples of distributed graph processing applications are GraphLab [360] which utilizes asynchronous distributed shared memory, and PowerGraph [361], which focuses on efficient partitioning of graphs with power-law distributions.

*F. Big Data Streaming Applications:*

Applications related to e-commerce and social networking typically receive virtually unbounded data, also known as *streams* that are asynchronously generated by a huge number of sources or users at uncontrollable rates. Stream processing systems are required to provide low latency and efficient processing that copes with the arrival rate of the events within the streams. If they did not cope, the processing of some events is dropped leading to load shedding which is undesirable. Conceptually, a streaming systems model is composed of continuously arriving input data to almost static queries for processing, while the batch and RDBMS systems model is composed of different queries applied to static data [362]. Streams processing can be performed in batch systems, however this leads to several inefficiencies. Firstly, the implementation is not straightforward as it requires transforming the streams into partitions of batch data before processing. Secondly, considering relatively large partitions increases the processing latency while considering small partitions increases the overheads of segmenting, maintaining the inter-segment dependencies, in addition to fault-tolerance overheads caused by the frequent materialization.

Main Memory MapReduce (M3R) [363], MillWheel [364], Storm [346], Yahoo S4 [365], and Spark streaming [366] are examples of streams processing systems. M3R is introduced to provide high reliability for interactive and continuous queries over streamed data of terabytes in small clusters [363]. However, it does not provide adequate resilience guarantee as it caches the results totally in memory. MillWheel is a scalable fault-tolerant, low-latency stream processing engine developed at Google [364] that allows time-based aggregations and provides fine-grained check-pointing. Storm was developed at Twitter as a distributed and fault-tolerant platform for real-time processing of streams [346]. It utilized two primitives; *spout* and *bolts* to apply transformations to data streams, also named tuples, in a reliable and distributed manner. The spouts define the streams sources, while the bolts perform the required computations on the tuples, and emit the resultant modified tuples to other bolts. The computations in Storm are described as graphs where each node is either a spout or a bolt, and the vertices are the tuples routes. Storm relies on different grouping methods to specify the distribution of the tuples between the nodes. These include *shuffle* where streams are partitioned randomly, *field* where partitioning is performed according to a defined criteria, *all* where the streams are sent to all bolts, and *global* where all streams are copied to a single bolt.

Yahoo Simple Scalable Streaming System (S4) is a general-purpose distributed stream processing engine that provides low latency and fault-tolerance [365]. In S4, the computations are performed by identical Processing Elements (PEs) that are logically hosted in Processing Nodes (PNs). The coordination between PEs is performed by "ZooKeeper" which is an open-source cluster management application. The PEs either emit results to another PE or publish them, while PNs are responsible for listening to events, executing the required operations, dispatching events, and emitting output events. Spark streaming [366], which is extended from Spark, introduced a stream programming model named discretized streams (D-Streams) to provide consistency, fault tolerance, and efficient integration with batch systems. Two types of operations can be used which are transformation operations that produce new D-streams, and output operations that save resultant RDDs to HDFS. Spark Streaming allows users to seamlessly combine streaming, batch, and interactive queries. It also contains stateful operators such as windowing, incremental aggregation, and time-skewed joins.

### *G. The Lambda Architecture:*

The Lambda architecture is a hybrid big data processing system that provides concurrent arbitrary processing and querying for batch and real-time data in a single entity at the same time [367]. As depicted in Figure 6, it consists of three layers which are the batch layer, the speed layer, and the serving layer. The batch layer contains a batch system such as Hadoop or Spark to support the querying and processing of archived or stored data, while the speed layer contains a streaming system to provide low-latency processing to incoming input data in real-time. The input data is fed to both; the batch and speed layers in parallel to produce two sets of results that represent the batch and real-time view of the queries, respectively. The serving layer role is to combine both results and produce the finalized results. The Lambda architecture is typically integrated with general messaging systems such as Kafka to aggregate the queries of users. Kafka is a high performance, open-source messaging system developed at LinkedIn for the purpose of aggregating high-throughput log files [368]. Although the Lambda architecture introduces high level of flexibility by getting the benefits of both real-time, and batch systems, it lacks simplicity due to the need to maintain two systems. Moreover, the Lambda architecture encounters some limitations with event-oriented applications [367].

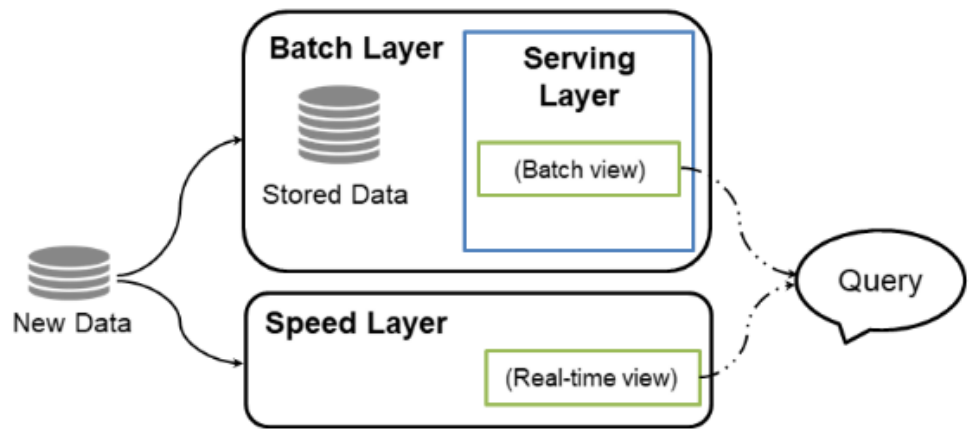

Fig. 6. Components of the Lambda architecture.

## III APPLICATIONS-FOCUSED OPTIMIZATION STUDIES

This Section discusses some big data applications optimization studies performed at the application-level. Optimizations at this level include tuning application parameters rather than using the default settings, or modifying the programming models themselves to obtain enhanced performance. The rest of this Section is organized as follows: Subsection III-A considers studies focusing on optimizing data and/or jobs and tasks placements. Subsection III-B summarizes studies that optimize jobs scheduling, while Subsection III-C describes and evaluates studies focusing on reducing the completion time of jobs. Finally, Subsection III-D provides a summary of big data benchmarking suites, traces, and simulators. Table I provides a summary of the studies presented in the first three Sections while focusing on their objectives, targeted platform, optimization tools, benchmarks used, and experimental setup and/or simulations environments.

*A. Optimized Jobs/Data Placements:*

The early work in [36] addressed the problem of optimizing data placement in Homogeneous Hadoop clusters. A dynamic algorithm that assigns data fragments according to nodes processing capacities was proposed to reduce data movement between over-utilized and under-utilized servers. However, the effects of replications were not considered. Typically, most of the nodes within a Hadoop cluster are kept active to provide high data availability which is energy inefficient. In [37], the energy consumption of Hadoop clusters was reduced by dynamically switching off unused nodes while ensuring that replicas of all data sets are contained in a covering subset which is kept always active. Although the energy consumption was reduced, the jobs running time was increased. Dynamic sizing and locality-aware scheduling in dedicated MapReduce clusters with three types of workloads namely; batch MapReduce jobs, web applications, and interactive MapReduce jobs, were addressed in [38]. A balance between performance and energy saving was achieved by allocating the optimum number of servers by delaying batch workloads while considering data locality and delay constraints of web and interactive workloads. A Markov Decision Process (MDP) model was considered to reflect the stochastic nature of jobs arrival and an event-driven simulator was used to empirically validate the proposed algorithms optimality. Energy savings between 30% and 59% were achieved over the no allocation strategy.

Graphs and iterative applications typically have highly dependent and skewed workloads as some portions of the datasets are more accessed for processing than others. The study in [39] proposed a Dependency Aware Locality for MapReduce (DALM) algorithm to process highly skewed and dependent input data. The algorithm was tested with Giraph and was found to reduce the cross-server traffic by up to 50%. Several studies addressed optimizing reduce tasks placement to minimize networking traffic by following greedy approaches. These approaches degrade the performance as they maximize the intermediate data locality for current reduce tasks without considering the effects on the input data locality for map tasks in following MapReduce jobs [40]. The theoretical and experimental study in [40] addressed optimizing reduce tasks data locality for sequential MapReduce jobs and achieved up to 20% improvement in performance compared to the greedy approaches.

The efficiency of indexing, grouping, and joint querying operations can be significantly affected by the used framework data placement strategy. In [41], a data block allocation approach was proposed to reduce job execution time and improve query performance in Hadoop clusters. The default random data placement policy of Hadoop 1.0.3 was modified by *using k-means clustering techniques* to co-allocate related data blocks on same clusters while considering the default replication factor. The results indicated a query performance improvement by up to 25%. The study in [42] addressed the poor resource utilization caused by misconfiguration in HBase. An algorithm named HConfig for semi-automating resources allocation and data bulks loading was proposed and an improvement between 2% and 3.7% was achieved compared to default settings.

Several recent studies considered jobs or data placement optimizations in Hadoop 2.x with YARN. The impact of data locality in YARN with delay scheduler on jobs completion time was addressed in [43]. To measure the data locality achieved, a YARN Location Simulator (YLocSim) tool was developed, validated, and compared with experimental results in a private cluster. Moreover, the effects of inherent resources allocation imbalance which increase with data locality, and the redundant I/O operations due to same data requests by different containers, were studied. Existing job schedulers ignore the partitioning skew caused by uneven shuffled intermediate data volumes to reducers. In [44], a framework to control resource allocation to reduce jobs, named Dynamic REsource Allocation technique for MapReduce with Partitioning Skew (DREAMS), was developed while considering data replication. Experimental results show an improvement by up to 2.29 times compared to native YARN. Inappropriate configurations in YARN lead to degraded performance as resource usage by tasks typically vary during the execution, while the resources initially assigned to them are fixed. In [45], JellyFish which is a self-tuning system based on YARN, was proposed to reduce the completion time and improve the utilization by considering elastic containers, performing online reconfigurations, and rescheduling for the idle resources. The study considered the most critical subset of YARN parameters and an average improvement of 65% was achieved compared to default YARN for repetitive jobs.

*B. Jobs Scheduling:*

Default scheduling mechanisms such as FIFO, capacity scheduler, and HFS, can lead to undesirable performance especially with mixtures of small interactive and long batch jobs [318]. Several studies suggested enhanced scheduling mechanisms to overcome the inefficiencies of default schedulers in terms of resource utilization, jobs makespan (i.e. the completion time of last task), or energy efficiency. The conflict between fairness in scheduling and data locality was addressed in [46]. To maximize data locality, some tasks assigned according to fairness scheduling were intentionally delayed until a cluster containing the corresponding data is available. An algorithm named delay scheduling was implemented on HFS and tested on commercial and private clusters based on Hive workloads from Facebook. It was found that only short delays were required to achieve almost 100% locality. The response times for small jobs were improved by about five times at the expense of moderately slowing down larger jobs. *Quincy*, a slot-based scheduler for Dryad, was implemented as a minimum-cost flow algorithm in [47] to achieve fairness and data locality for concurrent distributed jobs. The scheduling was implemented as a DAG, where the edges represent the competing demand of locality and fairness. Quincy provided better performance than queue-based schedulers and was capable of effectively reducing the network traffic.

In [48], a data replication-aware scheduling algorithm was proposed to reduce the network traffic and speculative executions. The proposed map jobs scheduler was implemented in two waves. Firstly, the empty slots in the cluster were filled based on the number of hosted maps and the replication schemes. Secondly, a run-time scheduler was utilized to increase the locality and balance the intermediate data distribution for the shuffling phase. A map scheduler that balances locality and load balancing was proposed in [49]. In this work, the computing cluster is modeled as a time-slotted system where job arrivals are modeled as Bernoulli random variables and the service time is modelled as a geometric random variable with different mean values based on data locality. The scheduler utilized 'Join the Shortest Queue' (JSQ) and the maximum weight policies and was proven to be throughput and delay optimal under heavy traffic conditions.

In [50], a Resource-aware Adaptive Scheduling (RAS) mechanism was proposed to improve resource utilization and achieve user-defined completion time goals. RAS utilizes offline job profiling information to enable dynamic adjustments for the number of slots in each machine while meeting the different requirements of the map and reduce phases. FLEX, is a flexible scheduling allocation scheme introduced in [51] to optimize the response time, makespan, and SLA. FLEX ensures minimum slot guarantees for jobs as in fair scheduler and introduces maximum slots guarantees. Experimental results showed that FLEX outperformed the fair scheduler by up to 30% in terms of the response time. The studies in [52], [53] scheduled MapReduce jobs according to apriori known job sizes information to optimize Hadoop clusters performance. HFSP in [52] focused on resource allocation fairness and the reduction of the response time among concurrent interactive and batch jobs by using the concepts of ageing function and virtual time. LsPS in [53] is a self-tuning two-tier scheduler, the first scheduler is used among multiple users and the second scheduler is for each user. It aimed to improve the average response time by adaptively selecting between fair and FIFO schedulers. Prior job information was also utilized in [54] for offline MapReduce scheduling while taking server assignments into consideration.

The authors in [55] reported that shuffling in MapReduce accounted for about a third of the overall jobs completion time. A joint scheduling mechanism for the map, shuffle, and reduce phases that considered their dependencies was implemented via linear programs and heuristics with precedence constraints. The study in [56] focused on improving the CPU utilization by dynamically scheduling waiting map tasks to nodes with high I/O waiting time. The execution time of I/O intensive jobs with the proposed scheduler was improved by 23% compared to FIFO. To reduce the energy consumption of performing MapReduce jobs, the work in [57] proposed an energy-aware scheduling mechanism while considering the Dynamic Voltage Frequency Scaling (DVFS) settings for the cluster. DVFS was experimentally utilized in [58] to reduce the energy consumption of computation extensive workloads. In [59], two heuristics namely EMRSA-I, and EMRSA-II were developed to assign map and reduce tasks to slots with the goal of reducing the energy consumption while satisfying SLAs. The energy consumption was reduced by about 40% compared to default schedulers at the expense of increased jobs makespan. In [60], a dynamic slot allocation framework; DynamicMR was proposed to improve the efficiency of Hadoop. Dynamic Hadoop Slot Allocation (DHSA) algorithm was utilized to relax the strict slot allocation to either map or reduce tasks by enabling their reallocation to achieve a dynamic map to reduce ratio. DynamicMR outperformed YARN by 2% - 9%, and considerably reduced the network contention caused by the high number of over-utilized reduce tasks. Unlike scheduling at the task-level, the work in [61] introduced PRISM which is a fine-grained resource aware scheduler. PRISM, which divides tasks into phases each with its own resources usage profile, provided 1.3 times reduction in the running time compared to default schedulers.

Several recent studies considered optimizing MapReduce scheduling under YARN. HaSTE in [62] utilized the fine-grained CPU and RAM resources management capabilities of YARN to schedule map and reduce tasks under dependency constrains without the need for prior knowledge of tasks execution times. HaSTE improved the resources utilization and the makespan even for mixed workloads. The study in [63] suggested an SLA-aware energy-efficient scheduling scheme based on DVFS for Hadoop with YARN. The scheme, which was applied to the per-application AM, reduced the CPU frequency in the cases of tasks finishing before their expected completion time. In [64], a priority-based resource scheduler for a streaming system named Quasit was proposed. The scheduler dynamically modifies the data paths of lower priority tuples to allow faster processing for higher priority tuples for vehicular traffic streams.

*C. Completion Time:*

The reduction of jobs overall completion time was considered in several studies with the aim of improving the SLA or reducing the power consumption in underlying clusters. Numerical evaluations were utilized in [65] to test two power efficient resource allocation approaches in a pool of MapReduce clusters. The algorithms developed aimed to reduce the end-to-end delay or the energy consumption while considering the availability as an SLA metric. In an effort to provide predictable services to deadline-constrained jobs, the work in [66] suggested a Resource and Deadline-aware Hadoop scheduler (RDS). RDS is based on online optimization and a self-learning completion time estimator for future tasks which makes it suitable for dynamic Hadoop clusters with mixture of energy sources and workloads. The work in [67] focused on reducing the completion time of small jobs that account for the majority of the jobs in production Hadoop clusters. The proposed scheduler; Fair4S achieved an improvement by a factor of 7 compared to fair scheduler where 80% of small jobs waited less than 4 seconds to be served. In [68], a Bipartite Graph MapReduce scheduler (BGMRS) was proposed for deadline-constrained MapReduce jobs in clusters with heterogeneous nodes and dynamic job execution time. By optimizing scheduling and resources allocations, BGMRS reduced the deadline miss ratio by 79% and the completion time by 36% compared to fair scheduler.

The authors in [69] developed an Automatic Resource Inference and Allocation (ARIA) framework based on jobs profiling to reduce the completion time of MapReduce jobs in shared clusters. A Service Level Objective (SLO) scheduler was developed to utilize the predicted completion times and determine the schedule of resources allocation for tasks to meet soft deadlines. The study in [70] proposed a Dynamic Priority Multi-Queue Scheduler (DPMQS) to reduce the completion time of map tasks in heterogeneous environments. DPMQS increased both the data locality and the priority of map jobs that are near to completion. Optimizing the scheduling of mixed MapReduce-like workloads was considered in [71] through offline and online algorithms that determine the order of tasks that minimize the weighted sum of the completion time. The authors in [72] also emphasized the role of optimizing jobs ordering and slots configurations in reducing the total completion time for offline jobs and proposed algorithms that can improve non-optimized Hadoop by up to 80%. The work in [73] considered optimizing four NoSQL databases (i.e. HBase, Cassandra, and Hive) by reducing the Waiting Energy Consumption (WEC) caused by idle nodes waiting for job assignments, I/O operations, or results from other nodes. RoPE was proposed in [73] to reduce the response time of relational queries performed as cascaded MapReduce jobs in SCOPE which is a parallel processing engine used by Microsoft. A profiler for code and data properties was used to improve future invocations of the same queries. RoPE achieved 2× improvements in the response time for 95% of Bing's production jobs while using 1.5× less resources.

Jobs failures lead to huge increase in their completion time. To improve the performance of Hadoop under failures, a modified MapReduce work flow with fine-grained fault-tolerance mechanism called BEneath the Task Level (BeTL) was proposed in [75]. BeTL allows generating more files during the shuffling to create more checkpoints. It improved the performance of Hadoop under no failures by 6.6% and under failures by up to 51%. The work in [76] proposed four multi-queue size-based scheduling policies to reduce jobs slowdown variability which is defined as the idle time to wait for resources or I/O operations. Several factors such as parameters sensitivity, load unbalance, heavy-traffic, and fairness were considered. The work in [77] optimized the number of reduce tasks, their configurations, and memory allocations based on profiling the intermediate results size. The results indicated a complete disregard for job failures due to insufficient memory and a reduction in the completion time by up to 88.79% compared to legacy memory allocation approaches. To improve the performance of MapReduce in memory-constrained systems, Mammoth was proposed in [78] to provide global memory management. In Mammoth, related map and reduce tasks were launched in a single Java Virtual Machine (JVM) as threads that share the memory at run time. Mammoth actively pushes intermediate results to reducers unlike Hadoop that passively pulls from disks. A rule-based heuristic was used to prioritize memory allocations and revocations among map, shuffle, and reduce operations. Mammoth was found to be 5.19 times faster than Hadoop 1 and to outperform Spark for interactive and iterative jobs when the memory is insufficient [78]. An automatic skew mitigation approach; SkewTune was proposed and optimized in [79]. SkewTune detects different types of skew, and effectively re-partitions the unprocessed data of the stragglers to process them in idle nodes. The results indicated a reduction by a factor of 4 in the completion time for workloads with skew and minimal overhead for workloads without skew.

TABLE I
SUMMARY OF APPLICATIONS-FOCUSED OPTIMIZATIONS STUDIES

| Ref | Objective | Application | Tools | Benchmarks/workloads | Experimental Setup/Simulation environment |
|---|---|---|---|---|---|
| [36]* | Optimize data placement in heterogeneous clusters | Hadoop 1.x | Dynamic algorithm | Grep, WordCount | 5 heterogeneous nodes with Intel's (Core 2 Duo, Celeron, Pentium 3) |
| [37]* | Reduce energy consumption by scaling down unutilized clusters | Hadoop 0.20.0 | Modification to Hadoop and defining covering subset | Webdata_sort, data_scan from Gridmix benchmark (16-128 GB) | 36 nodes (8 CPU cores, 32GB RAM, Gigabit NIC, two disks), 48-port HP ProCurve 2810-48G switch |
| [38]* | Balance energy consumption and performance | - | locality-aware scheduler via Markov Decision Process | Geometric distribution for tasks arrival rate | Event-driven simulations |
| [39]* | Dependency-Aware Locality for MapReduce (DALM) | Hadoop 1.2.1, Giraph 1.0.0 | Modification to HDFS replication scheme and scheduling algorithm | 3.5 GB Graph from Wikimedia database, 2.1 GB of public social networks data | 4 nodes (Intel i7 3.4 GHz Quad, 16GB DDR3 RAM, 1 TB disk, 1 Gbps NIC, NETGEAR 8-port switch |
| [40]* | Optimizing reduce task locality for sequential MapReduce jobs | Hadoop 1.x | Classical stochastic sequential assignment | Grep (10 GB), Sort (15 GB, 4.3 GB) | 7 slave nodes (4 cores 2.933 GHz CPU, 32KB cache, 6GB RAM, 72GB disk) |
| [41]* | Optimizing data placements for query operations | Hadoop 1.0.3, with Hive 0.10.0 | k-means clustering for data placement, HDFS extension to support customized data placement | 920 GB business ad-hoc queries from TPC-H | 10 nodes; 1 NameNode (6 2.6 GHz CPU cores, 16GB RAM, 1TB SATA), 9 DataNodes (Intel i5, 4GB RAM, 300GB disk), 1 Gbps Ethernet |
| [42]* | HConfig: Configuring data loading in HBase clusters | HBase 0.96.2, Hadoop 2.2.0 | Algorithms to semi-automate data loading and resource allocation | YCSB Benchmark | 13 nodes (1 manager, 3 coordinators, 9 workers), and 40 nodes (1 manager, 3 coordinators, 36 workers) with (AMD Opeteron CPU, 8GB RAM, 2 SATA 1TB disks), 1 Gigabit Ethernet |
| [43]* | Impact of data locality in YARN on completion time and I/O operations | Hadoop 2.3.0, Sqoop, Pig, Hive | Modify Rumen and YARN Scheduler Load Simulator (SLS) to report data locality | 8GB of synthetic text for Grep, WordCount, 10GB TPC-H queries | One node (4 CPU cores, 16GB RAM), 16 nodes (2 CPU cores, 8GB RAM), 1 Gigabit Ethernet, YARN Location Simulator (YLocSim) |
| [44]* | DREAMS: dynamic reduce tasks resources allocation | YARN 2.4.0 | Additional feature in YARN | WC, Inverted Index, k-means, classification, DataJoin, Sort, Histo-movies (5GB, 27GB) | 21 Xen-based virtual machines (4 2GHz CPU cores, 8GB RAM, 80GB disk) |
| [45]* | JellyFish: Self tuning system based on YARN | Hadoop 2.x | Parameters tuning, resources rescheduling via elastic containers | PUMA benchmark (TeraSort, WordCount, Grep, Inverted Index) | 4 nodes (dual Intel 6 cores Xeon CPU, 15MB L3 cache, 7GB RAM, 320GB disk) Gigabit Ethernet |
| [46]° | Delay scheduling: balance fairness and data locality | Hadoop 0.20 | Algorithms implemented in HDFS | Facebook traces (text search, simple filtering selection, aggregation, join) | 100 nodes in Amazon EC2 (4 2 GHz core, 4 disks, 15GB RAM, 1 Gbps links), 100 nodes (8 CPU cores, 4 disks), 1 Gbps Ethernet |
| [47]° | Quincy: Fair Scheduling with locality and fairness | Dryad | Graph-based algorithms | Sort(40,160,320) GB, Join (11.8, 41.8) GB, PageRank(240) GB, WordCount (0.1-5) GB | 243 nodes in 8 racks (16GB RAM, 2 2.6 GHz dual core AMD) 48-port Gigabit Ethernet per rack |
| [48]° | Maestro: Replica-aware Scheduling | Hadoop 0.19.0, Hadoop 0.21.0 | Heuristic | GridMix (Sort, WordCount) (2.5,1,12.5,200) GB | 20 virtual nodes in local cluster, 100 Grid5000 nodes (2 GHz dual core AMD, 2GB RAM, 80GB disk) |
| [49]° | Map task scheduling to balance data locality and load balancing | - | Join Shortest Queue, Max Weight Scheduler in heavy traffic queues | Parameters from search jobs in databases | Simulations for a 400 machines cluster |
| [50]° | Resource-aware Adaptive Scheduling (RAS) | Hadoop 0.23 | Scheduler and job profiler | Gridmix Benchmark (Sort,Combine,Select) | 22 nodes (64-bit 2.8 GHz Intel Xeon, 2GB RAM), Gigabit Ethernet |
| [51]° | FLEX: flexible allocation scheduling scheme | Hadoop 0.20.0 | Standalone plug-in or add-on module in FAIR | 1.8 TB synthesized data, and GridMix2 | 26 nodes (3 GHz Intel Xeon, 13 4-core blades in one rack and 13 8-core blades in second rack |
| [52]° | HFSP: size-based scheduling | Pig, Hadoop 1.x | Job size estimator and algorithm | PigMix (1,10,100 GB and 1TB) | 20 workers with TaskTracker (4 CPU cores, 8GB RAM) |
| [53]° | LsPS: self-tuning two-tiers scheduler | Hadoop 1.x | Plug-in scheduler | WordCount, Grep, PiEstimator, Sort | EC2 m1.large (7.5GB RAM, 850GB disk), 11 nodes (1 master, 10 slaves), trace-driven simulations |
| [54]° | Joint scheduling of MapReduce in servers | Hadoop 1.2.0 | 3-approximation algorithm, heuristic | WordCount (43.7 GB Wikipedia document package) | 16 nodes EC2 VM (1 GHz CPU, 1.7GB memory, 160GB disk) |
| [55]° | Joint Scheduling of processing and shuffle | - | Linear programming, heuristics | Synthesized workloads | Event-based simulations |
| [56]° | Dynamic slot scheduling for I/O intensive jobs | Hadoop 1.0.3 | Dynamic algorithm based on I/O and CPU statistics | Sort (3,6,9,12,15) GB | 2 masters (12 CPU cores 1.9 GHz AMD, 32GB RAM), 4,8,16 slaves (Intel Core i5 1.9 GHz, 8GB RAM) |
| [57]° | Energy-efficient Scheduling | - | Polynomial time constant-factor approximation algorithm | Synthesized workloads | MATLAB simulations |
| [58]° | Energy efficiency for comp-utation intensive workloads | Hadoop 0.20.0 | DVFS control in source code and via external scheduler | Sort, CloudBurst, Matrix Multiplication | 8 nodes (AMD Opteron quad-core 2380 with DVFS support, 2 64 kB L1 caches) Gigabit Ethernet |
| [59]° | Energy-aware Scheduling | Hadoop 0.19.1 | Scheduling algorithms EMRSA-I, EMRSA-II | TeraSort, Page Rank, k-means | 2 nodes (24GB RAM,16 2.4 GHz CPU cores, 1TB Disk), 2 nodes (16GB RAM,16 2.4 GHz CPU cores, 1TB Disk) |
| [60]° | DynamicMR: slot allocation in shared clusters | Hadoop 1.2.1 | Algorithms: PI-DHSA, PD-DHSA, SEPB, slot pre-scheduling | PUMA benchmark | 10 nodes (Intel's X5675, 3.07 GHz, 24GB RAM, 56GB disk) |
| [61]° | PRISM: Fine-grained resource aware scheduling | Hadoop 0.20.2 | Phase-level scheduling algorithm | Gridmix 2, PUMA benchmark | 10 nodes; 1 master, 15 slaves (Quad-core Xeon E5606, 8GB RAM, 100GB disk), Gigabit Ethernet |
| [62]° | HaSTE: scheduling in YARN to reduce makespan | Hadoop YARN 2.2.0 | dynamic programming algorithm | WordCount (14,10.5) GB, Terasort 15GB, WordMean 10.5GB, PiEstimate | 8 nodes (8 CPU cores, 8GB RAM) |
| [63]° | SLA-aware energy efficient scheduling in Hadoop YARN | - | DVFS in the per-applications master | Sort, Matrix Multiplications | CloudSim-based simulations for 30 servers with network speeds between 100-200 Mbps |
| [64]° | Priority-based resource scheduling for Distributed Stream Processing Systems | Quasit | Meta-scheduler | 40 million tuples of realistic 3 hours vehicular traffic traces | 5 nodes; 1 master, 4 workers (AMD Athlon64 3800+, 2GB RAM) |
| [65]† | Power-efficient resources allocation and mean end-to-end delay minimization | - | Algorithms | - | Arena simulator |
| [66]† | Resource and Deadline-aware scheduling in dynamic Hadoop clusters | Hadoop 1.2.1 | Receding horizon control algorithm, self-learning completion time estimator | PUMA benchmark (WordCount, TeraSort, Grep) | 21 Virtual Machines, 1 master and 20 slaves (1 CPU core, 2GB RAM) |
| [67]† | Fair4S: Modified Fair Scheduler | Hadoop 0.19 | Scheduler and modi-fication to JobTracker | Synthesized workloads generated by Ankus | Discrete event-based MapReduce simulator |
| [68]† | Deadline-Constrained MapReduce Scheduling | Hadoop 1.2.1 | Bipartite Graph Modelling to perform BGMRS scheduler | WordCount, Sort, Grep | 20 virtual machines in 4 physical machines (Quad core 3.3 GHz, 32 GB RAM, 1TB disk), Gigabit Ethernet. MATLAB simulations for 3500 nodes |
| [69]† | ARIA (Automatic Resource Inference and Allocation) | Hadoop 0.20.2 | Service level Objective scheduler to meet soft deadlines | WordCount, Sort, Bayesian classification, TF-IDF from Mahout, WikiTrends, twitter workload | 66 nodes (4 AMD CPU cores, 8GB RAM, 2 160GB disks) in two racks, Gigabit Ethernet |
| [70]† | Scheduling for improved response time | Hadoop 1.2.4 | Dynamic Priority multi-queue scheduler as a plug-in | Text search, Sort, WordCount, Page Rank | Cluster-1: 6 nodes (dual core CPU 3.2 GHz, 2GB RAM, 250GB disk), Cluster-2: 10 nodes (dual core CPU 3.2 GHz, 2GB RAM, 250GB disk), Gigabit Ethernet |
| [71]† | Scheduling for fast completion in MapReduce-like systems | - | 3-approximation algorithms; OFFA, Online heuristic ONA | Synthesized workload | Simulations |
| [72]† | Dynamic Job Ordering and slot configuration | Hadoop 1.x | Greedy algorithms based on Johnson's rule for 2-stage flow shop | PUMA Benchmark (WordCount, Sort, Grep), synthesized Facebook traces | 20 nodes EC2 Extra Large instances (4 CPU cores, 15GB RAM, 4 420GB disks) |
| [73]† | Reduction of idle energy consumption due to waiting in NoSQL | HDFS 0.20.2, HBase 0.90.3, Hive 0.71, Cassandra 1.0.3, HadoopDB 0.1.1 | Energy consumption model | Loading, Grep, Selection, Aggregation, join | 12 nodes ( Intel i5-2300, 8GB RAM, 1TB disk) |
| [74]† | RoPE: reduce response time of relational querying | SCOPE | Query optimizer that piggyback job execution | 80 jobs from major business groups | Bing's Production cluster (tens of thousands of 64 bit, multi-core, commodity servers) |
| [75]† | BEneath the Task Level (BeTL) fine-grained checkpoint strategy | Hadoop 2.2.0 | Algorithms to modify MapReduce workflow | Hibench Benchmark (WordCount, Hive queries) | 16 nodes in Windows Azure (1 CPU core, 1.75GB RAM, 60GB disk), Gigabit Ethernet |
| [76]† | Job Slowdown Variability reduction | Hadoop 1.0.0 | Four algorithms; FBQ, TAGS, SITA, COMP | SWIM traces, Grep, sort, WordCount | 6 nodes (24 dual-quad core, 24GB RAM, 50TB storage), 1 Gigabit Ethernet and 20 Gbit/s InfiBand, Mumak simulations |
| [77]† | Memory-aware reduce tasks configurations of Reduce tasks | Hadoop 1.2.1 | Mnemonic mechanism to determine number | PUMA benchmarks (InvertedIndex, Ranked InvertedIndex, Self Join, Sequence Count, WordCount) | 1 node (4 CPU cores, 7GB RAM) |
| [78]† | SkewTune: mitigating skew in MapReduce applications | Hadoop 0.21.1 | Modification to job tracker and task tracker | Inverted Index, PageRank, CloudBurst | 20 nodes cluster (2 GHz quad-core CPU, 16GB RAM, 2 750GB disks) |
| [79] † | Mammoth: global memory management | Hadoop 1.0.1 | Memory management via the public pool | WordCount, Sort, WordCount with Combiner | 17 nodes (2 8 CPU cores 2.6 GHz Intel Xeon E5-2670, 32GB memory, 300GB SAS disk) |

*Jobs/data placement, °Jobs Scheduling, †Completion time.

*D. Benchmarking Suites, Production Traces, Modelling, Profiling Techniques, and Simulators for Big Data Applications:*

***1) Benchmarking Suites:***

Understanding the complex characteristics of big data workloads is an essential step toward optimizing the configurations for the frameworks parameters used and identifying the sources of bottlenecks in the underlying clusters. As with many legacy applications, MapReduce and other big data frameworks are supported by several standard benchmarking suites such as [80]-[88]. These benchmarks have been widely utilized to evaluate the performance of big data applications in different infrastructures either experimentally, analytically or via simulations as summarized for the optimization studies in Tables I, III, V, and VI. Moreover, they can be used in production environments for initial tuning and debugging purposes, in addition to stress-testing and bottlenecks analysis before the actual run of intended commercial services.

The workloads contained in these benchmarks are typically described by a semantic that run on previously collected or randomly generated datasets. Examples are text retrieval-based (e.g. Word-Count (WC), Word-Count with Combiner (WCC), and Sort), and web search-based (e.g. Grep, Inverted Index, and Page Rank) workloads. WC, and WCC calculate the occurrences of each word in large distributed documents. WC and WCC differ in that the reduction is performed totally at the reduce stage in WC, and is done partially at the map stage with the aid of a combiner at the map stage in WCC. Sort generates alphabetically sorted output from input documents. Grep finds the match of regular expressions (regex) in input files, while Inverted Index generates a word-to-document indexing for a list of documents. PageRank is a link analysis algorithm that measures the popularity of web pages based on their referral by other websites. In addition to the above examples, computations related to graphs and to machine learning such as *k*-means clustering are also considered in benchmarking big data applications. These workloads vary in being I/O, memory, or CPU intensive. For example, PageRank, Grep, and sort are I/O intensive, while WC, Page rank, and *k*-means are CPU intensive [8]. This should be considered in optimization studies to correctly address bottlenecks and targeted resources. For a detailed review of benchmarking different specialized big data systems, the reader is referred to the survey in [317], where workloads generation techniques, input data generation, and assessment metrics are extensively reviewed. Here, we briefly describe some examples of the widely used big data benchmarks and their workloads as summarized in Table II and listed below:

1) GridMix [80]: GridMix is the standard benchmark included within the Hadoop distribution. GridMix, has three versions and provides a mix of workloads synthesized from traces generated by Rumen [369] from Hadoop clusters.
2) HiBench[1] [81]: HiBench is a comprehensive benchmark suite provided by Intel for big data. HiBench provides a wide range of workloads to evaluate computations speed, systems throughput, and resources utilization.
3) HcBench [82]: HcBench is a Hadoop benchmark provided by Intel that includes workloads with a mix of CPU, storage, and network intensive jobs with Gamma inter-job arrival times for realistic clusters evaluations.
4) PUMA [83]: PUMA is a MapReduce benchmark developed at Purdue University that contains workloads with different computational and networking demands.
5) Hive Benchmark [84]: The Hive benchmark contains 4 queries and targets comparing Hadoop with Pig.
6) PigMix [85]: PigMix is a benchmark that contains 12 queries types to test the latency and scalability performance of Apache Pig and MapReduce.
7) BigBench [86]: BigBench is an industrial-based benchmark that provides quires with structures, semi-structured and unstructured data.
8) TPC-H [87]: The TPC-H benchmark, provided by the Transaction Processing Performance Council (TPC), allows generating realistic datasets and performing several business oriented ad-hoc queries. Thus, it can be used to evaluate NoSQL and RDBMS scalability, processing power, and throughput.
9) YCSB[2] [88]: The Yahoo Cloud Serving Benchmark (YCSB) targets testing the inserting, reading, updating and scanning operations in database-like systems. YCSB contains 20 records data sets and provides a tool to create the workloads.

10) TABLE II

11) EXAMPLES OF BENCHMARKING SUITES FOR BIG DATA APPLICATIONS AND THEIR WORKLOADS.

| Benchmark | Application | Workloads |
|---|---|---|
| GridMix [80] | Hadoop | Synthetic loadjob, Synthetic sleepjob |
| HiBench [81] | Hadoop, SQL, Kafka, Spark Streaming | Micro Benchmarks (Sort, WordCount, TeraSort, Sleep, enhanced DFSIO to test HDFS throughput), *Machine Learning* (*Bayesian Classification, k-means, Logistic Regression, Alternating Least Squares, Gradient Boosting Trees, Linear Regression, Latent Dirichlet Allocation, Principal Components Analysis, Random Forest, Support Vector Machine, Singular Value Decomposition*), SQL(Scan, Join, Aggregate), Websearch Benchmarks (PageRank, Nutch indexing), Graph Benchmark (NWeight), Streaming Benchmarks (Identity, Repartition, Stateful WC, Fixwindow) |
| HcBench [82] | Hadoop, Hive, Mahout | Telco-CDR (Call Data Records) interactive queries, Hive workloads (PageRank URLs, aggregates by source, average of PageRank), k-means Clustering iterative jobs in machine learning, and Terasort |
| PUMA [83] | Hadoop | Micro Benchmarks (Grep, WordCount, TeraSort), term vector, inverted index, self-join, adjacency-list, k-means, classification, histogram, histogram ratings, sequence count, ranked inverted index |
| Hive [84] | Hive | Grep selection, Ranking selection, user visits aggregation, user visits join |
| PigMix [85] | Pig | Explode, fr join, join, distinct agg, anti-join, large group by key, nested split, group all, order by 1 field, order by multiple fields, distinct + union, multi-store |
| BigBench [86] | RDBMS, NoSQL | Business queries (cross-selling, customer micro-segmentation, sentiment analysis, enhancing multi-channel customer experience, assortment and pricing optimization, performance transparency, return analysis, inventory management, price comparison) |
| TPC-H [87] | RDBMS, NoSQL | Ad-hoc queries for New Customer Web Service, Change Payment Method, Create Order, Shipping, Stock Management Process, Order Status, New Products Web Service Interaction, Product Detail, Change Item |
| YCSB [88] | Cassandra,HBase, Yahoo's PNUTS | Update heavy, Read heavy, Read only, Read latest, Short ranges |

### *2) Production Traces:*

[1] Available at: https://github.com/intel-hadoop/hibench
[2] Available at: https://github.com/brianfrankcooper/YCSB

As the above-mentioned benchmarks are defined by their semantics where the codes functionalities are known and the jobs are submitted by a single user to run on deterministic datasets, they might be incapable of fully representing production environments workloads where realistic mixtures of workloads with different data sizes and inter-arrival times for multi users coexist. Information about such realistic workloads can be provided in the form of traces collected from previously submitted jobs in production clusters. Although sharing such information for production environments is hindered by confidentiality, legal and business restrictions, several companies have provided archived jobs traces while normalizing the resources usage. Examples of realistic evaluations based on publicly available production traces were conducted by Yahoo [89], Facebook [90], [91], Cloudera and Facebook [92], Google [93], [94], IBM [95], and in clusters with scientific [96], or business-critical workloads [97]. These traces describe various job features such as number of tasks, data characteristics (i.e. input, shuffle and output data sizes and their ratios), completion time, in addition to jobs inter arrival times and resources usage without revealing information about the semantics or users sensitive data. Then, synthesized workloads can be generated by running dummy codes to artificially generate shuffle and outputs data sizes that match the trace on randomly generated data. To accurately capture the characteristics in the trace while running in testing clusters with a scale smaller than production clusters, effective sampling should be performed.

Traces based on ten month log files for data intensive MapReduce jobs running in the M45 supercomputing production Hadoop cluster for Yahoo were characterized in [89] according to utilization, job patterns, and sources of failures. Facebook traces collected from a 600 nodes for 6 months and Yahoo traces were utilized in [90] to provide comparisons and insights for MapReduce jobs in production clusters. Both traces were classified by *k*-means clustering according to the number of jobs, input, shuffle, and output data sizes, map and reduce tasks, and jobs durations into 10 and 8 bins for Facebook and yahoo traces, respectively. Both traces indicated input data sizes ranging between kBytes and TBytes. The workloads are labeled as small jobs which constitute most of the jobs, load jobs with only map tasks, in addition to expand, aggregate, and transformation jobs based on input, shuffle, and output data sizes. A framework that properly sample the traces to synthesize representative and a scaled down workloads for use in smaller clusters is proposed and sleep requests in Hadoop to emulate the inter-arrivals of jobs were utilized. Facebook traces from 3000 machines with total data size of 12TBytes for MapReduce Workloads with Significant Interactive Analysis were utilized in [91] to evaluate the energy efficiency of allocating more jobs to idle nodes in interactive job clusters. Six Cloudera traces from e-commerce, telecommunications, media, and retail users' workloads and Facebook traces collected over a year for 2 Million jobs were analyzed in [92]. Several insights for production jobs such as the weekly time series, and the burstiness of submissions were provided. These traces are available in a public repository which also contains a synthesizing tool; SWIM[3] which is integrated with Hadoop. Although previously-mentioned traces contain rich information about various jobs characteristics, the lack of per job resources utilization information makes them partially representative for estimating workloads resources demands.

Google workloads traces were characterized in [93] and [94] using *k*-means based on their duration and CPU and memory resources usage per task to aid in capacity planning, forecasting demands growth, and for improving tasks scheduling. Insights such as "most tasks are short", "duration of long and short tasks follows a Bimodal distribution", and that "most of the resources are taken by few extensive jobs" were provided. The traces[4] were collected from 12k machines cluster in 2011 and include information about scheduling requests, taken actions, tasks submission times and normalized resources usage, and machines availability [94]. However, disk and networking resources were not covered. The traces collected from IBM-based private clouds for banking, communication, e-business, production, and telecommunication industries in [95] further considered disk and file system usage in addition to CPU and memory. The inter-dependencies between the resources were measured and disk and memory resources were found to be negatively correlated indicating the potential benefit of co-locating memory intensive and disk intensive tasks.

A user-centric study was conducted in [96] based on traces[5] collected from three research clusters; OPENCLOUD, M45, and WEB MINING. Workloads, configurations, in addition to resources usage and sharing information were used to address the gaps between data scientists needs and systems design. Evaluations for preferred applications and the penalties of using default parameters were provided. Two large-scale and long term traces[6] collected from distributed data centers for business-critical workloads such as financial simulators were utilized in [97] to provide basic statistics, correlations, and time-pattern analysis. Full characteristics for the provisioned and actual usage of CPU, memory, disk I/O, and network I/O throughput resources were presented. However, no information about the inter arrival times were provided. Ankus MapReduce workload synthesizer was developed as part of the study in [67] based on e-commerce traces collected from Taobao which is a 2k nodes production Hadoop cluster. The inter arrival times in the traces were found to be Poisson.

---

[3] Available at: https://github.com/SWIMprojectucb/swim/wiki
[4] Available at: https://github.com/google/cluster-data
[5] Available at: http://www.pdl.cmu.edu/HLA/

***3) Modelling and Profiling Techniques:***

Statistical-based characterization and modelling were considered for big data workloads based on production clusters traces as in [98] or benchmarks as in [99], and [101]. Also, different profiling and workloads modelling studies such as in [102]-[105] were conducted with the aim of automating clusters configurations or estimating different performance metrics such as the completion time based on the selected configurations and resources availability. A statistical-driven workloads generator was developed in [98] to evaluate the energy efficiency of MapReduce clusters under different scales, configurations, and scheduling policies. A framework to ease publishing production traces anonymously based on inter-arrival times, jobs mixes, resources usage, idle time, and data sizes was also proposed. The framework provide non-parametric statistics such as the averages and standard deviations and 5 numbers percentile summaries ($1^{st}$, $25^{th}$ , $50^{th}$, $75^{th}$, and $99^{th}$) for inter-arrival times and data sizes. Statistical modelling for GridMix, Hive, and HiBench workloads based on principal component analysis for 45 metrics and regression models was performed in [99] to provide performance estimations for Hadoop clusters under different workloads and configurations. BigDataBench[7] was utilized in [100] to examine the performance of 11 representative big data workloads in modern superscale out-of-order processors. Correlation analysis was performed to identify the key factors that affect the Cycles Per Instruction (CPI) count for each workload. Keddah[8] in [101] was proposed as a toolchain to profile based on end host or switches traffic measurements, empirically characterize, and reproduce Hadoop traffic. Flow-level traffic models can be derived by Keddah for use with network simulators with varied settings that affect networking requirements such as replication factor, cluster size, split size, and number of reducers. Results based on TeraSort, PageRank and *k*-means workloads in dedicated and cloud-based clusters indicated high correlation with Keddah-based simulation results.

To automate Hadoop clusters configurations, the authors in [102], [103] developed a cost-based optimization that utilize an online *profiler* and a *What-if Engine*. The *profiler* uses the *BTrace* Java-based dynamic instrumentation tool to collect job profiles at run time for the data flows (i.e. input data and shuffle volumes) and to calculate the cost based on the program, input data, resources, and configuration parameters at tasks granularity. The *What-if Engine* contains a cost-based optimizer that utilizes a task scheduler simulator and model-based optimization to estimate the costs if different combination of cost variables are used. This suits just-in-time configurations for the computations that run periodically in production clusters, where the profiler can trial the execution on a small set of the cluster to obtain the cost function, then the *What-if Engine* obtains the best configurations to be used for the rest of the workload. In [104], a Hadoop performance model is introduced. It estimates completion time and resources usage based on Locally Weighted Linear Regression and Langrage Multiplier, respectively. The non-overlapped and over-lapped phases of shuffling and the number of reduce waves were carefully considered. Polynomial regression is used in [105] to estimate the CPU usage in clocks per cycle for MapReduce jobs based on the number of map and reduce tasks.

***4) Simulators:***

Tuning big data applications in clusters requires time consuming and error-prone evaluations for a wide range of configurations and parameters. Moreover, the availability of a dedicated cluster or an experimental setup is not always guaranteed due to their high deployment costs. To ease tuning the parameters and to study the behaviour of big data applications in different environments, several simulation tools were proposed [106]-[119], and [148]. These simulators differ in their engines, scalability, flexibility with parameters, level of details, and the support for additional features such as multi disks and data skew. Mumak [106] is an Apache Hadoop simulator included in its distribution to simulate the behaviour of large Hadoop clusters by replaying previously generated traces. A built-in tool; Rumen [369] is included in Hadoop to generate these traces by extracting previous jobs information from their log files. Rumen collects more than 40 properties of the tasks. In addition, it provides the topology information to Mumak. However, Mumak simplifies the simulations by assuming that the reduce phase starts only after the map phase finishes, thus it does not provide accurate modelling for shuffling and provides rough completion time estimation. SimMR is proposed in [107] as a MapReduce simulator that focuses on modelling different resource allocation and scheduling approaches. SimMR is capable of replaying real workloads traces as well as synthetic traces based on the statistical properties of the workloads. It relies on a discrete event-based simulator engine that accurately emulates Job Tracker decisions in Hadoop for map/reduce slot allocation, and a pluggable scheduling policy engine that simulates decisions based on the available resources. SimMR was tested on a 66-node cluster and was found to be more accurate and two orders of magnitude faster than Mumak. It simplifies the node modelling by assuming several cores but only one disk. MRSim[9] is a discrete event-based MapReduce simulator that relies on SimJava, and GridSim to test workloads behaviour in terms of completion time and utilization [108]. The user provides cluster topology information and jobs specifications such as the number of map and reduce tasks, the data layout (i.e locations and replication factor), and algorithms description (i.e. number of CPU instructions per record and average record size) for the simulations. MRSim focuses on modelling multi-cores, single disk, network traffic, in addition to memory, buffers, merge, parallel copy, and sort parameters. The results of MRSim were validated on a single rack cluster with four nodes.

[6] Available at: http://gwa.ewi.tudelft.nl/datasets/Bitbrains
[7] Available at: http://prof.ict.ac.cn/BigDataBench
[8] Available at: https://github.com/deng113jie/keddah

The authors in [109] proposed MR-cloudsim as a simulation tool for MapReduce in cloud computing data centers. MR-cloudsim is based on the widely used open-source cloud systems event-driven simulator; CloudSim [110]. In MR-cloudsim, several simplifications such as assigning one reduce per map, and allowing the reduce phase to start only after the map phase finishes are assumed. To assist with MapReduce clusters design and testing, MRPerf[10] is proposed in [111] to provide fine-grained simulations while focusing on modelling the activities inside the nodes, the disk and data parameters, and the inter and intra rack networks configurations. MRPerf is based on Network Simulator-2 (ns-2)[11] which is a packet-level simulator, and DiskSim[12] which is an advanced disk simulator. A MapReduce heuristic is used to model Hadoop behaviour and perform scheduling. In MRPerf, the user provides the topology information, the data layout, and the job specifications, and obtains a detailed phase-level trace that contains information about the completion time and the volume of transferred data. However, MRPerf needs several minutes per evaluation and has the limitation of modelling only one replica and a single disk per node. Also it does not enable speculative execution modeling and simplifies I/O and computations processes by not overlapping them. MRemu[13] in [112] is an emulation-based framework based on Miminet 2.0 for MapReduce performance evaluations in terms of completion time in different data center networks. The user can emulate arbitrary network topologies and assign bandwidth, packet loss ratio, and latency parameters. Moreover, network control based on SDN can be emulated. A layered simulation architecture; CSMethod is proposed in [113] to map the interactions between different software and hardware entities at the cluster level with big data applications. Detailed models for JVMs, the name nodes, data nodes, JT, TT, and scheduling were developed. Furthermore, the diverse hardware choices such components, specifications, and topologies were considered. Similar approaches were also used in [114], and [115] to simulate NoSQL and Hive applications, respectively. As part of their study, the authors in [136] developed a network flow level discrete-event simulator named PurSim to aid with simulating MapReduce executions in up to 200 virtual machines.

Few recent articles considered the modelling and simulations of YARN environments [116]-[119]. Yarn Scheduler Load Simulator (SLS) is included in the Hadoop distribution to be used with Rumen to evaluate the performance of YARN different scheduling algorithms with different workloads [116]. SLS utilizes a single JVM to exercise a real RM with thread-based simulators for NM and AM. However, SLS ignores simulating the network effects as the NM and AM simulators interact with the RM only via heartbeat events. The authors in [117] suggested an extension for SLS with an SDN-based network emulator; MaxiNet and a data center traffic generator DCT[2] to add realistic modelling for network and traffic in YARN environments and to emulate the interactions between jobs scheduling and flow scheduling. Real-Time ABS language is utilized in [118] to develop ABS-YARN simulator that focuses on prototyping YARN and modelling job executions. YARNsim in [119] is a parallel discrete-event simulator in YARN that provides comprehensive protocol-level accuracy simulations for task executions and data flow. A detailed modeling for networking, HDFS, data skew, and I/O read and write latencies is provided. However, YARNsim simplifies scheduling policies and fault tolerance. A comprehensive set of Hadoop benchmarks in addition to bioinformatics clustering applications were utilized for experimental validation and an average error of 10% was achieved.

---

[9] Available at: http://code.google.com/p/mrsim
[10] Available at: https://github.com/guanying/mrperf
[11] Available at: http://www.isi.edu/nsnam/ns
[12] Available at: http://www.pdl.cmu.edu/DiskSim/
[13] Available at: https://github.com/mvneves/mremu

## IV. Cloud Computing Services, Deployment Models, and Related Technologies:

Cloud computing aims to enable seamless access for multi-users or tenants to a pool of computational, storage, and networking resources that typically reside within and between several geographically distributed data centers. Unlike traditional IT services that are limited by localized resources inaccessible by remote computing units, cloud computing services allow dynamic outsourcing to software and/or hardware resources. Hence, they can provide scalable and large-scale computational solutions while increasing the resources utilization. Moreover, cloud computing considerably reduces both; the capital expenditures (CAPEX) and operational expenses (OPEX) of software and hardware and increases the resilience. Consequently, it is continuing to encourage wide deployments of large-scale Internet-based services by various organizations and enterprises [370].

Although MapReduce and many other big data frameworks were originally provisioned for use in local clusters under controlled environments, there is an increasing number of cloud computing-based big data applications realizations and services despite the incurred overheads and challenges. This Section provides a brief overview of cloud computing concepts and discusses the challenges and implications of deploying big data applications in cloud computing environments. Moreover, it reviews some related emerging technologies that support computing and networking resource provisioning in cloud computing infrastructures and can hence improve the performance of big data applications. These related technologies aim mainly to virtualize and softwarize networking and computing systems at different levels to provide agile and flexible future-proof solutions. The rest of this Section is organized as follows: Subsection IV-A reviews cloud computing services and deployment models while Subsection IV-B reviews machine and network virtualization, Network Function Virtualization (NFV), containers, and Software-Defined Networking (SDN) technologies. Subsection IV-C discusses the requirements and the challenges of deploying big data applications in cloud computing infrastructures, while Subsection IV-D illustrates the options for deploying big data application in geo-distributed clouds. Finally, Subsection IV-E presents the implications of big data on cloud networking infrastructures.

### *A. Cloud Computing Services and Deployment Models:*

The ability to share hardware, software, development platforms, network, or applications resources between multi-users enables cloud computing systems to provide what can be described as anything-as-a-service (XaaS). Cloud computing services can be categorized according to the outsourced resources and end-users' privileges into Software-as-a-Service (SaaS), Platform-as-a-Service (PaaS), and Infrastructure-as-a-Service (IaaS) [371]. SaaS provides on-demand Internet-based services and applications to end-users without providing the privilege of controlling or accessing the hardware, network, operating system, or the development platforms resources. Examples of SaaS are Salesforce, Google Apps, and Microsoft Office 365. In the PaaS model, end-users have access privilege to the platform which enables them to develop, control, and upgrade their own cloud applications, but not to the underlying hardware. Thus, more flexibility is provided without the need for owning, operating and maintaining the hardware. Microsoft Azure, AWS Elastic Beanstalk, and Google App Engine are examples of PaaS provided as a pay-as-you-go service. The IaaS model provides end-users with extra provisioning privileges that allow them to fully control the hardware, the operating systems, and the applications development platforms [372]. Examples of IaaS are Amazon Elastic Compute Cloud (EC2), Rackspace, and Google Compute Engine.

Based on their accessibility and ownership, cloud infrastructures have four deployment models which are public, private, hybrid, and community clouds [313]. Providers of public clouds allow users to rent cloud resources so they can accelerate the launching of their businesses, services, and applications. The advantages include the elimination of CAPEX and OPEX costs, and the reduction of risks especially for start-up companies. The disadvantage is however the lack of fine-grained control over data storage, network, or security settings. In the private cloud model, the infrastructure is exclusive for use by the owners. Although the costs are high, the full control of the infrastructure typically leads to guaranteed performance, reliability, and security metrics. In the hybrid clouds model, a private cloud may offload or burst some of its workload to a public cloud due to exceeding its capacity, for costs reduction, or to achieve other performance metrics [121], [142]. Hybrid clouds combine the benefits of both private and public clouds but at the expense of additional controlling overhead. Finally, community clouds are owned by several organizations to serve customers with shared and specific interests and needs [313].

*B. Cloud Computing and Big Data Related Technologies:*

***1) Machine Virtualization:***

Machine virtualization abstracts the CPU, memory, network, and disk resources and provides isolated interfaces to allow several Virtual Machines (VMs), also called instances, to coexist in the same physical machine while having different Operating Systems (OS) [373]. Cloud computing relies heavily on machine virtualization as it allows dynamic resources sharing among different workloads, applications, or tenants while isolating them [374]. Moreover, VMs are widely considered in the billing structures of pay-as-you-go cloud services as they can be leased with different prices based on pre-defined sets of resources and configurations. For example, IaaS is typically provided in the form of on-demand VMs such as on EC2 [375] and Google Compute engine [376]. Examples of EC2 offerings are the general purpose t2 VMs with nano, micro, small, medium, large, xlarge, and 2xlarge options, in addition to compute, memory, or storage optimized offers [375]. Google Compute engine offers standard machines (e.g. n1-standard-64), high memory, and high CPU machines [376]. Figure 7(a) illustrates the components of a physical machine with VMs.

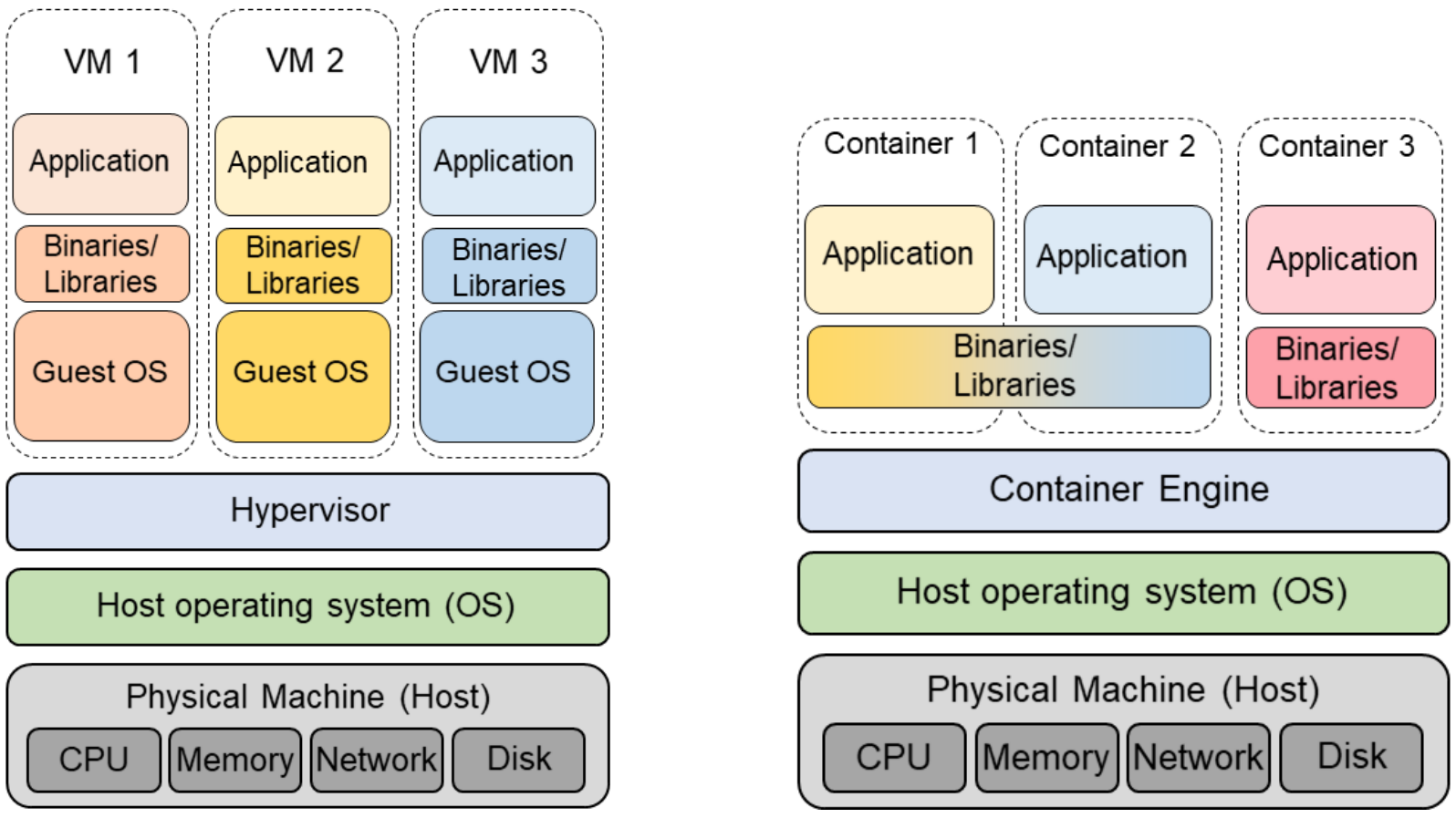

(a) (b)

Fig. 7. Physical machines with (a) VMs and (b) containers.

Virtualization increases the utilization and energy efficiency of cloud infrastructures as with proper management, VMs can be efficiently assigned and seamlessly migrated while running and can thus be consolidated in fewer physical machines. Then, more machines can be switched to low-power or sleep modes. Managing and monitoring VMs in cloud data centers have been realized through several management platforms (i.e. hypervisors) such as Xen [377], KVM [378], and VMWare [379]. Also, virtual infrastructures management tools such as Open Nebula, and VMWare VSphere, were introduced to support the management of virtualized resources in heterogeneous environments such as hybrid cloud [374]. However, and compared to ``bare-metal'' implementations (i.e. native use of physical machines), virtualization can add performance overheads related to the additional software layer, memory management requirements, and the virtualization of I/O and networking interfaces [380], [381]. Moreover additional networking overheads are encountered when migrating VMs between several clouds for load balancing and power saving purposes through commercial or dedicated inter data center networks [382]. These migrations, if done concurrently by different service providers, can lead to increased network congestion, exceeding delay limits, and hence services performance fluctuation, and violations to SLAs [383]. In the context of VMs, several studies considered optimizing their placements and resource allocation to improve big data applications performance. Also, overcoming the overheads of using VMs with big data applications is considered as detailed in Subsection V-B.

### 2) *Network Virtualization:*

To complement the benefits of the mature VMs technologies used in cloud computing systems, virtualizing networking resources have also been considered [384], [385]. Network virtualization enables efficient use of cloud networking resources by allowing multiple heterogeneous Virtual Networks (VNs), also known as slices, composed of virtual links and nodes (i.e. switches and routers) to coexist on the same physical network (substrate network). As with VMs, VNs can be created, updated, migrated, and deleted upon need and hence, allow customizable and scalable on-demand allocations. The assignments and mapping of the VNs on the physical network resources can be performed through Virtual Network Embedding (VNE) offline or online algorithms [386]. VNE can target different goals such as maximizing revenue and resilience through effective links remapping and virtual node migrations, in addition to energy efficiency by consolidating over-provisioned resources [387], [388]. Figure 8 illustrates the concept of VNs and VNE where three different VNs share the physical network. Several challenges can be encountered with network virtualization due to the large scale and the heterogeneous and autonomous nature of cloud networking infrastructures [384]. Also, different infrastructure providers (InP) can have conflicting goals and un-unified QoS measures for their network services.

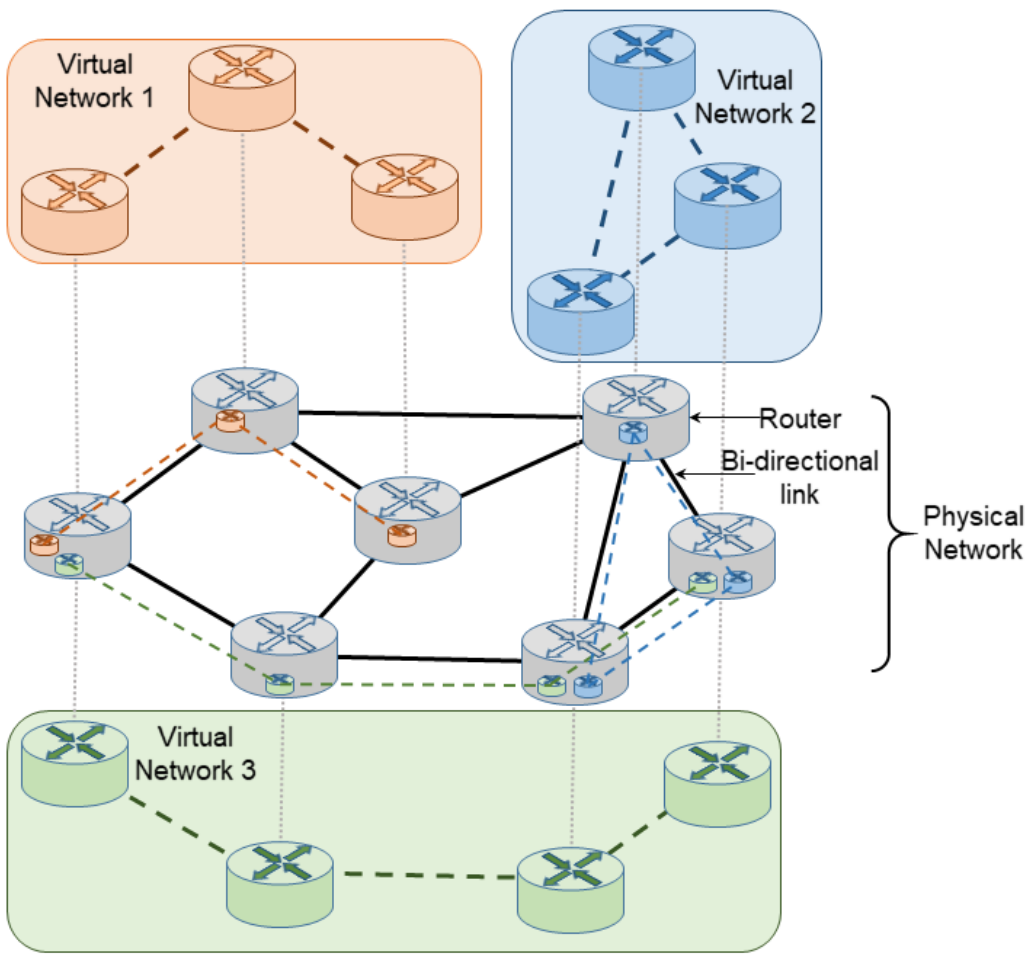

Fig. 8. The concept of virtual networks and VNE.

### 3) *Network Function Virtualization (NFV):*

Network Function Virtualization (NFV) is the detachment of networking functions (NFs) from their proprietary equipment (i.e. middleboxes and appliances) [389] - [393]. Various NFs such as routing, load balancing, firewalls, multimedia caching, QoS monitoring, gateways, proxies, Deep Packet Inspection (DPI), Dynamic Host Configuration Protocol (DHCP), and Network Address Translator (NAT), in addition to mobile networking and telecom operators services such as BaseBand Units (BBUs) and Mobility Management Entities (MMEs) can then be provided as software instances virtualized in data centers, servers, or high capacity edge networking devices [394], [395]. The state-of-the-art implementations of such functions in proprietary hardware are characterized by relatively high OPEX and CPEX, and slow per-device basis upgrades and development cycles. NFV allows rapid development of each NF independently which supports the release and upgrades of new, and existing networking and telecom services in a cost-effective, scalable, and agile manner. Such virtualized NF instances can be created, replicated, terminated, and migrated as needed to elastically improve the handling of the evolving functions and the increasing big data traffic to be acquired, transferred, cached or, processed by those functions. Moreover, NFV can increase the energy efficiency for example by consolidating and scaling resource usage of BBU in cloud environments and virtualized MME pools in base stations [396]-[399].

A key effort to standardize NFV was initiated in 2012 by the European Telecommunication Standard Institute (ETSI) which was selected by seven leading telecom operators to provide the requirements, deployment recommendations, and unified terminologies for NFV. According to ETSI, the components of a Virtual Network Function (VNF) architecture are illustrated in Figure 9 and are listed below [393], [394]:

- The networking services which are delivered as VNFs.
- The NFV Infrastructure (NFVI) composed of the software and hardware to virtualize the computing and networking infrastructure required to realize and host the VNFs.

- The Management and Orchestration (MANO) which manages the life-cycle of VNFs and their resources usage.

NFV aided by cloud computing infrastructures is considered a key enabler for future 5G networks that target innovation, agility, and programmability in networking services for multi-tenant usage with heterogeneous requirements and demands [393], [400]. To achieve this, telecom infrastructures are required to transform their infrastructure design and management for example by integrating VNF with Cloud Radio Access Networks (C-RANs) in the front-haul network (e.g. virtualized BBUs that support multiple Remote Radio Heads (RRHs) and MMEs) [401], [402], and with virtualized Evolved Packet Core networks (vEPC) in the back-haul network [390], [391]. Moreover, and to provide isolated and unified end-to-end network slices in 5G, wireless and spectrum resources in access [403] and wireless networks [404], [405] are also virtualized and integrated with cloud infrastructures and VNFs.

Among the challenges with NFV deployments are the need for the coexistence of NFV-enabled and legacy equipment, and the sensitivity of VNFs to underlying NFVI specifications [406]. Also, further challenges are encountered in the optimized allocation of NFVI resources to VNFs while keeping the cost and energy consumption low, and the composition of end-to-end chained services where traffic has to pass through a certain ordered set of NFVs to deliver the required functionality, [394].

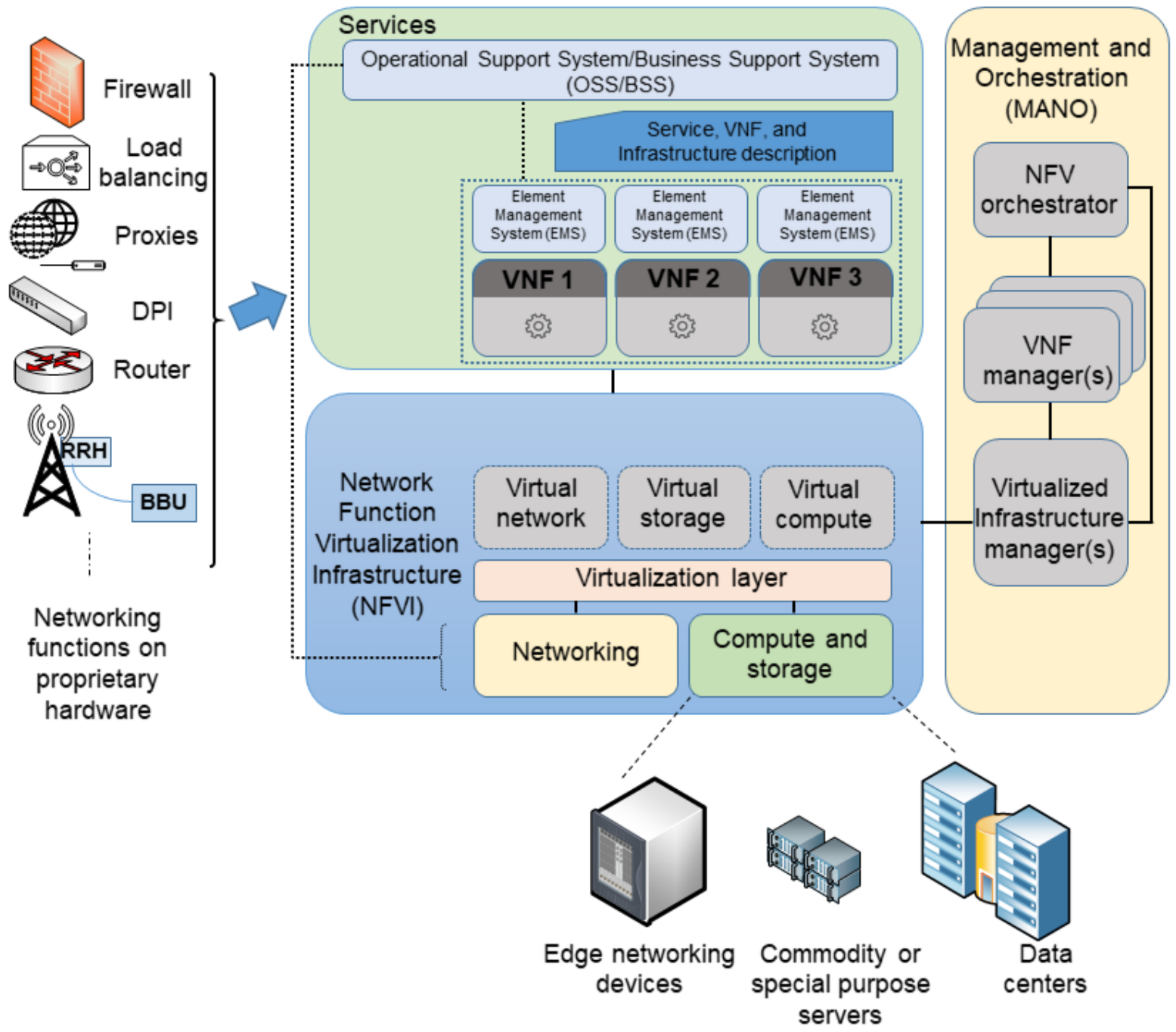

Fig. 9. The concept of NFV and the ETSI-NFV architecture [393].

#### 4) *Container-based Virtualization:*

Container-based virtualization is a recently introduced technology for cloud computing infrastructures proposed to reduce the overheads of VMs as containers can help provide up to five times better performance [407]. While hypervisors perform the isolation at the hardware level and require an independent guest OS for each VM, containers perform the isolation at the OS level and thus, can be regarded as a lightweight alternative to VMs [408]. In each physical machine, the containers share the OS kernel and provide isolation through having different user-spaces by using Linux Kernel containment (LKC) user-space interfaces for example. Figure 7 compares the components of physical machines when hosting VMs 7(a) and containers 7(b). In addition to sharing the OS, containers can also share the binary and library files of the applications running on them. With these features, containers can be deployed, migrated, restarted, and terminated faster and can be deployed in larger numbers in a single machine compared to VMs. However, and due to sharing the OS, containers can be less secure than VMs. To improve their security, containers can be deployed inside VMs and share the Guest OS [409] at the cost of reduced performance.

Some examples of Linux-based container engines are Linux-Vserver, OpenVZ and Linux containers (LXC). The performance isolation, ease of resources management, and overheads of these systems for MapReduce clusters have been addressed in [410] and near bare-metal performance was reported. Docker [407] is an open-source and widely-used containers manager that extends LKC with the kernel and application APIs within the containers [411]. Docker containers have been used to launch the containers within YARN (e.g. in Hadoop 2.7.2) to provide better software environment assembly and consistency and elasticity in assigning the resources to different components within YARN (e.g. map or reduce containers) [412]. Besides YARN, other cloud resources management platforms have successfully adopted containers for resources management such as Mesos and Quasar [160]. Similar to YARN, V-Hadoop was proposed in [413] to enable the use of Linux containers for Hadoop. V-Hadoop scales the number of containers used according to the resource usage and availability in cloud environments.

***5) Software-defined Networking (SDN):***

Software-defined networking (SDN) is an evolving networking paradigm that separates the control plane, which generates the rules for where and how data packets are forwarded, from the data plane, which handles the packets received at the device according to agreed rules, in networking infrastructures [414]. Legacy networks as depicted in Figure 10(a), contain networking devices with vendor-specific integrated data and control planes that are hard to update and scale, while software-defined networks introduce a broader level of programmability and flexibility in networking operations while providing a unified view for the entire network. SDN architectures can have centralized or semi-centralized controllers distributed across the network [415] to monitor and operate networking devices such as switches and routers while considering them as simple forwarding elements [416] as illustrated in Figure 10(b).

Software defined networks have three main layers namely; infrastructure, control, and application. The infrastructure layer which is composed of SDN-enabled devices interacts with the control layer through a Southbound Interface (SBI), while the application layer connects to the control layer through a Northbound Interface (NBI) [415], [417]. Several protocols for SBI such as OpenFlow were developed as an effort to standardize and realize SDN architectures [418] - [420]. OpenFlow performs the switching at the flow granularity (i.e. group of sequenced packets with common set of header fields) where each forwarding element contains a flow table that receives rules updates dynamically from the SDN controllers. Examples of commercially available OpenFlow centralized controllers are NOX, POX, Trema, Ryu, FloodLight, Beacon, and Maestro, and of distributed controllers are Onix, ONOS, and HyperFlow [420] - [422]. Software-based OpenFlow switches such as Open vSwitch (OVS) are also introduced to enable SDN in virtualized environments and ease the interactions between hypervisors, container engines, and various application elements and the software-based SDN controller while connecting several physical machines [423]. An additional tool that aids the control in SDN at a finer granularity is the Programmable Protocol-independent Packet Processor (P4) high-level language that enables arbitrary modifications to packets in forwarding elements at the line rate [424].

The flexibility and agility benefits that result from adopting SDN in large-scale networks have been experimentally validated in several testbeds [425] and in commercial Wide Area Networks (WANs) that interconnect geo-distributed cloud data centers such as Google B4 WAN [426], and Microsoft Software-driven WAN (SWAN) [427]. SDN in WAN supports various Traffic Engineering (TE) operations and improves their congestion management, load balancing, and fault tolerance. Also, SDN relaxes the over-provisioning requirements of traditional WANs which are typically 30-60-% utilized to increase the resilience and performance [421], [422].

The programmability of SDN enables fine-grained integration of adaptive routing and flow aggregation protocols with additional resource allocation and energy-aware algorithms across different network layers [428]. This allows dynamic configuration and ease of implementation for networking applications with scalability requirements and dynamic traffic. The concepts have found wide spread use in cloud computing services and big data applications [429], [430], NFVs [391]-[393] and wireless networks [405]. This is in addition to intra data center networking [421], [431] as will be discussed in Subsection VI-E.

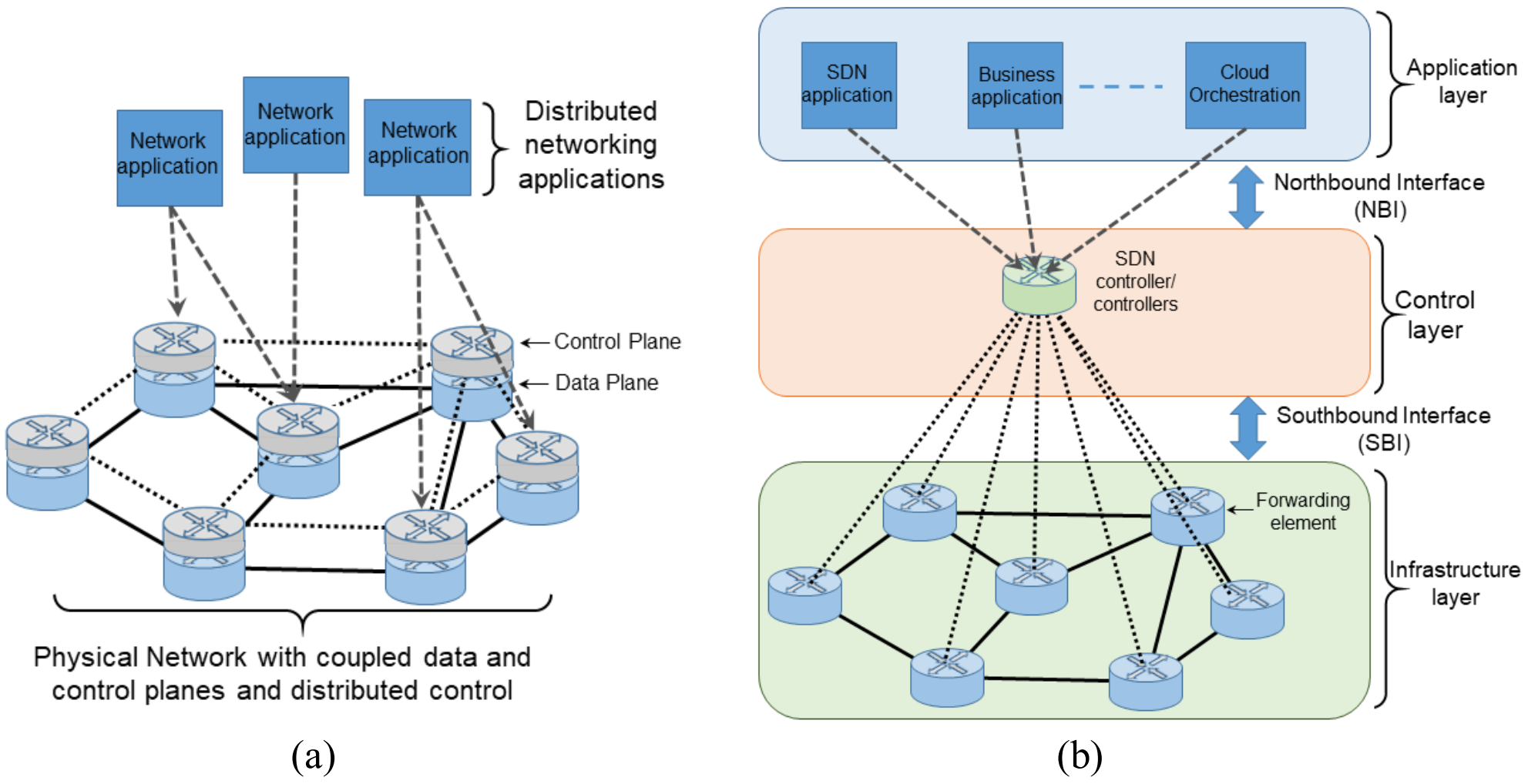

Fig. 10. Architectures of (a) Legacy, and (b) SDN-enabled networks.

#### *6) Advances in Optical Networks:*

Transport networks, as depicted in Figure 1, are composed of a core WAN, typically connected as a mesh to link cities in a country or across continents, Metropolitan Area Network (MAN) connected in ring or star topologies, and access networks mostly with Passive Optical Networks (PONs) technologies and topologies [432]. Core networks use fiber optic communication technologies due to their high capacity, reach, and energy efficiency. High-end Internet Protocol (IP) routers in core networks are widely integrated with optical switches to form IP over Wavelength-Division-Multiplexing (WDM) architectures. The IP layer aggregates the traffic from metro and access networks and connects with the optical layer through short reach interfaces. The optical layer is composed of optical switching devices, mainly Reconfigurable Optical Add Drop Multiplexers (ROADM) realized by Wavelength Selective Switches (WSS), or Optical Cross Connect (OXC) switches to drop wavelengths of terminating traffic, insert wavelengths of newly received traffic, and convert wavelengths, if necessary, to groom transit traffic. In addition, transponders for demodulation and modulation and Optical-Electrical-Optical (O/E/O) conversations are required [433]. Between the nodes, fiber links spanning up to thousands of kilometers are utilized. Due to physical impairments in fibers causing optical power losses and pulse dispersion, amplification is required mainly with Erbium-Doped Fiber Amplifiers (EDFAs) at fixed distances. Reamplification, Reshaping, and Retiming (3R) regenerators can also be installed at distances depending on the line rate. Such networks have sophisticated heterogeneous configurations and are typically configured to be static for long periods to reduce labor risk and costs. However, the increasing aggregated traffic due to Internet-based services with big data workloads makes legacy infrastructures incapable of serving future demands efficiently. This calls for improvements at different layers [434]-[437].

Legacy Coarse or Dense WDM (CDMA or DWDM) systems, standardized by the International Telecommunication Union (ITU), utilized transponders based on On Off Keying (i.e. intensity) modulation and direct detection with carriers in the Conventional band (C-band) (1530-1565 nm) with 50 GHz spacing between channels and with up to 96 channels [438]. Typically, Single-Mode Fiber (SMF) links with a Single Line Rate (SLR) at 10 Gbps or 40 Gbps per fiber are utilized. To increase the capacity with existing fiber plants and to improve the spectral efficiency, the use of Mixed Line Rates (MLR) and multiple modulation formats with more than 1 bit per symbol such as duobinary and Differential Quadrature Phase Shift Keying (DQPSK) were proposed [438]. The rates in MLR systems can be adjusted to transport traffic with short/long reach with high/low rates to reduce regeneration requirements. With the advances in digital coherent detection enabled with high sampling rate Digital-to-Analogue Converters (DACs), Digital Signal Processors (DSP), pre and post digital Chromatic Dispersion (CD) compensation, and Forward Error Correction (FEC) [439]-[441], the use of coherent higher order modulation formats such as QPSK, Quadrature Amplitude Modulation (QAM), in addition to Polarization Multiplexing were also proposed to enhance the spectral efficiency. DSP can realize matched-filters for detection and hence can enable transmission at the Nyquist limit for Inter Symbol Interference (ISI)-free transmission and relaxes the 50 GHz guard bands requirements between adjacent WDM channels allowing more channels in the C-band [440]. The use of the Long-band (L-band) (1565-1625 nm), and the Short-band (S-band) (1460–1530 nm) was also proposed to accommodate more channels, however, costly and complex amplifiers such as Raman amplifiers are required in addition to careful impairments and non-linearities compensation approaches [440], [442]. Space Division multiplexing (SDM)-based solutions that enable wavelengths reuse in the fiber are also considered to increase links' capacity. However, these solutions call for the replacement of SMFs with Multi Mode Fibers (MMFs) (i.e. with large core diameter to accommodate several lightpaths propagating at different modes), or Multi Core Fiber (MCFs) (i.e. with several cores within the fiber) to spatially separate lightpaths that use the same carrier [442].

To further improve the spectral efficiency and allow dynamic bandwidth allocation, the concept of superchannels in Elastic Optical Networks (EONs) is also introduced [443]-[446]. A superchannel is composed of bundled spatial or spectral channels with variable bandwidths at the granularity of 12.5 GHz, as defined by the ITU-T SG15/G.694.1 FlexGrid. These channels can be transmitted as a single entity with guard bands only between superchannels. Such channels can be constructed with Nyquist WDM or with coherent Optical Orthogonal Frequency Division Multiplexing (O-OFDM) which overlaps adjacent subcarriers [447]-[449]. Such flexibility in bandwidth assignments in FlexGrids requires programmable and adaptive networking equipment such as Bandwidth-Variable Transponders (BV-Ts), Bandwidth-Variable Wavelength Selective Switches (BV-WSSs), and Contentionless, Directionless, and Colorless (CDC) Reconfigurable Optical Add Drop Multiplexers (ROADMs) as detailed in [435], [436], and [442]. The modulation format, line rate, and bandwidth of each channel can then be dynamically configured with SDN control [436] to provide programmable and agile Software-Defined Optical Networks with fine-grained control at the optical components and IP layer levels [188], [450].

### *C. Challenges and requirements for Big Data Applications deployments in Cloud Environments:*

Although Hadoop and other big data frameworks were originally designed and provisioned to run in dedicated clusters under controlled environments, several cloud-based services, enabled by leasing resources from public, private or hybrid clouds, are offering big data computations to public users aiming to increase the profit and utilization of cloud infrastructures [451]. Examples of such services are Amazon Elastic MapReduce (EMR)[14], Microsoft's Azure HDInsight[15], VMWare Serengeti's project [452], Cloud MapReduce [120], and Resilin [453]. Cloud-based implementations are increasingly considered as cost-effective, powerful, and scalable delivery models for big data analytics and can be preferred over cluster-based deployments especially for interactive workloads [454]. However, to suit cloud computing environments, conventional big data applications are required to adopt several changes in their frameworks [124], [148]. Also, to suit big data applications and to meet the SLAs and QoS metrics for the heterogeneous demands of multi-users, cloud computing infrastructures are required to provide on-demand elastic and resilient services with adequate processing, storage, memory, and networking resources [126]. Here we identify and discuss some key challenges and requirements for big data applications in cloud environments:

***Framework modifications:*** Single site cluster implementations typically couple data and compute nodes (e.g. name nodes and JVMs coexist in same machines) to realize data locality. In virtualized cloud environments and because elasticity is favored for computing, while resilience is favored for storage, the two entities are typically decoupled or splitted [142]. For example, VMWare follow this split Hadoop architecture, and Amazon EMR utilizes S3 for storage and EC2 instances for computing. This requires an additional data loading step into computing VMs before running jobs [148]. Such transfers are typically free of charge within the same geographical zone to promote the use of both services, but are charged if carried between different zones. For cloud-based RDBMs and as discussed in Section II-C, the ACID requirements are replaced by the BASE requirements, where either the availability or the consistency are relaxed to guarantee partition-tolerance which is a critical requirement in cloud environments [455]. Another challenge associated with the scale and heterogeneity in cloud-implementations is that applications debugging cannot be performed in smaller configurations as in single-site environments and requires tracing in the actual scale [313].

***Offer selection and pricing:*** The availability of large number of competing Cloud Service Providers (CSP) each offering VMs with different characteristics can be confusing to users whom should define their policies, and requirements. To ease optimizing Hadoop-based applications for un-experienced users, some platforms such as

[14] Available at: https://aws.amazon.com/emr/
[15] Available at: https://azure.microsoft.com/en-gb/services/hdinsight/

AmazonEC2[16] provide a guiding script to aid in estimate computing requirements and tuning of configurations. However, unified methods for estimation and comparison are not available. CSP are also challenged with resources allocation while meeting revenue goals as surveyed in [456] at different networking layers where different economic and pricing models are considered.

***Resources allocation:*** Cloud-based deployments for big data applications require dynamic resource allocation at run-time as newly generated data arrivals and volumes are unprecedented and production workloads might change periodically, or irregularly. Fixed resource allocations or pre-provisioning can lead to either under-provisioning and hence QoS violations or over-provisioning which lead to increased costs [120]. Moreover, cloud-based computations might experience performance variations due to interference caused by the share of resources usage. Such variations require careful resources allocations especially for scientific workflow, which also require elasticity in resources assignments as the requirements of different stages vary [457]. Data intensive applications and scientific workflows in clouds have data management challenges (i.e. transfers between storage and compute VMs) and data transfer bottlenecks over WANs [458].

***QoS and SLA guarantees:*** In cloud environments, guaranteeing and maintaining QoS metrics such as response time, processing time, trust, security, failure-rate, and maintenance time to meet SLAs, which are the agreements between CSPs and users about services requirements and violation penalties, is a challenging task due to several factors as discussed in the survey in [459]. With multiple tenants sharing the infrastructure, QoS metrics for each user should be predictable and independent of the co-existence with other users which requires efficient resource allocation and performance monitoring. A QoS metric that directly impacts the revenue is the response time of interactive services. It was reported in [460], and [461] that a latency of 100 ms in search results caused 1% loss in the sales of Amazon, while a latency of 500 ms caused 20% sales drop, and a speed up of 5 seconds resulted 10% sales increase in Google. QoE in video services is also impacted by latency. It was measured in [462] that with more than 2 seconds delay in content delivery, 60% of the users abandon the service. Different interactive services depending on big data analytics are expected to have similar impacts on revenue and customer behaviors.

***Resilience:*** Increasing the reliability, availability, data confidentiality, and ensuring the continuity of services provided by cloud infrastructures and applications against cyber-attacks or systems failures are critical requirements especially for sensitive services such as banking and healthcare applications. Resiliency in cloud infrastructures is approached at different layers including the computing and networking hardware layer, the middleware, and virtualization layer, and at the applications layer by efficient replication, checkpointing, and clouds collaboration as extensively surveyed in [463].

***Energy consumption:*** A drawback of cloud-based applications is that the energy consumption at both network and end-user devices can exceed the energy consumption of local deployments [464], and [465].

*D. Options for Big Data Applications deployment in Geo-distributed Cloud Environments:*

CSPs place their workloads and content in geo-distributed data centers to improve the quality of their services, and rely on multiple Internet Service Providers (ISPs), or dedicated WANs [426], [427] to connect these data centers. Advantages such as load balancing, increasing the capacity, availability and resilience against catastrophic failures, and reducing the latency by being close to the users are attained [440]. Such environments attract commercial and non-commercial big data analytics as they match the geo-distributed nature of data generation and provide scales beyond single-cluster implementations. Thus, there is a recent interest in utilizing geo-distributed data centers for big data applications despite the challenging requirements as surveyed in [323], and suggested in [466]. Figure 11 illustrates different scenarios for deploying big data applications in geo-distributed data centers and compares single-site clusters (case **A**) with various geo-distributed infrastructures with big data including bulk data transfers (case **B**), cloud bursting (case **C**), and different implementations for geo-distributed big data frameworks (case **D**) as outlined in [177]. Case **B** refers to legacy and big data bulk data transfers between geo-distributed data centers including data backups, content replications, bulk data transfers, and VMs migrations which can reach between 330TB and 3.3PB a month [164] requiring cost and performance optimized scheduling and routing. Case **C** is for hybrid frameworks, where a private cloud bursts some of its workloads to public clouds for various goals such as reducing costs, or accelerating completion time. Such scenarios require interfacing with public cloud frameworks in addition to profiling for performance estimations to ensure the gain from bursting compared to localized computations.

[16] Available at: https://wiki.apache.org/hadoop/AmazonEC2

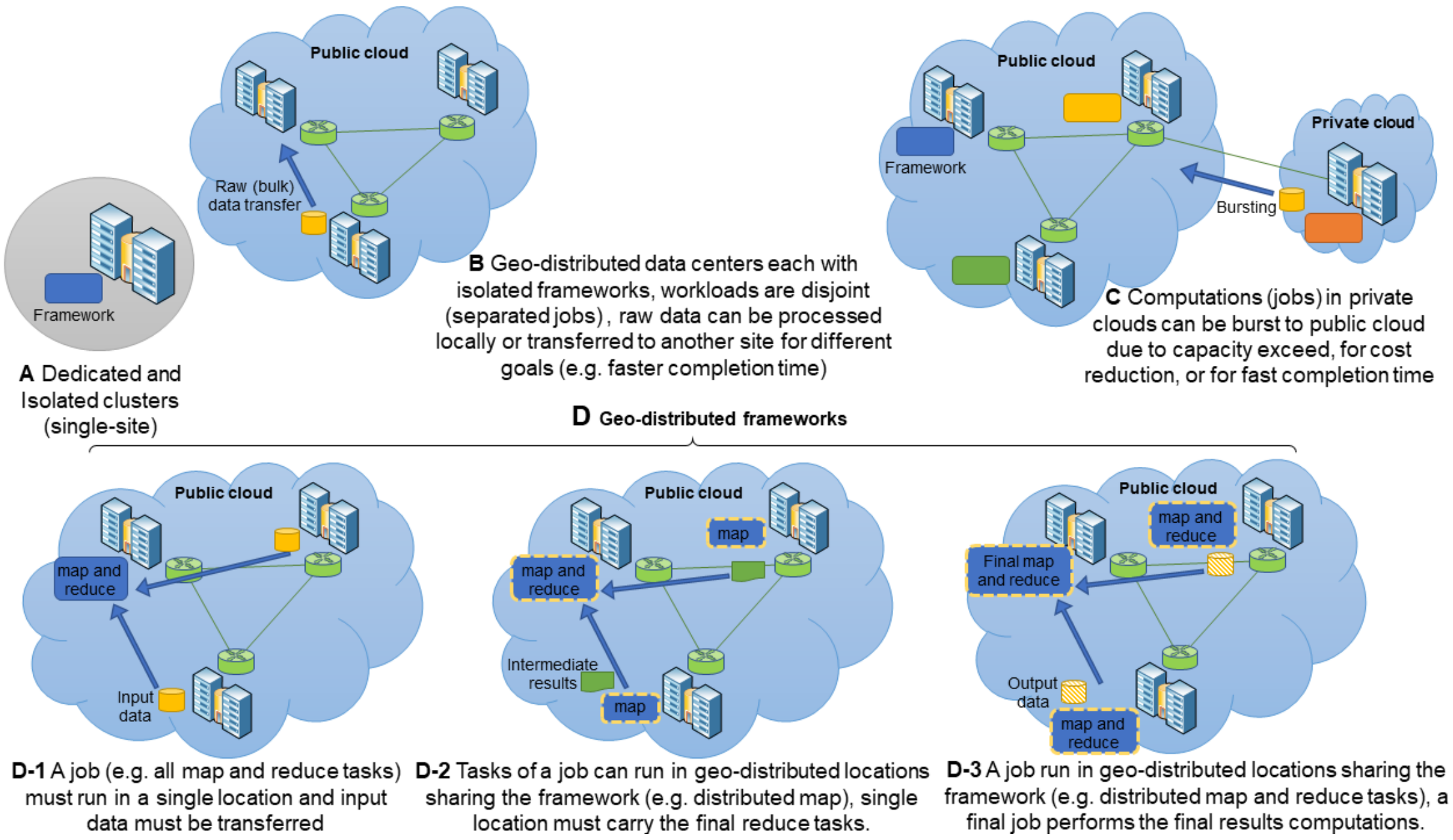

Fig. 11. Different scenarios for deploying big data applications in geo-distributed cloud environments.

Geo-distributed big data frameworks in public cloud have three main deployment modes [177]. Case **D-1** copies all input data into a single data center prior to processing. This case is cost-effective and less complex in terms of framework management and computations but encounters delays due to WAN bottlenecks. Case **D-2** distributes the map-like tasks of a single job in geo-distributed locations to locally process input data. Then, intermediate results are shuffled geo-distributively to a single location for the final reduce-like tasks for final output computations. This arrangement suits workloads with intermediate data sizes much less than the input data. Although networking overheads can be reduced compared to case **D-1**, this case requires complex geo-aware control for the distributed framework components and experiences task performance fluctuations under heterogeneous clouds computing capabilities. The last case, **D-3**, creates a separate job in each location and transmits the individual output results to a single location to launch a final job for results aggregation. This realization relaxes the fine-grained control requirements of the distributed framework in **D-2**, but is considered costly due to the large number of jobs. Also, it only suits associative workloads such as 'averages' or 'counting' where the computations can be performed progressively in stages.

Although such options provide flexibility in implementing geo-distributed big data applications, some unavoidable obstacles can be encountered. Due to privacy and governance regulation reasons, copying data beyond their regions might be prohibited [467]. Also, due to HDFS federation (i.e. data nodes cannot register in other locations governed by a different organization), and the possibility of using different DFSs, additional codes to unify data retrieval are required.

*E. Big Data Implications on the Energy Consumption of Cloud Networking Infrastructures:*

Driven by the economic, environmental, and social impacts of the increased CAPEX, OPEX, Global Greenhouse Gas (GHG) emission, and carbon footprints as a result of the expanding demands for Internet-based services, tremendous efforts have been devoted by industry and academia to reduce the power consumption and increase the energy efficiency of transport networks [468]-[472]. These services empowered by fog, edge, and cloud computing, and various big data frameworks, incur huge traffic loads on networking infrastructures and computational loads on hosting data centers which in turn, increase the power consumption and carbon footprints of these infrastructures [1], [473]-[475]. The energy efficiency of utilizing different technologies for wireless access networks has been addressed in [476]-[481], while for wired PONs and hybrid access networks in [482]-[488]. Core networks, that interconnect cloud data centers with metro and access networks containing IoT and users devices, transport huge amounts of aggregated traffic. Therefore, optimizing core networks plays an important role in improving the energy efficiency of the cloud networking infrastructures challenged by big data. The reduction of energy consumption and carbon footprint in core networks, mainly IP over WDM networks, have been widely considered in the literature by optimizing the design of their systems, devices, and/or routing protocols [432], [489]-[506], utilizing renewable energy sources [507]-[513], and by optimizing the resources assignment and contents placement in different Internet-based applications [387], [514]-[527].

The early positioning study in [489] to green the Internet addressed the impact of coordinated and uncoordinated sleeps (for line cards, crossbars, and main processors within switches) on the switching protocols such as Open Shortest Path First (OSPF) and Internal Broader Gateway Protocol (IBGP). Factors such as how, when, and where to cause devices to sleep, and the overheads of redirecting the traffic and awakening the devices were addressed. The study pointed out that energy savings are feasible but are challenging due to the modification required in devices and protocols. In [490], several energy minimization approaches were proposed such as Dynamic Voltage Scaling (DVF) and Dynamic Frequency Scaling (DFS) at the circuit level, and efficient routing based on equipment with efficient energy profiles at the network level. The consideration of switching off idle nodes and rate adaptation have also been reported in [491]. The energy efficiency of the bypass and non-bypass virtual topologies and traffic grooming schemes in IP over WDM have been assessed in [492] through Mixed Integer Linear Programming (MILP) and heuristic methods. The non-bypass approach requires O/E/O conversation to lightpaths (i.e. traffic carried optically in fiber links and optical devices) in all intermediate nodes, to be processed electronically in the IP layer and routed to following lightpaths. On the other hand, the bypass approach omits the need for O/E/O conversation in intermediate nodes, and hence reduces the number of IP router ports needed, and achieves power consumption savings between 25% and 45% compared to the non-bypass approach. In [493], a joint optimization for the physical topology of core IP over WDM networks, the energy consumption and average propagation delay is considered under bypass or non-bypass virtual topologies for symmetric and asymmetric traffic profiles. Additional 10% saving was achieved compared to the work in [492].

Traffic-focused optimizations for IP over WDM networks were also considered for example in [494]-[499]. Optimizing static and dynamic traffic scheduling and grooming were considered in [494]-[497] in normal and post-disaster situations to reduce the energy consumption and demands blocking ratio. Techniques such as utilizing excess capacity, traffic filtering, protection path utilization, and services differentiation were examined. To achieve lossless reduction for transmitted traffic, the use of Network Coding (NC) in non-bypass IP over WDM was proposed in [498], [499]. Network-coded ports encode bidirectional traffic flows via XOR operations, and hence reduce the number of router ports required compared to un-coded ports. The energy efficiency and resilience trade-offs in IP over WDM networks were also addressed as in [500]-[503]. The impact on the energy consumption due to restoration after link cuts and core node failures was addressed in [500]. An energy efficient NC-based 1+1 protection scheme was proposed in [501], and [502] where the encoding of multiple flows sharing protection paths in non-bypass IP over WDM networks was optimized. MILP, heuristics, in addition to closed form expressions for the networking power consumption as a function of the hop count, network size, and demands indicated power saving by up to 37% compared to conventional 1+1 protection. The authors in [503] optimized the traffic grooming and the assignment of router ports to protection or working links under different protection schemes while considering the sleep mode for protection ports and cards. Up to 40% saving in the power consumption was achieved.

Utilizing renewable energy resources such as solar and wind to reduce non-renewable energy usage in IP over WDM networks with data centers was proposed in [507], and [508]. Factors such as renewable energy average availability and their transmission losses, regular and inter data center traffic, and the network topology were considered to optimize the locations of the data centers and an average reduction by 73% in non-renewable energy usage was achieved. The work in [509] considered periodical reconfiguration to virtual topologies in IP over WDM networks based on a “follow the sun, follow the wind” operational strategy. Renewable energy was also considered in IP over WDM networks for cloud computing to green their traffic routing [510], content distribution [511], services migration [512], and for VNE assignments [513].

The energy efficiency of Information-Centric Networks (ICNs) and Content Distribution Networks (CDNs) were extensively surveyed in [512] and [514] respectively. CDNs, are scalable services that cache popular contents throughout ISP infrastructures, while ICNs support name-based routing to ease the access to contents. The placement of data centers and their content in IP over WDM core nodes were addressed in [516] while considering the energy consumption, propagation delay, and users upload and download traffic. An early effort to green the Internet [517] suggested distributing Nano Data Centres (NaDa) next to home gateways to provide various caching services. In [518], the energy efficiency of Video-on-Demand (VoD) services was examined by evaluating five strategic caching locations in core, metro, and access networks. The work in [519]-[521] addressed the energy efficiency of Internet Protocol Television (IPTV) services by optimizing video content caching in IP over WDM networks while considering the size and power consumption of the caches and the popularity of the contents. To maximize caches hit rate, the dynamics of TV viewing behaviors throughout the day were explored. Several optimized content replacement strategies were proposed and up to 89% power consumption reduction was achieved compared to networks with no caching. The energy efficiency of Peer-to-Peer (P2P) protocol-based CDNs in IP over WDM networks was examined in [522] while considering the network topology, content replications, and the behaviors of users. In [523], the energy efficiency and performance of various cloud computing services over non-bypass IP over WDM networks under centralized and distributed computing modes were considered. Energy-aware MILP models were developed to optimize the number, location, capacity and contents of the clouds for three cloud services namely; content delivery, Storage-as-a-Service (SaaS), and virtual machines (VM)-based applications. An energy efficient cloud content delivery heuristic (DEER-CD) and a real-time VM placement heuristic (DEERVM) were developed to minimize the power consumption of these services. The results showed that replicating popular contents and services in several clouds yielded 43% power saving compared to centralized placements. The placement of VMs in IP over WDM networks for cloud computing was optimized in [524] while considering their workloads, intra-VM traffic, number of users, and replicas distribution and an energy saving of 23% was achieved compared to one location placements. The computing and networking energy efficiency of cloud services realized with VMs and VNs in scenarios using a server, a data center, or multiple geo-distributed data centers were considered in [387]. A Real-time heuristics for Energy Optimized Virtual Network Embedding (REOViNE) that considered the delay, clients locations, load distribution, and efficient energy profiles for data centers was proposed and up to 60% power savings were achieved compared to bandwidth cost optimized VNE. Moreover, the spectral and energy efficiencies of O-OFDM with adaptive modulation formats and ALR power profile were examined.

To bridge the gap between traffic growth and networking energy efficiency in wired access, mobile, and core networks, GreenTouch[17], which is a leading Information and Communication Technology (ICT) research consortium composed of fifty industrial and academic contributors, was formed in 2010 to provide architectures and specifications targeting energy efficiency improvements by a factor of 1000 in 2020 compared to 2010. As part of the GreenTouch recommendations, and to provide a road map to ISP operators for energy efficient design for cloud networks, the work in [504], [506] proposed a combined consideration for IP over WDM design approaches, and the cloud networking-focused approaches in [387], and [523]. The design approaches jointly consider optical bypass, sleep modes for components, efficient protection, MLR, optimized topology and routing, in addition to improvements in hardware where two scenarios; Business-As-Usual (BAU), and BAU with GreenTouch improvements are examined. Evaluations on AT&T core network with realistic 7 data centers locations and 2020 projected traffic, based on Cisco Visual Network Index (VNI) forecast and a population-based gravity model, indicated energy efficiency improvements of 315x compared to 2010 core networks. Focusing on big data and its applications, the work in [526], [527] addressed improving the energy efficiency of transport networks while considering different "5V" characteristics of big data and suggested progressive processing in intermediate nodes as the data traverse from source to central data centers. A tapered network that utilizes limited processing capabilities in core nodes in addition to 2 optimally selected cloud data centers is proposed in [527] and energy consumption reduction by 76% is achieved compared to centralized processing. In [527], the effects of data volumes on the energy consumption is examined. The work in the above mentioned two papers is extended in [173], [174] and is further explained is Section V with other energy efficiency and/or performance-related studies for big data applications in cloud networking environments.

## V. CLOUD NETWORKING-FOCUSED OPTIMIZATION STUDIES

This Section addresses a number of network-level optimization studies that target big data applications deployed partially or fully in virtualized cloud environments. Unlike in dedicated clusters implementations, cloud-based applications encounter additional challenges to meet QoS and SLA requirements such as the heterogeneity of infrastructures and the uncertainty of resources availability. Thus, a wide range of optimizations objectives, and utilizations for cloud technologies and infrastructures are proposed. This Section is organized as the follows: Subsection V-A focuses on cloud resources management and optimization for applications, while Subsection V-B addresses VMs and containers placement and resources allocation optimizations studies. Subsection V-C discusses optimizations for big data bulk transfer between geo-distributed data centers and for inter data center networking with big data traffic. Finally, Subsection V-D summarizes studies that optimize big data applications while utilizing SDN and NFV. The studies presented in this Section are summarized in Tables III, and IV for generic cloud-based, and specific big data frameworks, respectively.

[17] Available at: www.greentouch.org

### *A. Cloud Resource Management:*

To avoid undesired under or over-provisioning in cloud-based MapReduce systems, the authors in [120] provided a theoretical-based mechanism for resource scaling at run-time based on three sufficient conditions that satisfy hard-deadline QoS metrics. A dynamic resources allocation policy was examined in [121] to meet soft-deadlines of map tasks in hybrid clouds while minimizing their budget. The policy utilized local resources in addition to public cloud resources to allow concurrent executions and hence reduce the completion time. In [122], a Hyper-Heuristic Scheduling Algorithm (HHSA) was proposed to reduce the makespan of Hadoop map tasks in virtualized cloud environments. HHSA was found to outperform FIFO and Fair schedulers for jobs with heterogeneous map tasks. To jointly optimize resource allocation and scheduling for MapReduce jobs in clusters and clouds while considering data locality and meeting earliest start time, execution time, and deadline SLAs, the work in [123] proposed a MapReduce Constraint Programming-based Resource Management (MRCP-RM) algorithm. An average reduction by 63% in deadlines missing is achieved compared to existing schedulers. Cloud RESource Provisioning (CRESP) was proposed in [124] to minimize the cost of running MapReduce workloads in public clouds. A reliable cost model that allocates MapReduce workloads while meeting deadlines is obtained by quantifying the relations between the completion time, the size of input data, and the required resources. However, a linear relation was assumed between data size and computation cost which do not suit intensive computations. The energy efficiency of running a mix of MapReduce and video streaming applications in P2P and community clouds was considered in [125]. Results based on simulations that estimate the processing and networking power consumption indicated that community clouds are more efficient than P2P clouds for MapReduce workloads.

Due to the varying prices, in addition to the heterogeneity in hardware configurations and provisioning mechanisms among different cloud providers, choosing a service that ensures adequate and cost-efficient resources for big data applications is considered a challenging task for users and application providers. In [126], a real-time QoS-aware multi-criteria decision making technique was proposed to ease the selection of IaaS clouds for applications with strict delay requirements such as interactive online games, and real-time mobile cloud applications. Factors such as the cost, hardware features, location, in addition to WAN QoS were considered. Typically, cloud providers adopt resource-centric interfaces, where the users are fully responsible for defining their requirements and selecting corresponding resources. As these selections are usually non-optimal in terms of performance, and utilization, *Bazaar* was alternatively proposed in [127] as a job-centric interface. Users only provide high-level performance and cost goals and then the provider allocates the corresponding computing and networking resources. Bazaar achieved accepted request rates increase between 3% and 14% as it argued that several sets of resources lead to similar performance and hence provide more flexibility in resource assignments. In [128], an accurate projection-based model was proposed to aid the selection of VMs and network bandwidth resources in public clouds with homogeneous nodes. The model profiles MapReduce applications on small clusters and projects the resources it predicts into larger clusters requirements. The concurrency of in-memory sort, external merge-sort, and disk read/write speeds were considered. To address the heterogeneity in cloud systems caused by regular server replacements due to failures, the work in [129] proposed Application to Datacenter Server Mapping (ADSM) as a QoS-aware recommendation system. Results with 10 different server configurations indicated up to 2.5x efficiency improvement over a heterogeneity-oblivious mapping scheme with minimal training overhead. The authors in [130] considered a strategy for resource placement to provide high availability cloud applications. A multi-constrained MILP model was developed to minimize the total down time of cloud infrastructures while considering the interdependencies and redundancies between applications, the mean time to failure (MTTF), and mean time to recover (MTTR) of underlying infrastructures and applications components. Up to 52% performance improvement was obtained compared to OpenStack Nova scheduler.

The studies in [131]-[134] focused on the management of bandwidth resources in cloud environments. The authors in [131] proposed *CloudMirror* to guarantee bandwidth allocation for interactive and multi-tier cloud applications. A network abstraction that utilized tenants' knowledge of applications bandwidth requirements based on their communication patterns was utilized. Then, a workload placement algorithm that balances locality and availability and a runtime mechanism to enforce the bandwidth allocation were used. The results indicated that CloudMirror can accept more requests while improving the availability from 20% to 70%. *Falloc* was proposed in [132] to provide VM-level bandwidth guarantees in IaaS environments. First, a minimum base bandwidth is guaranteed for each VM, then a proportional share policy through a Nash bargaining game approach is utilized to allocate excess bandwidth. Compared to fixed bandwidth reservation, an improvement by 16% in jobs completion time was achieved. *ElasticSwitch* in [133] is a distributed solution that can be implemented in hypervisor to provide minimum bandwidth guarantees and then dynamically allocate residual bandwidth to VMs in a work-conserving manner. Small testbed-based results showed successful guarantees even for challenging traffic patterns such as one-all traffic patterns; and improved links utilization to 75% to 99%. A practical distributed system to provide fine-grained bandwidth allocation among tenants, *EyeQ*, was examined in [134]. To provide minimum bandwidth guarantees and resolve contention at the edge, EyeQ maintains at each host a Sender EyeQ Module (SEM) that control the transmitting rate and a Receiver EyeQ Module (REM) that measures the received rate. These modules can be implemented in the codebase of virtual switches in untrusted environments or as a shim layer in non-virtualized trusted environments. For traffic with different transport protocols, EyeQ maintained the $99.9^{th}$ percentile of latency equivalent to that of dedicated links.

The authors in [135] addressed optimizing NoSQL databases deployments in hybrid cloud environment. A cost-aware resources provisioning algorithm that considers the existence of different SLAs among multiple cloud tiers is proposed to meet queries QoS requirements while reducing their cost. In [136], the need for dynamic provisioning of cloud CPU and RAM resources was addressed for real-time streaming applications such as Storm. Jackson open queuing networks theory was utilized for modelling arbitrary operations such as loops, splits, and joins. The fairness of assigning Spark-based jobs in geo-distributed environments was addressed in [137] through max-min fairness scheduling. For up to 5 concurrent sort jobs and compared to default Spark scheduler, job completion time improvements by up to 66% was achieved. The work in [138] outlined that the memoryless fair resource allocation strategies of YARN are not suitable for cloud computing environments as they violate their good properties. To address this, two fair resource allocation mechanisms: Long-Term Resource Fairness (LTRF) for single-level and Hierarchical LTRF (H-LTRF) were proposed and implemented in YARN. The results indicated improvements compared to existing schedulers in YARN, however, only RAM resources were considered. While considering CPU, memory, storage, network and data resources management in multi-tenancy Hadoop environment, the authors in [139] utilized a meta-data scheme, meta-data manager, and a multi-tenant scheduler to introduce multi-tenancy features to YARN and HDFS. Kerberos authentication was also proposed to enhance the security and access control of Hadoop.

### *B. Virtual Machines and Containers Assignments:*

The efficiency of VMs for big data applications relies heavily on the resources allocation scheme used. To ease sizing the virtualized resources for cloud MapReduce clusters and meeting the requirements of non-expert users, the authors in [140] suggested *Elastisizer* which automates the selection of both VM sizes and jobs configurations. *Elastisizer* utilizes Starfish detailed profiling [103], in addition to white-box and black-box techniques to estimate the virtualized resource needs of workloads. *Cura* was proposed in [141] to provide cost-effective provisioning for interactive MapReduce jobs by utilizing Starfish profiling, VM-aware scheduling, and online VMs reconfigurations. A reduction of 80% in costs and 65% in job completion time were achieved but at the cost of increasing the risks for cloud providers as they fully decide on resource allocation and satisfying users QoS metrics. Cloud providers typically over-provision resources to ensure meeting SLAs, however, this leads to poor utilization. The work in [142] utilized the over-provisioning of interactive transactional applications in hybrid virtual and native clusters by concurrently running MapReduce batch jobs to increase the utilization. *HybridMR*, a 2-phase hierarchical scheduler, was proposed to dynamically configure the native and virtual clusters. The first phase categorizes the incoming jobs according to their expected virtualization overhead, then decides on placing them in the physical or virtual machine. The second phase is composed of dynamic resource manager (DRM) at run time, and an Interference Prevention System (IPS) to reduce the interference between tasks in co-hosted VMs and tasks co-located in the same VM. The results indicated an improvement of 40% in completion time compared with fully virtualized systems, and by 45% in utilization compared to native systems with minimum performance overhead. The differences in the requirements of long and short duration MapReduce jobs were utilized in [143] through fine-grained resources allocations and scheduling. A trade-off between scheduling speed and accuracy was considered where accurate constrained programming was utilized to maximize the revenue for long jobs, and two heuristics; first-fit and best-fit algorithms were utilized to schedule small jobs. As a result, the jobs accommodated have increased by 18% leading to an improvement of 30% in revenue. However, simplifications such as a job must run only in one machine, no preemption, and homogeneous VMs were assumed.

The elasticity of VM assignments was utilized in several studies to reduce the energy consumption of running big data applications in cloud environments [144]-[147]. In [144] an online Energy-Aware Rolling-Horizon (EARH) scheduling algorithm was proposed to dynamically adjust the scale of VMs to meet the real-time requirements of aperiodic independent big data jobs while reducing the energy consumption. Cost and Energy-Aware Scheduling (CEAS) was proposed in [145] to reduce the cost and energy consumption of workflows while meeting their deadline through utilizing DVFS and 5 sub algorithms. The work in [146] considered spatial and temporal placement algorithms for VMs that run repeated batch MapReduce jobs. It was found that spatio-temporal placements outperform spatial-only placement in terms of both energy saving and job performance. The work in [147] considered energy efficient VM placement while focusing on the characteristics of MapReduce workloads. Two algorithms; TRP as a trade-off between VM duration and resources utilization, and VCS to enhance the utilization of active servers, were utilized to reduce the energy consumption. Savings by about 16%, and 13% were obtained compared to Recipe Packing (RP), and Bin Packing (BP) heuristics, respectively.

Several studies addressed data locality improvements in virtualized environments [148]-[151]. To reduce the impact of sudden increases in shuffling traffic on the performance of clouds, *Purlieus* was proposed in [148] as a cloud-based MapReduce system. The system suggests using dedicated clouds with preloaded data and utilizes separate storage and compute units, which is different from conventional MapReduce clouds, where the data must be imported before the start of the processing. Purlieus improved the input and intermediate data locality by coupling data and VM placements during both; map and reduce phases via heuristics that consider the type of MapReduce jobs. The challenge of seamlessly loading data into VMs is addressed by loop-back mounting and VM disk attachment techniques. The results indicated an average reduction in job execution time by 50% and in traffic between VMs by up to 70%. However, it assumed knowledge about loads on datasets, job arrival rate and mean execution time. The headroom CPU resources that cloud providers typically preserve to accommodate the peaks in demands were utilized in [149] to maximize data locality of Hadoop tasks. These resources were dynamically adjusted at run-time via hot-plugging. If unstarted tasks in a VM have the data locally, headroom CPU resources are reserved to it while removing equivalent resources from other VMs to keep the price constant for the users. For the allocation and deallocation of resources, two Dynamic Resource Reconfiguration (DRR) schedulers were proposed; synchronous DRR that utilize CPU headroom resources only when they are available, and queue-based DRR that delay the schedule of tasks with locality-matched VM until headroom resources are available. An average improvement in the throughput by 15% was achieved. DRR differs from Purlies in that it assumes no prior knowledge and schedules dynamically, however, the locality for reduce tasks was not considered.

To tackle resources usage interference between VMs in virtual MapReduce clusters that can cause performance degradation, the authors in [150] proposed an interference and locality aware (ILA) scheduling algorithm based on delay scheduling and task performance prediction models. Based on observations that the highest interference is in the I/O resources, ILA was tested under different storage architectures. ILA achieved a speed up by 1.5-6.5 times for individual jobs, and 1.9 times improvement in the throughput compared with 4 state-of-art schedulers, interference-aware only, and locality aware only scheduling. The authors in [151] proposed *CAM* which is a cloud platform to support cloud MapReduce by determining the optimal placement of new VMs requested by users while considering the physical location and the distribution of workloads and input data. CAM exposes networking topology information to Hadoop and integrates IBM ISAAC protocol with GPFS-SNC for the storage layer which allows co-locating related data blocks to increase the locality of VM image disks. As a result, the cross-rack traffic related to operating system and MapReduce was reduced by 3 times and the execution duration was reduced by 8.6 times compared to state-of-art resources allocation schemes.

To increase the revenue and reliability for business critical workloads in virtualized environments, the authors in [152] proposed *Mnemos* which is a self-expressive resources management and scheduling architecture. Mnemos utilized existing system monitoring and VM management tools, in addition to a portfolio scheduler and Nebu topology-aware scheduler to adapt the policies according to workload types. Mnemos was found to respond well to workload changes, and to reduce costs and risks. Similarly, and to provide automated sizing for NoSQL clusters in real-time based on workloads and user defined policies, an open-sourced framework for use in IaaS platforms, *TIRAMOLA* was suggested in [153]. TIRAMOLA integrates VM-based monitoring with Markov Decision Processing (MDP) to decide on the addition and removal of VMs while monitoring the client and server sides. The results here showed a successful VMs resizing that tracks the typical sinusoidal variations in commercial workloads while using different load smoothing techniques to avoid VM assignments oscillations (i.e. being continuously added and removed without being useful to workloads).

To enhance current Xen-based VM schedulers that do not fully support MapReduce workloads, a MapReduce Group Scheduler (MRG) was proposed in [154]. MRG aims to improve the scheduling efficiency by aggregating the I/O requests of VMs located in the same physical machine to reduce context switching and improve the fairness by assigning weights to VMs. MRG was tested with speculative execution and under multiprocessor, and an improvement by 30% was achieved compared to Xen credit default scheduler. In [155], a large segment scheme and Xen I/O ring mechanism between front-end and back-end drivers were utilized to optimize block I/O operations for Hadoop. As a result, CPU usage was reduced by a third during I/O, and the throughput was improved by 10%.

Few recent studies addressed Hadoop improvements in Linux containers-based systems [156]-[160]. *FlexTuner* was proposed in [156] to allow users and developers to jointly configure the network topology and routing algorithms between Linux containers, and to analyze communication costs for applications. Compared with Docker containers that are connected through a single virtual Ethernet bridge, FlexTuner used virtualized Ethernet links to connect the containers and the hosts (i.e. each container has its own host name and IP address). The authors in [157] utilized a modified *k*-nearest neighbour algorithm to select the best configurations for newly submitted jobs based on information from similar past jobs in Hadoop clusters based on Docker containers. 28% gain in the performance was obtained compared to default YARN configurations. An automated configuration for Twister, which is a dedicated MapReduce implementation in container-driven clouds for iterative computations workloads, was proposed in [158]. The traditional client/server communication model (e.g. RPC) was replaced with a publish/subscribe model via NaradaBrokering open-source messaging infrastructure. Negligible difference in jobs completion time was found between physical and container-based implementations. To improve the performance of networking in Linux container-based Hadoop in cloud environments, the authors in [159] proposed the use of IEEE 802.3ad that enables separate interfaces to instances and then aggregates the traffic at the physical interface. Completion time results indicated a comparable performance with bare-metal implementation and improvement by up to 33.73% compared to regular TCP. EHadoop in [160] tackled the impact of saturated WAN resources on the completion time and costs of running MapReduce in elastic cloud clusters. A network-aware scheduler the profile jobs and monitor network bandwidth availability is proposed to schedule the jobs in optimum number of containers so that the costs are reduced and network contentions are avoided. Compared to default YARN scheduler, 30% reduction in costs was achieved.

### *C. Bulk Transfers and Inter Data Center Networking (DCN):*

Several studies considered optimizing bulk data transfers for backups, replication, and data migration between geo-distributed data centers in terms of links utilization, cost, completion time, or energy efficiency [161]-[170]. The early work in [161] proposed *NetStitcher* to utilize the unused bandwidth between multiple data centers for bulk transfers. NetStitcher used future bandwidth availability information (e.g. fixed free slot 3-6 AM), and intermediate data centers between the source and destination in a store-and-forward fashion to overcome availability mismatch between non-neighbouring nodes. Moreover, data bulks were split and routed over multipath to overcome intermediate nodes storage constraints. *Postcard* in [162] further considered different file types, multi-source-destination pairs, and proposed a time slotted model to consider the dynamics of storing and forwarding. An online optimization algorithm was developed to utilize the variability in bandwidth prices to reduce the routing costs. *GlobeAny* was proposed in [163] as an online profit-driven scheduling algorithm for inter data center data transfer requested by multiple cloud applications. Request admission, store-and-forward routing, and bandwidth allocation were considered jointly to maximize the cloud providers profit. Furthermore, differentiated services can be provided by adjusting applications weights to satisfy different priority and latency requirements.

*CLoudMPcast* is suggested in [164] to optimize bulk transfers from a source to multi geo-distributed destinations through overlay distribution trees that minimize costs without affecting transfer times while considering full mesh connections and store-and-forward routing. The charging models used by public cloud service providers which are characterized by flat cost depending on location and discounts for transfers exceeding a threshold in the range of TBytes were utilized. Results indicated improved utilization and savings for customers by 10% and 60% for Azure and EC2, respectively compared to direct transfers. Similarly, the work in [165] optimized big data broadcasting over heterogeneous networks by constructing a pipeline broadcast tree. An algorithm was developed to select the optimum uplink rate and construct an optimal LockStep Broadcast Tree (LSBT). The authors in [166], and [167] focused on improving bulk transfers while considering the optical networking and routing layer. In [166], joint optimization for the backup site and the data transmission paths was proposed to accelerate regular backups. The results indicated that it is better to select the sites and the paths separately. The spectrum management agility of elastic optical networks, realized by Bandwidth-Variable transponders and wavelength selective switches, was utilized in [167] to handle the uncertainty and heterogeneity of big data traffic by adaptively routing backups which were modeled as dynamic anycasts.

End systems Solutions besides networking solutions were proposed in [168] to reduce the energy consumption of data transfers between geo-distributed data centers. Factors such as file sizes, and TCP pipelining, parallelism, and concurrency, were considered. Moreover, data transfer power models based on power meters measurements, that consider CPU, memory, hard disk, and NIC resources, were developed. To emulate geo-distributed transfers in testbeds, artificial delays obtained by setting the round-trip time (RTT) were considered. A similar approach was also utilized in [169], and [170]. To assist users to get the best configurations, the work in [169] investigated the effects of the distance on the delay and the traffic characteristics in geo-distributed data centers while considering different block sizes, locality, and replicas. The authors in [170] experimentally addressed the cost and speed efficiency of data movements into a single data center for MapReduce workloads. Two online algorithms; Online Lazy Migration (OLM), and Randomized Fixed Horizon Control (RFHC) were proposed and the authors generalized that MapReduce computation are better if processed in single data center.

On the other hand, several other studies have utilized the fact that for most production jobs, the intermediate data generated is much smaller than the input data and they proposed alternative solutions based on distributed computations [171]-[174]. To reduce the computational and networking costs of processing globally collected data, the authors in [171] proposed a parallel algorithm to jointly consider data centric job placement and network coding based routing to avoid duplicate transmissions of intermediate data. An improvement by 70% was achieved compared to simple data aggregation and regular routing. To decrease MapReduce inter data centers traffic while providing predictable completion time, the authors in [172] also addressed joint optimization of input data routing and tasks placement. The bandwidth between data centers and the data input to intermediate ratio were determined by profiling, and the global allocation of map and reduce tasks was performed so that the maximum time required for input data, and intermediate data transmissions is minimized. A reduction by 55% was achieved compared to the centralized approach. The authors in [173], [174] improved the energy efficiency of bypass IP over WDM transport networks while considering the different 5V attributes of big data. The use of distributed Processing Nodes (PNs) in the sources and/or intermediate core nodes was suggested to progressively process the data chunks before reaching cloud data centers, and hence reduce the network usage. In [173], a joint optimization of the processing in PNs or cloud data centers and the routing of data chunks with different volumes and processing requirement yielded up to 52% reduction in the power consumption. The capacities and power usage effectiveness (PUE) of the PNs and the availability of the required processing frameworks were also examined. The impact of the velocity of big data on the power consumption of the network is further considered in [174] where two processing modes were examined; expedited (i.e. real-time) which is delay optimized, and relaxed (i.e. batch) which is energy optimized. Power consumption savings ranging between 60% for fully relaxed mode and 15% for fully expedited mode due to additional processing requirements were achieved compared to single location processing without progressive processing.

Several studies addressed detailed optimizations for the configurations and scheduling of MapReduce framework in geo-distributed environments for example to reduce the costs [175]-[177], the completion time [178], [179], and for hybrid clouds [180], [181], while few addressed optimizing geo-distributed data streaming, querying, and graph processing applications [182]-[187]. An inter data center allocation framework was proposed in [175] to reduce the costs of running large jobs while considering the variabilities in electricity and bandwidth prices. The framework considered jobs indivisibility (i.e. ability to run in data centers without the need for inter communications), data centers processing capacities, and time constraints. A StochAstic power redUction schEme (SAVE) was proposed in [176] to schedule incoming batch jobs into a front-end server to geo-distributed back-end clusters while trading-off power costs and delay. Scheduling in SAVE was based on two time scales, the first is to allocate jobs and activate/deactivate servers, and the second, which is finer, is to manage service rates by controlling CPUs power levels. To reduce the cost and execution time of MapReduce sequential jobs in geo-distributed environments, *G-MR* in [177] utilized data Transformation Graphs (DTG) to select an optimal execution sequence detailed by data movements and processing locations. G-MR used approximating functions to estimate completion times for map and reduce tasks, and sizes of intermediate and output data. Replicas were considered by constructing a DTG for each and comparing. Moreover, the effects of choosing different providers (i.e. different inter data center bandwidths, and networking and compute costs) with data centers in the US east coast, US west coast, Europe and Asia were considered. The work in [178] addressed some geo-distributed Hadoop limitations through prediction based job localization, map input data pre-fetching to balance completion times of local and remote map tasks, and by modifying HDFS data placement policy to be sub-cluster aware to reduce output data writing time. An improvement by 48.6% was achieved for the completion time of reduce tasks. The authors in [179] addressed the placement of MapReduce jobs in geo-distributed clouds with limited WAN bandwidth. A Community Detection-based Scheduling (CDS) algorithm was utilized to reduce the WAN data transmission volume and hence reduce the overall completion time while considering the dependencies between tasks, workloads, and data centers. Compared to a hyper-graph partition based scheduling and a greedy algorithm, the transmitted data volume was reduced by 40.7% and the completion time was reduced by 35.8%.

A performance model was proposed in [180] to estimate the speed up in completion time for iterative MapReduce jobs when considering cloud bursting. The model focused on weak links between on-premise and off-permise VMs and suggested asynchronous rack-aware replica placement and replica-aware scheduling to avoid additional data transmission to off-premise VMs if a task is scheduled before its data is replicated there. Experimental results with up to 15 on-premise VMs and 3 off-premise VMs indicated between 1-10% error in the model-based estimation. Compared to ARIA [69], up to one order of magnitude higher accuracy was achieved. *BStream* was proposed in [181] as cloud bursting framework that optimally couples batch processing in limited resources internal clouds (IC) with stream processing in external clouds (EC). To account for the delays incorporated with data uploading to ECs, BStream matched the stream processing rate with the data upload rate by allocating enough EC resources. Storm was used for stream processing with the aid of Nimbus, Zookeeper, and LevelDB for coordination, and Kafka servers to temporary store data in EC. BStream assumed homogeneous resources and no data skew and suggested an approach that performs partial reduction in IC if the tasks are associative. Alternatively, the approach runs final reduce tasks only in IC. Two checkpointing strategies were developed to optimize the reduce phase by overlapping the transmission of reduce results back to IC with input data bursting and processing in EC. However, users need to write two different codes (i.e. for Hadoop and for Storm) to express the same tasks.

*SAGE* was proposed in [182] as an efficient service-oriented geo-distributed data streaming framework. SAGE suggested the use of loosely coupled cloud queues to manage and process streams locally and then utilized independent global aggregator services for final results. Compared to single location processing, a performance improvement by a factor of 3.3 was achieved. Moreover, an economical model was developed to optimize the resource usage and costs of the decomposed services. However, the impact of limited WAN bandwidths on the performance was not addressed. To address this, the authors in [183] proposed *JetStream* which utilizes structured Online Analytical Processing (OLAP) data cubes to aggregate data streams in edge nodes, and proposed to adaptively filter the data (i.e. degrade) to match the instantaneous capacity of the WAN in exchange for the analysis quality. The work in [184] aimed to reduce the inter data centers networking cost while meeting SLAs of big data stream processing in geo-distributed data centers by jointly optimizing VM placements and flow balancing. Each task is replicated in multiple VMs and the distribution of data streams between them was optimized. The results revealed improvements compared to single VM tasks allocation. The computational and communication costs of streaming workflows in geo-distributed data centers was considered in [185]. An Extended Streaming Workflow Graph (ESWG)-based allocation algorithm that accounts for semantic types (i.e. sequential, parallel, and choice for incoming and outgoing patterns) and price diversity of computational and communication tasks was constructed. The algorithm assumes that each task is allocated a VM in each data center, and divides the graph into sub-graphs to simplify the allocation solutions. The results indicated improved performance and reduced costs compared to greedy, computational-cost only, and communication-cost only approaches. To optimize queries processing latency and the usage of limited bandwidth and varied price WANs in geo-distributed Spark-like systems, *Iridium* was proposed in [186]. An online heuristic that iteratively optimizes the placement of data prior queries arrival based on their value per byte, and the processing for a complex DAGs of tasks was used. Iridium was found to speed the queries by 3-19 times and reduce WAN bandwidth usage by 15-64% compared to single location processing and unmodified Spark. The work in [187] proposed *G-Cut* for efficient geo-distributed static and dynamic graphs partitioning. Two optimization phases were suggested to account for the usage constraints and the heterogeneity of WANs. The first phase improves the greedy vertex-cut partitioning approach of PowerGraph to be geo-aware and hence reduces inter data center data transfers. The second phase adopts a refinement heuristic to consider the variability in WAN bottlenecks. Compared to unmodified PowerGraph, reductions by up to 58% in the completion time, and 70% in WAN usage were achieved.

*D. SDN, EON, and NFV-based Optimization Studies:*

Static transport circuit-switched networks cannot provide bandwidth guarantees to applications as they have low visibility to transported traffic. The early work in [188], utilized OpenFlow as a unified control plane for the packet-switched, and circuit-switched layers to enable applications-aware routing that can meet advanced requirements such as low latency or low jitter. OpenFlow was shown to enable differentiated services and resilience for applications with different priorities. To provide latency and protection guarantees to applications, the Application Centric IP/Optical Networks Orchestration (ACINO) project was proposed in [189]. ACINO facilitates applications with the ability to specify reliability and latency requirements, and then groom the traffic of applications with similar requirements into specific optical services. Moreover, an IP layer restoration mechanism was introduced to assist optical layer restoration. A Zero Touch Provisioning, Operation and Management (ZTP/ZTPOM) model proposed in [190]. It utilized SDN and NFV to support Cloud Service Delivery Infrastructures (CSDI) by automating the provisioning of storage and networking resources of multiple cloud providers. However, these studies considered generalized large scale web, voice and video applications, and did not focus on big data applications. The authors in [191] focused on big data transmission and demonstrated the benefits of using Nyquist superchannels in software-defined optical networks for high speed transmission. Routing, Modulation Level, and Spectrum Allocation (RMLSA) optimizations were carried to select paths and modulation levels that improve the transmission speed while ensuring flexible spectrum allocation and high spectral efficiency. Compared to traditional Routing and Spectrum Assignments (RSA), an improvement by 77% was achieved. However, only bandwidth requirements were considered.

To control geo-distributed shuffling traffic that typically saturates local and wide area network switches, the work in [192] suggested an OpenFlow (OF) over Ethernet enabled Hadoop implementation to dynamically control flows paths and define OF-enabled queues with different priorities to accelerate jobs. Experiments with sort workloads demonstrated that OF-enabled Hadoop outperformed standard Hadoop. *Palantir* was proposed in [193] as an SDN service in Hadoop to ease obtaining adequate network proximity information such as rack-awareness within the data center and data center-awareness in geo-distributed deployments, without revealing security sensitive information. Two optimization algorithms were proposed; the first for data center-aware replica placement and tasks scheduling, and the second for proactive data prefetching to accelerate tasks with non-local data. To automate provisioning of computing and networking resources in hybrid clouds for big data applications, *VersaStack* was proposed in [194] as an SDN-based full-stack model driven orchestration. VersaStack computes models for virtual cloud network, manages VM placement, computes L2 paths, performs Single Root Input/Output Virtualization (SR-IOV) stitching, and manages Ceph which is a networked block storage.

Enhancing bulk data transfers has been considered with the aid of SDN and elastic optical inter data center networking in [195]-[199]. To overcome the limitations of current inter data centers WANs with pre-allocate static links and capacities for big data transfers, the authors in [195] proposed Open Transport Switch (OTS) which is a virtual light-weight switch that abstracts the control of transport networks resources for hybrid clouds. OpenFlow was extended to manage packet-optical cross connect (XCON) through additional messaging capabilities. OTS allows applications to optimally compute end-to-end paths across multiple domains or to just submit bandwidth requirements to OTS. Two modes of operations were suggested; implicit and explicit. The former allows the SDN controller to communicate only with edge switches which are located between different transport domains (e.g. Ethernet, Multiprotocol Label Switching (MPLS), OTN), and the later unifies the control for all the components across different domains. Malleable reservations for data bulks transfers in EONs were proposed in [196]. While accounting for flow-oriented reservations with fixed bandwidth allocations, the RSA of the bulk transfers was adjusted dynamically to increase the utilization of the EON and to recycle 2-D spectrum fragmentation. The authors in [197] considered dynamic routing for different data chunks in bulk data transfers between geo-distributed data centers by utilizing an SDN centralized controller and distributed gateway servers in the data centers. Tasks admission control, routing, and the store-and-forward mechanism were considered to maximize the utilization of dedicated links between the data centers. To boost high priority transfers when the networking resources are insufficient, transmission of lower priority chunks can be paused. These chunks were then temporarily stored in intermediate gateway servers and forwarded later at a carefully computed time. To further enhance the transfer performance, three dynamic algorithms were proposed which are bandwidth reserving, dynamic adjustment, and future demand friendly. *Owan* was proposed in [198] as an approach that can help reduce the completion time of deadline-constrained business-oriented bulk transfers in WANs by utilizing a centralized SDN controller that controls ROADMs and packet switches. Network throughput was maximized by joint optimizations of the network topology, routing, and transfer rates while assuming prior knowledge for demands. Topology reconfigurations were achieved by changing the connectivity between router and ROADM ports for example to double the capacity for certain links. Simulated annealing was utilized to update the network gradually and hence ease the reconfigurations. Moreover, consistency was ensured while applying the updates to avoid packets dropping. Compared to packet layer control only, the results indicated 4.45 times faster transfers and 1.36 times lower completion time. Flexgrid optical technologies were proposed with SDN control to meet SLAs and increase revenue of inter data center transfers in [199]. Application-Based Network Operations (ABNO) architecture, which is a centralized architecture under the standardization of IEFT, for the control plane was utilized to manage applications-oriented semantics to identify data volumes and completion time requirements. Heuristics for the Routing and Scheduled Spectrum Allocation (RSSA) were proposed to adjust BV transponders, and OXC to ensure sufficient resources and hence meet deadlines. If workloads with high priority and nearest completion time lack resources, assigned resources for lowest priority workloads can be dynamically reduced.

Few recent studies addressed utilizing NFV for improving the performance of big data applications [200]-[203]. An NFV framework for genome-centric networking was proposed in [200] to address the challenges of exchanging both; genetic data with large file sizes (e.g. 3.2 GB) and the VM images that contain the processing software between data centers. To reduce the traffic, SDN and NFV-based caches were utilized to store parts of the files to be processed along the path. A signalling protocol with on and off path message distribution is proposed to discover the cache resources. NFV enables service chaining by using sequences of VNFs to perform fundamental networks operations such as firewalls and deep packet inspection (DPI) on traffic. The work in [201] addressed optimizing VNF instances placement to minimize the communication costs between connected VNF instances while considering the volumes of big data traffic. An extended network service chain graph was proposed and joint optimization of VNFs and the flows between them was considered. The ETSI NFV specifications were followed in [202] and [203] to enhance cloud infrastructures with big data video workloads. The authors in [202] implemented an elastic CDN as a NFV to deliver live TV and VoD services via the Standard MPEG Dynamic Adaptive Streaming over HTTP (MPEG- DASH). A centralized CDN manager was proposed to control the virtualized CDN functions such as requests admission, and intermediate and leaf cache contents and capacity while reducing the networking costs and ensuring high QoE. *NUBOMEDIA* was proposed in [203] as an elastic PaaS with integrated API stack to integrate Real-time Multimedia Communication (RTC) with multimedia big data processing such as video content analysis (VCA) and augmented reality (AR) with the aid of NFV. The results indicated that NUBOMEDIA scaled well with increasing requests while maintaining the QoS for WebRTC workloads. A summary of the cloud networking-focused optimization studies for big data applications discussed is given in Table III. Table IV, in contrast, focuses on generic cloud applications and provides a summary in a fashion similar to that in TABLE III.

TABLE III
SUMMARY OF CLOUD NETWORKING-FOCUSED OPTIMIZATION STUDIES FOR BIG DATA APPLICATIONS

| Ref | Objective | Application | Tools | Benchmarks/workloads | Experimental Setup/Simulation environment |
|---|---|---|---|---|---|
| [120]* | Resources scaling to avoid over and under provisioning | MapReduce | Theoretical analysis | HiBench (TeraSort and WordCount) | Emulator for IaaS and cloud-based MapReduce |
| [121]* | Dynamic resources allocation in hybrid clouds to meet soft deadlines | Hadoop 1.x | Policy applied In the master | WordCount | Aneka platform; 4 IBM X3200 M3 servers (4 CPU cores, 3 GB RAM) and a master (2 cores, 2 GB RAM) EC2 m1.small instances. |
| [122]* | Hyper-Heuristic Scheduling Algorithm (HHSA) for cloud computing systems | Hadoop 1.1.3 | simulated annealing, genetic algorithm, swarm, ant colony | fMRI, protein annotation, PSPLIB, Grep, TeraSort | CloudSim simulator (4 VMs in a data center), Single node (Intel i7, 64GB RAM) multi-node cluster with 4 VMs (8 cores, 16GB RAM) |
| [123]* | MapReduce jobs resource allocation and scheduling | Hadoop 1.2.1 | Matchmaking and scheduling algorithm (MRCP-RM) | Synthetic workloads, WordCount | 11 nodes cluster in EC2 (Intel Xeon CPU, 3.75 GB RAM), simulations |
| [124]* | Cloud RESource Provisioning (CRESP) to minimize cost or completion time | Hadoop 1.0.3 | Profiling and machine learning | WordCount, Sort, TableJoin, PageRank | 16 nodes cluster (4 Quad-core CPU, 16GB RAM, 2 500GB disks), EC2 small instances (1 CPU, 1.7GB RAM, 160GB disk) |
| [127]* | Bazaar: job-centric interface for data analytics applications in clouds | Hadoop 0.20.2 | Profiling+ greedy heuristic for resources selection | Sort, WordCount, GridMix, TF-IDF, LinkGraph | Prediction: 35 Emulab nodes (from Cloudera), evaluation on 26 Emulab nodes |
| [128]* | Public cloud resources allocation to meet completion time requirements | Hadoop 1.x | Naive and linear profiling | Sort, Word Count, GridMix | 33 nodes (4 cores, 4 GB RAM, 1TB disk for data), 1 Gbps Ethernet |
| [132]* | Falloc: VM-level bandwidth guarantees in IaaS environments | Hadoop 1.0.0 | Nash bargaining game, rate control | Random VM-VM traffic, WordCount, Sort, join | 16 nodes cluster with 64 VMs, 1 Gbps Trace-driven Mininet simulations |
| [135]* | Cost model for NoSQL in public and private clouds | Elasticsearch 0.20.6 | Profiling + look-ahead optimization | YCSB | 6 nodes with mixture of VMs with (2-4 CPU, 2.4-3.2 GHz, 3-8GB RAM) |
| [136]* | Dynamic resource scheduling for Data Stream Management Systems (DSMS) | Storm | DRS scheduler | Video logo detection, frequent pattern detection | 6 nodes (Intel quad core CPU, 8 GB RAM), LAN network |
| [137]* | Fairness scheduler for Spark jobs In geo-distributed data centers | Apache Spark | Linear program, heuristic | Sort workloads for multiple jobs | 12 EC2 instances in 6 geo-distributed data centers (2 CPU, 7.5GB RAM, 32GB SSD |
| [138]* | Long Term Fair Resource Allocation in cloud computing | Hadoop 2.2.0 | LTYARN scheduler In YARN | Facebook traces, Purdue suite, Hive workloads TPC-H, Spark ML | 10 nodes cluster (2 Intel X5675 CPUs,, 24GB RAM) to emulate EC2 t2.medium (2 CPU cores, 4 GB RAM) |
| [139]* | Introduce multi-tenancy features in Hadoop | Hadoop 2.6.0 | Modification to YARN and HDFS | WordCount, Grep, PiEstimator | 8 nodes cluster(2.4 GHz CPU, 3GB RAM, 200 GB HDD), 1 GB Ethernet |
| [140]° | Elastizer: automating virtualized cluster sizing for cloud MapReduce | Hadoop 1.x | Profiling, white and black-box methods | Link Graph, Join, TF-IDF, TeraSort, WordCount | Various EC2 instances |
| [141]° | Cura: cost-effective cloud provisioning for MapReduce | Hadoop 1.x | Profiling, VM-aware scheduling, online VM reconfiguring | Facebook-like traces generated by SWIM | 20 nodes KVM-based cluster (16 cores CPU, 16GB RAM) in two racks, 1 Gbps Ethernet, Java based simulations |
| [142]° | HybridMR: MapReduce scheduler in hybrid native/virtualized clusters | Hadoop 0.22.0 | 2-phase hierarchical scheduler, profiling | Sort, WC, PiEst, DistGrep, twitter, K-means, TestDFSIO | 24 nodes (dual core 2.4 GHz CPU, 4GB RAM), 48 VMs (1 CPU core, 1 GB RAM) |
| [146]° | Energy-aware MapReduce in clouds by spatio-temporal placement tradeoffs | Hadoop 1.x | BinCardinality, BinDuration, RandomFF, recipe | Sort, WordCount, PiEstimator | 6 nodes (dual core CPU, 2GB RAM, 250GB disk), 3 VM types, 1 Gbps Ethernet. |
| [147]° | Energy-efficient VM placement in IaaS cloud data centers running MapReduce | Hadoop 1.1.0 | Tight Recipe Packing (TRP), Virtual Cluster Scaling (VCS) | Sort, TeraSort, WordCount | 10 nodes Xen-based cluster (4 cores CPU, 6GB RAM, 600GB storage) Gigabit Ethernet, Java-based simulations |
| [148]° | Purlieus: optimizing data and VM placements for MapReduce in clouds | Hadoop 1.x | Heuristics based on k-club | Grep, Permutation generator, Sort | 2 racks, 20 nodes, 20 VMs/job (4 2 GHz CPUs, 4GB RAM), KVM as hypervisor, 1 Gbps network, simulations (PurSim) |
| [149]° | Locality-aware Dynamic Resource Reconfiguration (DRR) | Hadoop 0.20.2 | Hot-plugging, synchronous and queue-based scheduler | Hive workloads (Grep, select aggregation, join) | 100 nodes EC2 cluster (8 CPU, 7 GB RAM), 6 nodes cluster (6 CPU, 16GB RAM), With 5 VMs/node (2CPU, 2GB) |
| [150]° | Interference and locality-aware MapReduce tasks scheduling in virtual clusters | Hadoop 0.20.205 | Meta task scheduler | TeraSort, Grep RWrite, WCount, TeraGen | 12 physical machines in 2 racks forming 72-nodes Xen-based cluster (12 CPU cores, 32GB, 500GB disk), 1 Gbps Ethernet |
| [151]° | CAM: topology-aware resources manager for MapReduce in clouds | Vanilla Hadoop | Scheduler based on min-cost flow problem | Hive workloads | 4 machines (2 quad CPU, 48GB RAM), KVM hypervisor, 1 Gbps Ethernet, 23 VMs (2 CPU, 1GB RAM), simulations (PurSim) |
| [152]° | Self-expressive management for business critical workloads in clouds | Hadoop and Microsoft HPC | Mnemos: portfolio scheduler, Nedu: virtualization scheduler | Real-world business-critical traces from 1300 VMs | Simulations |
| [153]° | TIRAMOLA: open-source framework to automate VM resizing in NoSQL clusters | Hadoop 1.0.1, HBase 0.92.0 | Ganglia for monitoring, Markov Decision Process | YCSB benchmark | OpenStack cactus cluster with 20 clients VMs (2 CPU cores, 4GB RAM), and 16 server VMS (4 CPU cores, 8GB RAM) |
| [154]° | Enhanced Xen scheduler for MapReduce | Hadoop 0.20.2 | modification to Xen hypervisor, 2-level scheduling policy | Word Count, Grep, sort, Hive benchmarks | Machine type 1 (2.8 GHz 2 core CPU, 2MB L2 cache, 3GB RAM) Machine type 2 (3 GHz 2 core CPU, 6MB L2 cache, 6GB RAM) |
| [155]° | Enhancing Hadoop on Xen-based VMs | Hadoop 1.x | Optimizing Xen I/O ring mechanism | TestDFSIO | (Intel i7 CPU, 16GB RAM, SSD 128GB disk) |
| [157]° | Automatic configurations for Hadoop on Docker containers | Hadoop 2.x | Lightweight custom k-nearest neighbour heuristic | HiBench workloads (k-Means, Bayes, PageRank) | 5 nodes (2 Quad-core 2.8 GHz CPU, 12 MB L2 cache, 8GB RAM, 320GB disk), 1 Gbps Ethernet |
| [158]° | Iterative MapReduce on Docker containers | Twister | Pub/sub broker network via NaradaBrokering | k-means clustering | 2 Docker containers |
| [159]° | Improving networking performance in Container-based Hadoop | Hadoop 2.7.3 | Link aggregation via IEEE 802.3ad standard | TestDFSIO | 2 ProLiant DL385 G6 servers (2 6-cores AMD, 32GB RAM), IEEE 802.3ad-enabled Gigabit Ethernet |
| [160]° | EHadoop: Network-aware scheduling in elastic MapReduce clusters | YARN | Online profiling, FIFO and fair schedulers | HiBench (sort, WordCount, PageRank) | 30 nodes, 4 data and 20-25 compute (8 CPU cores, 8GB RAM), 800 Mbps network |
| [169]† | Improving Hadoop in geo-distributed data centers | Hadoop 2.2.0 | Statistical analysis of traffic log data | WordCount, TeraSort | 18 nodes cluster in 3 sub-clusters, 1 Gbps network, each sub-cluster contains 3 racks and 6 data nodes |
| [171]† | Data-centric cloud architecture for geo-distributed MapReduce | Hadoop 1.x | parallel algorithm for job scheduling and data routing | Aggregate, expand, transform, summary | Simulations for 12 nodes in EC2 |
| [172]† | MapReduce in geo-distributed data centers with predictable completion time | Hive | Chance-constrained optimization | Hive traces | Simulations for 12 data centers with network bandwidth uniformly distributed between 100Mbps and 1Gbps |
| [173]† | Balancing processing location and type, and energy consumption in big data networks | Hadoop 1.x, others | MILP, heuristic | Input data chunks (10 -330)GB with output sizes (0.01-330)GB | NSFNET, Cost239, and Italian networks with 14 processing nodes and 2 cloud data centers |
| [174]† | Balancing processing speed and energy consumption in big data networks | Hadoop 1.x | MILP | Input data chunks (10 -220)GB with output sizes (0.2-220)GB | NSFNET network with 14 processing nodes and 2 cloud data centers |
| [177]† | G-MR: MapReduce framework for Geographically-distributed data centers. | Hadoop 1.x | Data Transformation Graph (DTG) algorithm | CensusData, WordCount, KNN, Yahoo, Google traces | 10 large instances from 4 EC2 data centers (4 CPU cores, 7.5GB RAM), 2 VICCI clusters |
| [178]† | Hadoop system-level improvements in geo-distributed data centers | Hadoop 1.x | Sub-cluster scheduling and reduce output placement | WordCount, Grep | 20 nodes in 2 sub-clusters, cross access latency of 200 ms, and subcluster latency of 1 ms |
| [180]† | Performance model for iterative MapReduce with cloud bursting | Hadoop 2.6.0 | Profiling, rack-aware scheduler | Iterative GREP, k-means, HiBench, PageRank | 8 nodes, 4 (4 CPU cores, 500GB HDD, 4GB RAM), 4 (2_8 CPU cores, 1 TB HDD, 64GB RAM), KVM |
| [181]† | BStream: Cloud bursting framework for MapReduce | Hadoop 0.23.0, Storm 0.8.1 | Profiling, modification to YARN | WordCount, MultiWord-Count, InvertedIndex, sort | Cluster: 21 nodes (2 2GHz CPU, 4GB , RAM, 500GB disk), Cloud: 12 nodes (Quad 3.06GHz CPU, 8GB RAM) |
| [187]† | G-Cut: geo-distributed graph partitioning | PowerGraph | modification to PowerGraph partitioning algorithm | PageRank, Shortest Single Source Path (SSSP), Subgraph Isomorphism (SI) on 5 real-world graphs | EC2 instances, Simulations on Grid5000 |
| [192]‡ | OpenFlow (OF)-enabled Hadoop over Ethernet | Matsu, Hadoop 1.x | modified JT, Open vSwitch (OVS) | MalStone Benchmark, sort | Open Cloud Consortium Project infra-structure with 10 nodes, 3 data centers, OF-enabled switches, 10Gbps network |
| [193]‡ | Palantir: SDN service to obtain network proximity for Hadoop | Hadoop 1.x | Service module in Floodlight | Facebook traces | Testbed with 32 nodes in 4 data centers, each with 2 racks, Pronto 3290 OpenFlow core switch and OVS in Top-of-Rack (ToR) switches |

*Cloud resources management, °VM and containers assignments, †Bulk transfers and inter DCN, ‡SDN and NFV-based.

TABLE IV

SUMMARY OF CLOUD NETWORKING-FOCUSED OPTIMIZATION STUDIES FOR GENERIC CLOUD APPLICATIONS

| Ref | Objective | Tools | Benchmarks/workloads | Experimental Setup/Simulation environment |
|---|---|---|---|---|
| [125]* | Energy efficiency in cloud data | energy consumption | MapReduce synthesized | Simulations |
| | centers with different granularities | evaluation | workloads, video streaming | |
| [126]* | Cloud recommendation system based on multicriteria QoS optimization | Analytic Hierarchy Process (AHP) decision making | Information about cloud providers (e.g. prices, locations) | Single machine as master, server from NeCTAR cloud, small EC2 instance, andC3.8xlarge EC2 instance |
| [129]* | QoS and heterogeneity-aware Application to Datacenter Server Mapping (ASDM) | Lightweight controller in cluster schedulers | Single and multi-threaded std benchmarks, Microsoft workloads | Homogeneous 40 nodes clusters with 10 configurations, heterogeneous 40 nodes cluster with 10 machine types |
| [130]* | Highly available cloud applications and services | MILP model for availability-aware VM placements | Randomly generated MTTF and MTTR | 50 nodes cluster with total (32 CPU cores, 30GB RAM) |
| [131]* | CloudMirror: Applications-based network abstraction and workloads placements with high availability | TAG modeling placement algorithm | Empirical and synthesized workloads | Simulations for tree-based cluster with 2048 servers |
| [133]* | ElasticSwitch: Work-conserving minimum bandwidth guarantees for cloud computing | Guarantee Partitioning (GP), Rate Allocation (RA), OVS | Shuffling traffic | 100 nodes testbed (4 3GHz CPU, 8GB RAM) |
| [134]* | EyeQ: Network performance Isolation at the servers | Sender and receiver EyeQ modules | Shuffling traffic, Memcached traffic | 16 nodes cluster (Quad core CPU), 10 Gbps NIC packet-level simulations |
| [143]° | Multi-resource scheduling in cloud data centers with VMs | Constrained programming, first-fit, best-fit heuristics | Google cluster Traces | Trace-driven simulations for (1024 nodes, 3-tier tree topology) |
| [144]° | Real-time scheduling for tasks in virtualized clouds | Rolling-horizon optimization | Google cloud Tracelogs | Simulations via CloudSim toolkit |
| [145]° | Cost and energy aware scheduling For deadline constraint tasks in clouds | VM selection/reuse, tasks with merging/slaking heuristics | Montage, LIGO, SIPHT, CyberShake | Simulations via CloudSim toolkit |
| [156]° | FlexTuner: Flexible container-based tuning system | Modified Mininet, iperf tools | MPICH2 version 1.3, NAS Parallel Benchmarks | Simulation for one VM (2GB RAM, 1 CPU core) |
| [161]† | NetStitcher: system for inter data center bulk transfers | Multipath and multi-hop, store-and-forward algorithms | Video and content sharing traces | Equinix topology, 49 CDN (Quad Xeon CPU, 4GB RAM, 3TB disk), 1 Gbps links |
| [162]† | Costs reduction for inter-data cener traffic | Convex optimizations, time-slotted model | Uniform distribution for file sizes | Simulations for inter DCN with 20 data centers |
| [163]† | Profit-driven traffic scheduling for inter DCN with multi cloud applications | Lyapunov optimizations | Uniform distribution for arrival rate | Simulations for 7 nodes in EC2 with 20 different cloud applications |
| [164]† | CloudMPcast: Bulk transfers in multi data centers with CSP pricing models | ILP and Heuristic based on Steiner Tree Problem | Data backup, video distribution | Trace-driven simulations for 14 data centers from EC2 and Azure |
| [165]† | Reduce completion time of big data broadcasting in heterogeneous clouds | LockStep Broadcast Tree (LSBT) algorithm | - | Numerical evaluations |
| [166]† | Optimized regular data backup in Geographically-distributed data centers | ILP and heuristics | Transfer of 1.35 PBytes | Simulations for US backbone network topology with 6 data centers |
| [167]† | Optimize migration and backups in EON-based geo-distributed DCNs | Greedy anycast algorithms | Poisson arrival demands, -ve exponential service time | Simulations for NSFNET network |
| [168]† | Energy saving in end-to-end data transfers | Power modelling by linear regression | 20 and 100GB data sets with different file sizes | Servers (Intel Xeon, 6GB RAM, 500 GB disk), (AMD FX, 10GB RAM, 2TB disk), Yokogawa WT210 power meter |
| [170]† | Reducing costs of migrating geo-dispersed big data to the cloud | An offline and 2 online algorithms | Meteorological data traces | 22 nodes to emulate 8 data centers, 8 gateways, and 6 user-side gateways, additional node for routing control |
| [175]† | Reduction of electricity and bandwidth costs in inter DCNs with large jobs | Distributed algorithms | 22k jobs from Google cluster traces | 4 data centers with varying electricity and bandwidth costs throughout the day |
| [176]† | Power cost reduction in distributed data centers for delay-tolerant workloads | Two time scales scheduling algorithm | Facebook traces | Simulations for 7 geo-distributed data centers |
| [179]† | Tasks scheduling and WAN bandwidth allocation for big data analytics | Community Detection based Scheduling (CDS) | 2000 file with different sizes | Simulations in China-VO network with 5 data centers |
| [182]† | SAGE: service-oriented architecture for data steaming in public cloud | Azure Queues | 5 streaming services with synthetic benchmark | Azure public cloud (North and West US and North and West Europe) |
| [183]† | JetStream: execution engine for data streaming in WAN | OLAP cubes, adaptive data filtering | 140 Million HTTP requests (51GB) | VICCI testbed to emulate Coral CDN |
| [184]† | Networking cost reduction for geo-distributed big data stream processing | MILP model, Multiple VM Placement algorithm | Streams with four different semantics | Simulations for NSFNET network |
| [185]† | Reduction of streaming workflows costs in geo-distributed data centers | MILP and 2 heuristics | 500 streaming workflows each with 100 tasks | Simulations for 20 data centers |
| [186]† | Iridium: low-latency geo-distributed data analytics | Linear program, online heuristic | Bing, Conviva, Facebook, TPC-DS queries, AMPLab Big-Data | EC2 instances in 8 worldwide regions, trace-driven simulations |
| [188]‡ | Application-aware Aggregation and TE in converged packet-circuit networks | OpenFlow, POX controller | Voice, video, and web traffic | 7 GE Quanta packet switched, 3 Ciena CoreDirector hybrid switch, 6 PCs with random traffic generators |
| [189]‡ | Application-Centric IP/optical Network Orchestration (ACINO) | SDN orchestrator | Bulk transfers, dynamic 5G services, security, CDN | Software for controller and use cases for ACINO infrastructure |
| [190]‡ | ZeroTouch Provisioning (ZTP) for managing multiple clouds | Network automation with SDN and NFV | Bioenformatics, UHD video editing | GEANT network and use cases for ZTP/ZTPOM |
| [191]‡ | Routing, Modulation level and Spec-trum Allocation for bulk transfers | MILP, NOX controller | Bulk data transfers | Emulation of NSFNET network with OF-enabled WSSs |
| [194]‡ | VersaStack: full-stack model-driven orchestrator for VMs in hybrid clouds | Resources modelling and orchestration workflow | Reading/writing from/to Ceph parallel storage | AWS and Google VMs, OpenStack-based VMs with SR-IOV interfaces, 100G SDN network |
| [195]‡ | Open Transport Switch (OTS) for Cloud bursting | OTS prototype based on OpenFlow | Bulk data transfers | ESnet Long Island Metropolitan Area Network (LIMAN) testbed |
| [196]‡ | Malleable Reservation for efficient bulk data transfer in EONs | MILP, dynamic programming | Bulk data transfers | Simulations on NSFNET |
| [197]‡ | SDN-based bulk data transfers orchestration | Dynamic algorithm, OpenFlow, Beacon | Bulk data transfers | 10 data centers (IBM BladeCenter HS23 cluster), 10 OF-enabled HP3500 switches, 10 server-based gateways |
| [198]‡ | Owan: SDN-based traffic management system for bulk transfers over WAN | Simulated annealing | Bulk data transfers | Prototype, testbed, simulations for ISP and inter data center WAN |
| [199]‡ | SDN-managed Bulk transfers in Flexgrid inter DCNs | ABNO, ILP heuristics | Bulk data transfers | OMNeT++ simulations for Telefonica (TEL), British Telecom (BT), and Deutsche Telekom (DT) networks |
| [200]‡ | Genome-Centric cloud-based networking and processing | NetServ, algorithms off-path signaling | Genome data | 3-server clusters with total (160 CPU cores, 498GB RAM, and 37TB storage), emulating GEANT Network |
| [201]‡ | VNF placement for big data processing | MILP, heuristics | Random DAGs for chained NFs | 10 nodes network with random communication cost values |
| [202]‡ | NFV-based CDN network for big data video distribution | OpenStack-based CDN manager | HD and full HD videos | Leaf cache on SYNERGY testbed (1 DC with 3 servers each with Intel i7 CPU, 16GB RAM, 1TB HDD) |
| [203]‡ | NUBOMEDIA: real-time multimedia communication and processing | PaaS-based APIs | WebRTC data | 3 media server on KVM-based instances |

*Cloud resources management, °VM and containers assignments, †Bulk transfers and inter DCN, ‡SDN and NFV-based.

# VI. DATA CENTERS TOPOLOGIES, ROUTING PROTOCOLS, TRAFFIC CHARACTERISTICS, AND ENERGY EFFICIENCY:

Data centers can be classically defined as large warehouses that host thousands of servers, switches, and storage devices to provide various data processing and retrieval services [528]. Intra Data Center Networking (DCN), defined by the topology (i.e. the connections between the servers and switches), links capacity, and the switching technologies utilized and routing protocols, is an important design aspect that impacts the performance, power consumption, scalability, resilience, and cost. Data centers have been successfully hosting legacy web applications but are challenged by the need to host an increasing number of big data and cloud-based applications with elastic requirements, multi-tenant, and heterogeneous workloads. Such requirements are contentiously challenging data center architectures to improve their scalability, agility, and energy efficiency while providing high performance and low latency. The rest of this Section is organized as follows: Subsection VI-A reviews electronic switching-based data centers, while Subsection VI-B reviews proposed and demonstrated hybrid electronic/optical and optical switching-based data centers. Subsection VI-C briefly describe HPC clusters, and disaggregated data centers. Subsection VI-D presents traffic characteristics in cloud data centers, while Subsection VI-E reviews intra DCN routing protocol and scheduling mechanisms. Finally, Subsection VI-F addresses the energy efficiency in data centers.

### *A. Electronic Switching Data Centers:*

Extensive surveys on the categorization and characterization of different data center topologies and infrastructures are presented in [529]-[535]. In what follows, we briefly review some of state-of-the-art electronic switching DCN topologies while emphasizing their suitability for big data and cloud applications. Servers in data centers are typically organized in "racks" where each rack typically accommodates between 16 and 32 servers. A Top-of-Rack (ToR) switch (also known as access or edge switch) is used to provide direct connections between the rack's servers and indirect connections with other racks via higher layer/layers switches according to the DCN topology. Most of legacy DCNs have a multi-rooted tree structure where the ToR layer is connected either to an upper core layer (two-tiers) or upper aggregation and core layers (three-tiers) [528]. For various improvement purposes, alternative designs based on Clos networks, flattened connections with high-radix switches, unstructured connections, and wireless transceivers were also considered. These architectures can be classified as switch-centric as the servers are only connected to ToR switches and the routing functionalities are exclusive to the switches. Another class of DCNs, known as server-centric, utilizes the servers/set of servers with multiport NIC and software-based routing to aid the process of traffic forwarding. A brief description of some electronic switching DCNs is provided below and some examples of small topologies are illustrated in Figure 12 showing the architecture in each case:

- **Three-tier data centers** [528]: Three-tier designs have access, aggregation, and core layers (Figure 12(a)). Different subsets of ToR/access switches are connected to aggregation switches which connect to core switches with higher capacity to ensure all-to-all racks connectivity. This increases the over-subscription ratio as the bisection bandwidth between different layers varies due to link sharing. Supported by firewall, load balancing, and security features in their expensive switches, three-tier data centers were well-suited for legacy Internet-based services with dominant south-north traffic and were widely adopted in production data centers.
- ***k*-ary Fat-tree** [536]: Fat-tree was proposed to provide 1:1 oversubscription and multiple equal-cost paths between servers in a cost-effective manner by utilizing commodity switches with the same number of ports ($k$) at all layers. Fat-tree organizes sets of equal edge and aggregation switches in *pods* and connects each pod as a complete bipartite graph. Each edge switch is connected to a fixed number of servers and each pod is connected to all core switches forming a folded-Clos network (Figure 12(b)). The Fat-tree architecture is widely considered in industry and research [529] indicating its efficiency with various workloads. However, its wiring complexities increase massively with scaling.
- **VL2** [537]: Is a three-tier Clos-based topology with 1:1 oversubscription proposed to provide performance isolation, load balancing, resilience, and agility in workload placements by using a flat layer-2 addressing scheme with address resolution. VL2 suits virtualization and multi-tenants, however, its wiring complexities are high.
- **Flattened ButterFLY (FBFLY)** [538]: FBFLY is a cost-efficient topology that flattens $k$-ary $n$-fly butterfly networks into $k$-ary $n$-flat networks by merging the $n$ switches in each row into a single high-radix switch. FBFLYs improve the path diversity of butterfly networks, and achieve folded-Clos network performance under load-balanced traffic with half the costs. However, with random adversarial traffic patterns, both load balancing and routing become challenging, thus, FBFLY is not widely considered for big data applications.
- **HyperX** [539]: HyperX is a direct network of switches proposed as an extension to hypercube and flattened butterfly networks. Further design flexibility is provided as several regular or general configurations are possible. For load-balanced traffic, HyperX achieved the performance of folded Clos with fewer switches. Like FBFLY, HyperX did not explicitly target improving big data applications.

- **Spine-leaf** ([e.g. [540], [541]]): Spine-leaf DCNs are folded Clos-based architectures that gained widespread adoption by industry as they utilize commercially-available high-capacity and high-radix switches. Spine-leaf allows flexibility in the number of spine, leaf, and servers per leaf and links capacities at all layers (e.g. in Figure 12(c)). Hence, controllable oversubscription according to cost-performance trade-offs can be attained. Their commercial usage indicates acceptable performance with big data and cloud applications. However, wiring complexities are still high.
- **BCube** [542] and **MDCube** [543]: BCube is a generalized hyper cube-based architecture that targets modular data centers with scales that fit in shipping containers. The scaling in BCube is recursive where the first building block "BCube0" is composed of $n$ servers and an $n$-port commodity switch and the $k^{th}$ level (i.e. BCube$_k$) is composed of $n$ BCube$_{k-1}$ and $n^k$ $n$-port switches. Figure 12(d) shows a BCube$_1$ with $n$=4. For its multipath routing and to provide low latency and high bisection bandwidth and fault-tolerance, BCube utilizes switches and servers equipped with multiple ports to connect with switches at different levels. BCube is hence, suitable for several traffic patterns such as 1-1, 1-many, one-all, and all-all which arise in big data workloads. However, with large scales, lower level to higher level bottlenecks increase and address space have to be overwritten. For larger scales, MDCube in [543] was proposed to interconnect BCube containers by high speed links in 1D or 2D connections.
- **CamCube** [544]: CamCube is a server-centric architecture that directly connects servers in a 3D torus topology where each server is connected to other neighbouring 6 servers. In CamCube, switches are not needed for intra routing which reduces costs and energy consumption. A Greedy key-based routing algorithm, Symbiotic routing, is utilized at the servers which enables applications-specific routing and arbitrary in-network functions such as caching and aggregation at each hop which can improve big data analytics [214]. However, CamCube might not suit delay sensitive workloads with high number of servers due to routing complexities, longer paths, and high store-and-forward delay.
- **DCell** [545]: DCell$_k$ is a recursively-scaled data center that utilizes a commodity switch per DCell$_0$ pod to connect its servers, and the remaining of the ($k$+1) ports in the servers for direct connections with servers in other pods of same level and in higher levels pods. Figure 12(e) shows a DCell$_1$ with 4 servers per pod. DCell provides high bandwidth, scalability, and fault-tolerance at low costs. In addition, under all-all, many-1, and 1-many traffic patterns, DCell achieves balanced routing, which ensures high performance for big data applications. However, as it scales, longer paths between servers in different levels are required.
- **FiConn** [546]: FiConn is a server-centric DCN that utilizes switches and dual port servers to recursively scale while maintaining low diameter and high bandwidth at reduced cost and wiring complexity compared to BCube and DCell. In FiConn$_0$, a port in each server is connected to the switch and in each level, half of the remaining ports in the pods are reserved for the connections with servers in the next level. For example, Figure 12(f) shows a FiConn$_2$ with 4 servers per FiConn$_0$. Real-time requirements are supported by employing a small diameter and hop-by-hop traffic-aware routing according to the network condition. This also improves the handling of bursty traffic of big data applications.
- **Unstructured data centers with random connections**: With the aim of reducing the average path lengths and easing incremental expansions, unstructured DCNs based on random graphs such as Jellyfish [547], Scafida [548], and Small-World Data Center (SWDC) [549] were proposed. Jellyfish [547] creates random connections between homogeneous or heterogeneous ToR switches and connect hosts to the remaining ports to support incremental scaling while achieving higher throughput due to low average path lengths. Scafida [548], is an asymmetric scale-free data center that incrementally scale under limits on the longest path length. In Scafida, two disjoint paths are assigned per switch pairs to ensure high resilience. SWDC [549] includes its servers in the routing and connects them in small-world-inspired distribution connections. Unstructured DCNs, however, have routing and wiring complexities and their performance with big data workloads is not widely addressed.
- **Data centers with wireless 60 GHz radio transceivers**: To improve the performance of tree-based DCNs without additional wiring complexity, the use of wireless transceivers at servers or ToR switches was also proposed [550], [551].

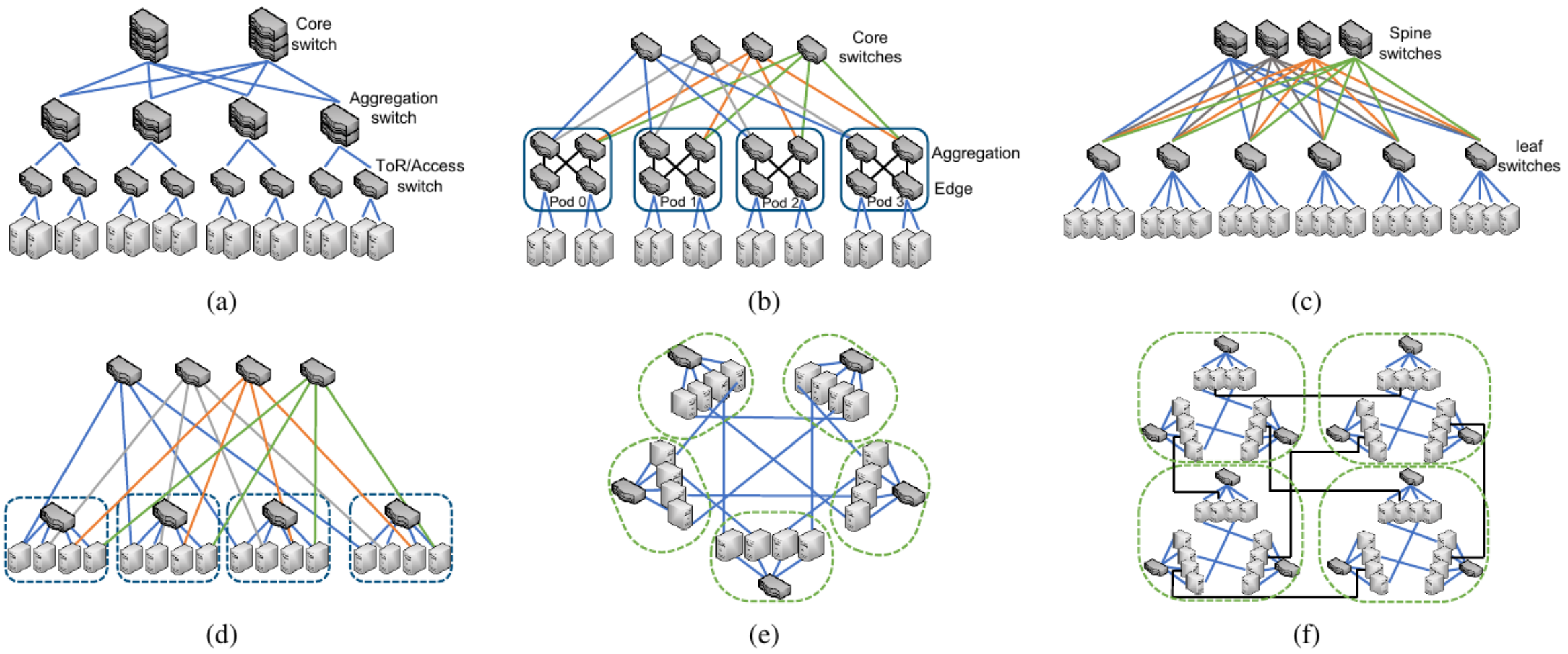

Fig. 12. Examples of electronic switching DCNs (a) Three-tier, (b) Fat-tree, (c) Spine-leaf, (d) BCube, (e) DCell, and (f) FiConn.

*B. Hybrid Electronic/Optical and All Optical Switching Data Centers:*

Optical switching technologies have been proposed for full or partial use in DCNs as solutions to overcome the bandwidth limitations of electronic switching, reduce costs, and to improve the performance and energy efficiency [552]-[557]. Such technologies eliminate the need for O/E/O conversion at intermediate hops and make the interconnections data-rate agnostic. Hybrid architectures add Optical Circuit Switching (OCS), typically realized with Micro-Electro-Mechanical System Switches (MEMSs) or free-space links, to enhance the capacity of an existing Electronic Packet Switching (EPS) network. To benefit from both technologies, bursty traffic (i.e. for mice flows) is offloaded to EPS while bulky traffic (i.e. for elephant flows) is offloaded to the OCS. MEMS-based OCS requires reconfiguration time in the scale of ms or μs before setting paths between pairs of ToR switches, and because packet headers are not processed, external control is needed for the reconfigurations. Another shortcoming of MEMS is their limited port count. WDM technology can increase the capacity of ports without huge increase in the power consumption [270], resolve wavelength contention, and reduce wiring complexities at the cost of additional devices for multiplexing, de-multiplexing, and fast tuning lasers and tuneable transceivers at ToRs or servers. In addition, most of the advances in optical networking discussed in Subsection IV-B6 have also been considered for DCNs such as OFDM [558], PONs technologies [559]-[573], and EONs [574]-[576].

In hybrid and all optical DCNs, both active and passive components were considered. The passive components including fibers, waveguides, splitters, couplers, Arrayed Waveguide Gratings (AWGs), and Arrayed Waveguide Grating Routers (AWGRs), do not consume power but have insertion loss, crosstalk, and attenuation losses. Active components include Wavelength Selective Switches (WSSs), that can be configured to route different sets of wavelengths out of a total of $M$ wavelengths in an input port to $N$ different output ports (i.e. $1\times N$ switch), MEMSs, Semiconductor Optical Amplifiers (SOAs) that can provide switching time in the range of ns, Tuneable Wavelength Converters (TWCs), and Mach-Zehnder Interferometer (MZI) which are external modulators based on controllable phase shifts in split optical signals. In addition to OCS, Optical Packet Switching (OPS) [577]-580] was also considered with or without intermediate electronic buffering. Examples of hybrid electrical/optical and all optical switching DCNs are summarized below, and some are illustrated in Figure 13:

- **c-Through** [581]: In c-Through, electronic ToR switches are connected to a two-tier EPSs network and a MEMS-based OCS as depicted in Figure 13(a). The EPS maintains persistent but low bandwidth connections between all ToRs and handles mice flows, while the OCS must be configured to provide high bandwidth links between pairs of ToRs at a time to handle elephant flows. As the MEMS used have ms switching time, c-Through was only proven to improve the performance of workloads with slowly varying traffic.
- **Helios** [582]: In Helios, electronic ToR switches are connected to a single tier containing arbitrary number of EPSs and MEMS-based OCSs as in Figure 13(b). Helios performs WDM multiplexing in the OCS links and hence requires WDM transceivers in the ToRs. Due to its complex control, Helios was demonstrated to improve the performance of applications with second-scale traffic stability.

- **Mordia** [583]: Mordia is a 24-port OCS prototype based on ring connection between ToRs each with 2D MEMS-based WSS that provides 11.5 µs reconfiguration time at 65% of electronic switching efficiency. Mordia can support unicast, multicast, and broadcast circuits, and enables both long and short flows offloading which makes it suitable for big data workloads. However, it has limited scalability as each source-destination needs a dedicated wavelength.
- **Optical Switching Architecture (OSA)** [584] / **Proteus** [585]: OSA and Proteus utilize a single MEMS-based optical switching matrix to dynamically change the physical topology of electronic ToRs connections. Each ToR is connected to the MEMS via an optical module that contains multiplexers/demultiplexers for WDM, a WSS, circulators, and couplers as depicted in Figure 13(c). This flexible design allows multiple connections per ToR to handle elephant flows and eliminates blocking for mice flows by enabling multi-hop connections via relaying ToRs. OSA was examined with bulk transfers and mice flows and minimal overheads were reported while achieving 60%-100% non-blocking bisection bandwidth.
- **Data center Optical Switch (DOS)** [586]: DOS utilizes an $N$+1-ports AWGR to connect $N$ ToR electronic switches through OPS with the aid of optical label extractors as shown in Figure 13(d). Each ToR is connected via a TWC to the AWGR to enable it to connect to one other ToR at a time. At the same time, each ToR can receive from multiple ToR simultaneously. The last ports on the AWGR are connected to an electronic buffer to resolve contention for transmitting ToRs. DOS suits applications with bursty traffic patterns, however, its disadvantages include the limited scalability of AWGRs and the power hungry buffering.
- **Petabit** [587]: Petabit is a bufferless and high-radix OPS architecture that utilizes three stages of AWGRs (Input, central, and output) in addition to TWCs as depicted in Figure 13(e). At each stage the wavelength can be tuned to a different one according to the contention at the next stage. However, electronic buffering at the ToRs and effective scheduling are required to achieve high throughput. Petabit can scale without impacting latency and thus can guarantee high performance for applications even at large scales.
- **Free-Space Optics (FSO)-based data centers**: Using FSO-based interconnection to link ToRs using mirrors in roofs in DCNs was proposed by several studies such as FireFly [588], and Patch Panels [589].

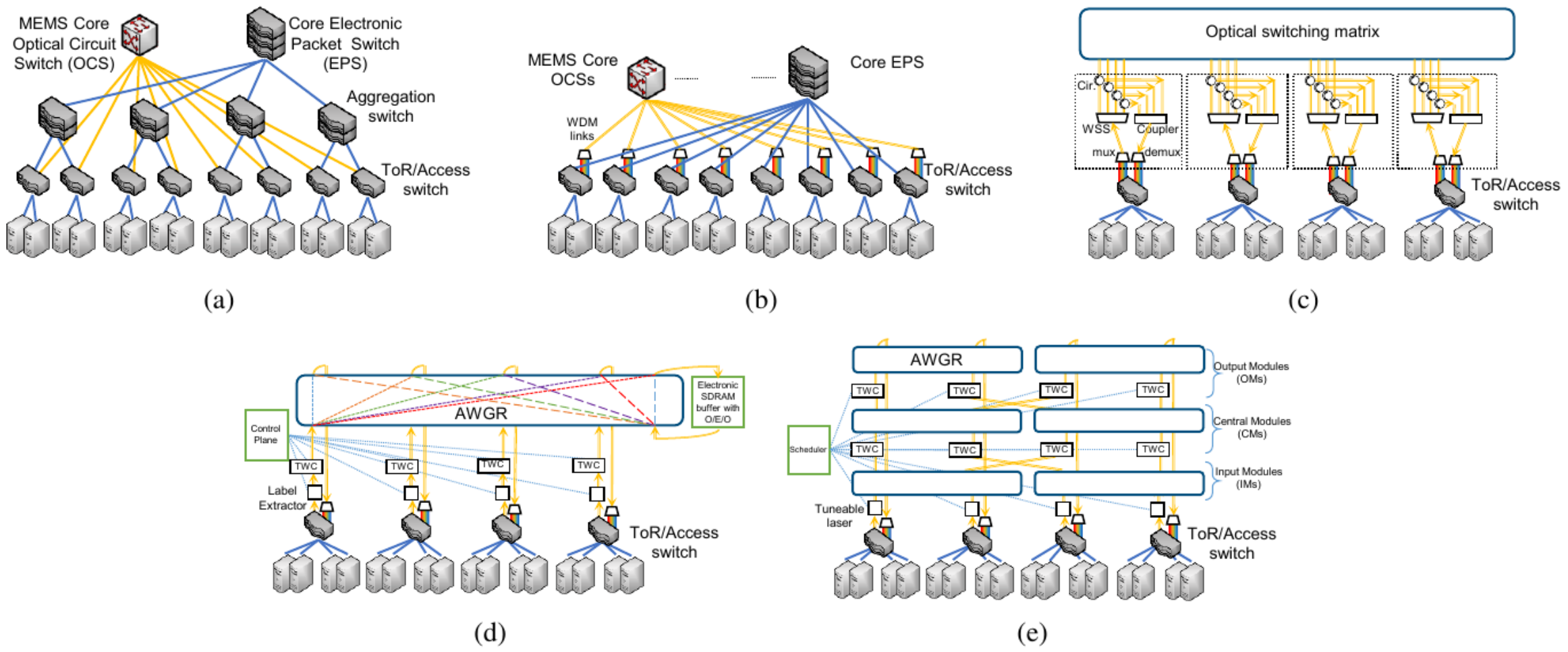

Fig. 13. Examples of hybrid/all optical switching DCNs (a) c-Through, (b) Helios, (c) OSA/Proteus, (d) DOS, and (e) Petabit.

### *C. HPC Clusters and Disaggregated Data Centers:*

The DCNs summarized in the previous two Subsections target production and enterprise data centers with general-purpose usage. Alternatively, HPC clusters target specific set of compute-intensive application and are thus designed with specialized servers, accelerators such as Graphical Processing Units (GPUs), in addition to having dedicated high speed interconnections to parallel storage systems such as InfiniBand and Myrinet, Torus or mesh topologies, and photonic interconnects for chip-chip, board-board, blade-blade, and rack-rack links [590]-[592]. According to the International Data Corporation (IDC), more than 67% of HPC facilities performed big data analytics [78]. Another variation to conventional data centers is disaggregating the CPU, memory, IO, and network resources at rack, pod, or entire data center level to achieve utilization and power efficiency advantages over legacy single-box servers [593]-[599]. Combinations of big data applications with uncorrelated resources demands can be deployed in disaggregated data centers with higher utilization, energy efficiency, and performance [593].

*D. Characteristics of Traffic inside Data Centers:*

Traffic characteristics within enterprise, cloud, social networking, and university campus data centers have been reported and analysed in [600]-[604] to provide several insights about traffic patterns, volume variations, and congestion, in addition to various flows statistics such as their duration, arrivals, and inter-arrival times. Intra data center traffic is mainly composed of different mixes of data center applications traffic including retrieval services and big data analytics, and provisioning operations such as data transfers, replication and backups. The first three pioneering studies by Microsoft Research [600]-[602] pointed that traffic monitoring tools used by ISPs in WANs do not suit data centers environments as their traffic characteristics are not statistically similar. The authors in [600] utilized low-overhead socket-level instrumentation at 1500 servers to collect application-aware traffic information. This cluster contained diverse workloads including MapReduce style workloads that generated 10GB per server per day. Two traffic patterns were defined; *Work-Seeks-Bandwidth*: for engineered applications with high locality, and *Scatter-Gather*: for applications that require servers to push or pull data from several others. It was found that 90% of the traffic stays inside the racks and that 80% of the flows last less than 10s, while less than 0.1% last more than 200s. The traffic patterns showed transient or stable spikes and the flows inter-arrivals were periodic short term bursts spaced by an average of 15ms and a maximum of 10s. In [601], an empirical study for traffic patterns in 19 tree-based two-tier and three-tier data centers for web-based services was carried based on coarse-grained measurements for links utilization and packets loss rate taken from Simple Network Management Protocol (SNMP) logs at routers and switches every five minutes. Average link loads were found the highest at the core switches and the highest packets losses were found at edge (i.e. ToR) switches. Additionally, fine-grained packet-level statistics at five edge switches in a smaller cluster showed a clear ON-OFF intensity and log-normal inter-arrivals during ON intervals. Reverse engineering methods to obtain fine-grained characteristics from SNMP logs were suggested for data center traffic generators and simulators.

SNMP logs were also utilized in [602] to study the traffic empirically while considering broader data centers usages and topologies. Those included 10 data centers with two-tier, three-tier, star-like, and Middle-of-Rack switches-based (i.e. connecting servers in several racks) topologies for university campuses, private enterprises, in addition to web-based services and big data analytics cloud data centers. The send/receive patterns of applications including authentication services, HTTP and secure HTTP-based, and local-use applications were examined at the flow-level to measure their effects on links utilization, congestion, and packets drop rate. It was found that 80% of the traffic stayed inside the racks in cloud data centers, while 40-90% left the racks in universities and enterprises data centers. Fine-grained packet traces from selected switches in 4 DCNs indicated that 80% of the flows are small (i.e. ≤ 10 kB), active flows were less than 10,000 per second per rack, and Inter-arrivals were less than 10 μs for 2-13% of the flows. The recent studies [603], and [604] presented the traffic characteristics inside Facebook's 4-tier Clos-based data center that hosts hundreds of thousands of 10 Gbps servers. In [603], wide monitoring tools and per-host packet-level traces were utilized to characterize the traffic while focusing on its locality, stability, and predictability. The practice in this architecture recommends assigning each machine to one role (e.g. cache, web server, Hadoop), localizing each type of workloads in certain racks, and varying the level of oversubscription according to the workload needs. Traffic was found to be neither fully rack-local nor all-all, and without ON-OFF behaviour. Also, it was found that servers communicated with up to 100s of servers concurrently, most of the flows were long-lived, and non-Hadoop packets were <200 Bytes. To capture fine-grained network behaviours such as μbursts (i.e. high utilization events lasting <1 ms), high-resolution measurements at the rack-level with granularity of tens/hundreds of μs were utilized in [604] for the same data center and application sets in the previous study. The measurements were based on a developed counter collection framework that polls packet counters and buffer utilization statistics every 25μs, and 50μs, respectively. It was found that high utilization events were short-lived as more than 70% of bursts lasted at most for tens of μs, and that load is very unbalanced as web and Hadoop racks had hot downlink ports while Cache had uplink hot ports. 90% of bursts lasted < 200μs for all application types, and < 50μs for Web racks, and the highest tail was recorded for Hadoop racks at 0.5 ms. It was noticed that the packets included in μbursts are larger than in the outside, μbursts were caused by application behavioural changes, and that the arrival rate of μbursts was not Poisson with 40% of inter-arrivals being > 100 μs for Cache and Web racks. Regarding the impact of μbursts on shared buffers, Hadoop racks had ports buffers at > 50% utilization, while web and cache racks had a maximum of 71% and 64% of their ports buffers at high utilization. Latency and packet loss measurements between VMs in different public clouds were presented and performed in [605] through a developed tool, PTPmesh, that aided cloud users in monitoring network conditions. Results for one way messaging delay between data centers in the same and different clouds were shown to range between μs and ms values. Specific traffic measurements for big data applications were presented in [279] and three traffic patterns; Single peak, repeated fixed-width peaks, varying heights and widths peaks were reported.

To forecast Hadoop traffic demands in cloud data centers, HadoopWatch which utilizes real-time file system monitoring tools was proposed in [606]. HadoopWatch monitors the meta-data and log files of Hadoop to extract accurate traffic information for importing, exporting, and shuffling data flows before entering the network by few seconds. Based on coflows information, a greedy joint optimization for scheduling and routing flows improved jobs completion time by about 14%. Several recent studies tackled improving the profiling, estimation, and generation of data center's traffic to aid in examining DCN TE, congestion control and load balancing methods [607]-[610]. The authors in [607] applied sketch-based streaming algorithms to profile data center traffic while considering its skewness among different services. To model spatial and temporal characteristics of traffic in large-scale DC systems, ECHO was developed in [610] as a scalable topology and workload-independent modeling scheme that utilizes hierarchical Markov Chains at the granularities of groups of racks, racks and servers. CREATE, a fine-grained Traffic Matrix (TM) estimation proposed in [608], utilized the sparsity of traffic between ToR switches to remove underutilized links. Then, the correlation coefficient between ToR switches in the reduced topology is extracted via SNMP counters and service placement logs to estimate the TM. In [609], random number generators for Poisson shot-noise processes were utilized to design a realistic TM generator at the flow-level between hosts in tree-based DCN while considering flows arrival rate, ratio in the same rack, duration, and size that can be used with packet-level traffic generators for DCN simulations.

### *E. Intra Data Centers Routing Protocols and Traffic Scheduling Mechanisms:*

Routing protocols, which define the rules for choosing the paths for flows or flowlets between source and destination servers, were extensively surveyed in [611]-[616]. Routing in DCNs can be static or adaptive where paths assignments can be dynamic according to criteria measured by a feedback mechanism. Adaptive routing or traffic scheduling can be centralized where a single controller is required to gather network-wide information and to distribute routing and rate decisions to switches and servers, or distributed where the decisions are taken independently by the switches or servers according to local decision based on partial view of the network. Centralized mechanisms provide optimal decisions but have limited scalability while distributed mechanisms are scalable but not always optimal.

Tree-based data centers such as three-tier designs typically utilize VLAN with Spanning Tree Protocol (STP), which is a simple Layer2 protocol that eliminates loops by disabling redundant links and forcing the traffic to route through core switches. Spine-leaf DCNs typically use improved protocols such as Transparent Interconnection of Lots of Links (TRILL) or Shortest Path Bridging (SPB) that enables the utilization of all available links while ensuring loop-free routing. CONGA [617] was proposed as distributed flowlets routing mechanism for spine-leaf data centers, and achieves load balancing by utilizing leaf-leaf congestion feedback. Improved tree-based DCNs such as Fat-tree and server-centric DCNs require designing their routing protocols closely with their topological properties to fully exploit the topology. For example, Fat-tree requires specific routing with two-level forwarding tables for servers with fixed pre-defined addresses [536]. To select paths according to network bandwidth utilization in Fat-tree, DARD was proposed in [618] as a host-based distributed adaptive routing protocol. For agility, VL2 uses two addresses for servers; a Location-specific Address (LA), and an Application-specific Address (AA) [537]. For packets forwarding, VL2 employs Equal Cost Multi-Path (ECMP), which is a static layer3 routing protocol that distributes flows to paths by hashing, and Valiant Load Balancing (VLB), that randomly selects intermediate nodes between a source and a destination. BCube employs a Source Routing protocol (BSR) [542], DCell adopts a distributed routing protocol (DFR) [545], and JellyFish [547] uses a $k$-shortest paths algorithm. c-Through uses the Edmond's algorithm to obtain the MEMSs configurations from the traffic matrix, then the ToR switches traffic is sent via VLAN-based routing into the OSC or the EPS [581]. Helios has a complex control scheme of three modules; Topology Manager (TM), circuit switch manager (CSM), and Pod Switch Manager (PSM) [582]. Mordia utilizes a Traffic Matrix Scheduling (TMS) algorithm that obtains effective short-lived circuit schedules, based on predicted demands, that can be applied to configure the MEMS and WSSs sequentially. OSA, and Proteus use the maximum-weight b-matching problem to enable the connection of multiple ToR switches, configure the WSS to match capacities and then use shortest path-based routing [584].

Using the Transmission Control Protocol (TCP) of Internet in data center environments has been proved to be inefficient [613] due to the difference in the nature of their traffic, the higher sensitivity to incast, and the key requirement in data center applications of minimizing Flow Completion Time (FCT). Thus, different transport protocols were proposed for DCNs. DCTCP [619] provides similar or better throughput than TCP and guarantees low Round Trip Time (RTT) by active control of queue lengths in the switches. MPTCP [620] splits flows to sub-flows and balances the load across several paths via linked congestion control. However, it might perform excessive splitting which requires extensive CPU and memory resources at end hosts. D2TCP [621] is a Deadline-aware Data center TCP protocol that considers single paths for flows and performs load balancing. For FCT reduction, D3 proposed in [622] uses flow deadline information to control the transmission rate. pFabric [623] and PDQ [624] enable the prioritirization of the flows closet to completion, and DeTail [625] splits flows, and performs adaptive load balancing based on queues occupancy to reduce the highest FCT. Alternatively, the centralized schedulers; Orchestra [258], Varys [262], and Baraat [263], which are further elaborated in Subsection VII-C, target reducing the completion time of coflows which are sets of flows with applications-related semantics such as intermediate data shuffling in MapReduce.

Virtualization in data centers was surveyed in [626] with a focus on routing and resources management and in [627] while focusing on the techniques for network isolation in multi-tenant data centers. SDN has been widely considered for data centers as it can improve the load balancing, congestion detection and mitigation [421], [431], [628]. To allow users to make bandwidth reservation in data centers for their VM-to-VM communications, a centralized controller is used to determine the rate and path for each user's flow. SecondNet was proposed in [629] and is such a controller. Hedera in [630], detects elephant flows and maximizes their throughput via a centralized SDN-based controller. ElasticTree in [631] improves the energy efficiency of Fat-tree DCNs by dynamically switching-off sets of links and switches while meeting demands and maintaining fault-tolerance.

### *F. Energy Efficiency in Data Centers:*

The energy consumption in data center is attributed to servers and storage, networking devices, in addition to cooling, powering, and lightning facilities with percentages of 26%, 10%, 50%, 11%, and 3% respectively of the total energy consumption [632]. As the energy consumption of servers is becoming proportional to loads, hence their energy efficiency is improving faster, the portion of the networking is expected to increase [633]. In [632], techniques for modeling the data centers energy consumption were comprehensively surveyed where they were divided into, hardware-centric, and software-centric approaches. In [634], green metrics including the Power Usage Effectiveness (PUE) (defined as the total facility power over the IT equipment power), and measurement tools that can characterise emissions were surveyed to aid in sustaining distributed data centers.

Several studies considered reducing the energy consumption and costs in data centers at different levels [635]-[639]. For the hardware, dynamically switching off the idle components, proposing efficient hardware with inherent higher efficiency components, DVFS, and utilizing optical networking elements were considered. For example, to improve the energy proportionality of Ethernet switches, the Energy Efficient Ethernet (EEE) standard [639] was developed. EEE enables three states for interfaces which are active, idle with no transmission, and low power idle (i.e. deep sleep). Although EEE have gained industrial adoption, its activation is not advised due to uncertainty with its impact on applications performance [283]. Placement of workloads and VMs into fewer servers, and scheduling tasks to shave peak power usage were also proposed to balance the power consumption and utilization in data centers.

## VII. DATA CENTERS-FOCUSED OPTIMIZATION STUDIES

This Section summarizes a number of big data applications optimization studies that consider the characteristics of their hosting data centers including details such as improving their design and protocols or analyzing the impact of their computing and networking parameters on the applications' performance. Subsection

VII-A addresses the performance, scalability, flexibility, and energy consumption improvements and tradeoffs for big data applications under various data centers topologies and design considerations. Subsection VII-B focuses on the studies that improve intra data centers routing protocols to enhance the performance of big data applications while improving the load balancing and utilization of the data centers. Subsection VII-C discusses flows, coflows, and jobs scheduling optimization studies to achieve different applications and data centers performance goals. Finally, Subsection VII-D addresses the studies that utilize advanced technologies to scale up big data infrastructures and improve their performance. The studies presented in this Section are summarized in Tables V, and VI.

### *A. Data Center Topology:*

Evaluating the performance and energy efficiency of big data applications in different data centers topologies was considered in [204]-[209]. The authors in [204] modeled Hadoop clusters with up to 4 ToR switches and a core switch to measure the influence of the network on the performance. Several simplifications such as homogeneous servers and uniform data distribution were applied and model-based and experimental evaluations indicated that Hadoop scaled well enough under 9 different clusters configurations. The MRPerf simulator was utilized in [205] to study the effect of the data center topology on the performance of Hadoop while considering several parameters related to clusters (e.g. CPU, RAM, and disk resources), configurations (e.g. chunk size, number of map and reduce slots), and framework (e.g. data placement and task scheduling). DCell was compared to star and double-rack clusters with 72 nodes under the assumptions of 1 replica and no speculative execution and was found to improve sorting by 99% compared to double-rack clusters. The authors in [206] extended CloudSim simulator [110] as CloudSimExMapReduce to estimate the completion time of jobs in different data center topologies with different workload distributions. Compared to hypothetically optimal topology for MapReduce with a dedicated link for each intermediate data shuffling flow, CamCube provided the best performance. Different levels of intermediate data skew were also examined and worse performance was reported for all the topologies. In [207], we examined the effects of the data center network topology on the performance and energy efficiency of shuffling operations in MapReduce with sort workloads in different data centers with electronic, hybrid and all-optical switching technologies and different rate/server values. The results indicated that optical switching technologies achieved an average power consumption reduction by 54% compared to electronic switching data centers with comparable performance. In [208], the Network Power Effectiveness (NPE) defined as the ratio between the aggregate throughput and the power consumption was evaluated for six electronic switching data center topologies under regular and energy-aware routing. The power consumption of the switches, the server's NIC ports and CPU cores used to process and forward packets in server centric topologies were considered. The results indicated that FBFLY achieved the highest NPE followed by the server-centric data centers, and that NPE is slightly impacted by the topology size as the number of switches scales almost linearly with the data center size for the topologies examined. Design choices such as link speeds, oversubscription ratio, and buffer sizes in spine and leaf architectures with realistic web search queries with Poisson arrivals and heavy-tail size distribution were examined by simulations in [209]. It was found that ECMP is efficient only at link capacities higher than 10 Gbps, where those resulted in 40% degradation in the performance compared to ideal non-blocking switch. Higher oversubscription ratios degraded the performance only at 60% and higher loads. Examining spine and leaf switches queue sizes revealed that it is better to maintain consistency and that additional capacities are beneficial at leaf switches.

Flexible Fat-tree is proposed in [210] as an improvement and generalization of the Fat-tree topology in [536] to achieve higher aggregate bandwidth and richer paths by allowing uneven number of aggregation and access switches in the pods. With more aggregation switches, shuffling results indicated about 50% improvement in the completion time. As a cost-effective solution to improve oversubscribed production data centers, the concept of *flyways*, where additional on-demand wireless or wired links between congested ToRs, was introduced in [211] and further examined in [212]. Under sparse ToR-to-ToR bandwidth requirements, the results indicated that few *flyways* allocated in the right ToR switches improved the performance by up to 50% bringing it closer to 1:1 DCNs performance. The *flyways* can be 802.11g, 802.11n, or 60 GHz wireless links, or random wired connections for subset of the ToR switches via commodity switches. The wired connections, however, cannot help if the congestion is between unconnected ToRs. A central controller is proposed to gather the demands and utilize MPLS to forward the traffic over the oversubscribed link or one of the *flyways*. In [213], a spectrum efficient and failure tolerant design for wireless data centers with 60 GHz transceivers was examined for data mining applications. A spherical rack architecture based on bimodal degree distribution for the servers' connections was proposed to reduce the hop count and hence reduce the transmission time compared to traditional wireless data center with cylindrical racks and same degree Cayley graph connections. Challenges related to interference, path loss, and the optimization of hub servers' selection were addressed to improve the data transmission rate. Moreover, the efficiencies of executing MapReduce in failure prone environments (due to software and hardware failures) were simulated.

Several big data frameworks that tailor their computations to the data center topology or utilize their properties were proposed as in [214]-[216]. Camdoop in [214] is a MapReduce-like system that run in CamCube and exploits its topology by aggregating the intermediate data along the path to reduce workers. A window-based flow control protocol and independent disjoint spanning trees with the same root per reduce worker were used to provide load balancing. CamCube achieved improvements over switch centric, Hadoop and Dryad, and over Camdoop with TCP and Camdoop without aggregation. In [215], network-awareness and utilization of existing or attached networking hardware were proposed to improve the performance of query applications. In-network processing in ToR switches with attached Network-as-a-Service (NaaS) boxes was examined to partially reduce the data and hence reduce bandwidth usage and increase the queries throughput. For API transparency, a shim layer is added to perform software-based custom routing for the traffic through the NaaS boxes. A RAM-based key-value store in BCube [542]; RAMCube was proposed in [216] to address false failure detection in large data centers caused by network congestion, entire rack blockage due to ToR switch failure, and the traffic congestion during recovery. A symmetric multi-ring arrangement that restricts failure detection and recovery to one hop in BCube is proposed to provide fast fault recovery. Experimental evaluation for the throughput under single switch failure with 1 GbE NIC cards in the servers indicated that a maximum of 8.3 seconds is needed to fully transmit data from a recovery to a backup server.

The topologies of data centers were also considered in optimizing VM assignments for various applications as in [217]-[220]. A Traffic-aware VM Placement Problem (TVMPP) to improve the scalability of data centers was proposed in [217]. TVMPP follows two-tier approximating algorithm that leverages knowledge of traffic demands and the data center topology to co-allocate VMs with heavy traffic in nearby hosts. First, the hosts and the VMs are clustered separately and a 1-to-1 mapping that minimizes the aggregated traffic cost is performed. Then, each VM within each cluster is assigned to a single host. The gain as a result of TVMPP compared to random placements for different traffic patterns was examined and the results indicated that multi-level architectures such as BCube benefit more than tree-based architectures and that heterogeneous traffic leads to more benefits. To tackle intra data center network performance variability in multi-tenant data centers with network-unaware VM-based assignments, the work in [218] proposed *Oktopus* as an online network abstraction and virtualization framework to offer minimum bandwidth guarantees. *Oktopus* formulates virtual or oversubscribed virtual clusters to suit different types of cloud applications in terms of bandwidth requirements between their VMs. These allocations are based on greedy heuristics that are exposed to the data center topology, residual bandwidths in links, and current VMs allocation. The results showed that allocating VMs while accounting for the oversubscription ratio improved the completion time and reduced tenant costs by up to 75% while maintaining the revenue. In [219], a communication cost minimization-based heuristic: Traffic Amount First (TAF) was proposed for VMs to PMs assignments under architectural and resources constraints and was examined in three data centers topologies. Inter VM traffic was reduced by placing VMs with higher inter traffic in the same PM as much as possible. A Topology-independent resources allocation algorithm namely; NetDEO was designed in [220] based on swarm optimizations to gradually reallocate existing VMs and allocate newly accepted VMs based on matching resources and availability. NetDEO maintains the performance during network and servers upgrades and accounts for the topology in the VM placements.

The performance of big data applications in SDN-controlled electronic and hybrid electronic/optical switching data centers topologies was considered in [221]-[228]. To evaluate the impact of networking configurations on the performance of big data applications in SDN-controlled data centers with multi-racks before deployments, a Flow Optimized Route Configuration Engine (FORCE) was proposed in [221]. FORCE emulates building virtual topologies with OVS over the physical network controlled by SDN to enable optimizing the network and enhance the applications performance at run-time and improvements by up to 2.5 times were achieved. To address big data applications and the need for frequent reconfigurations, the work in [222] examined a ToR-level SDN-based topology modification in a hybrid data center with core MEMS switch and electrical

Ethernet-based switches at run-time. Different topology construction and routing mechanisms that jointly optimize the performance and network utilization were proposed for single aggregation, shuffling, and partially overlapped aggregation communication patterns. The authors accounted for reconfiguration delays by starting the applications early and accounted for the consistency in routing tables updates. The work in [223] experimentally examined the performance of MapReduce in two hybrid electronic/optical switching data centers namely c-Through and Helios with SDN control. An "observe-analyze-act" control framework based on OpenFlow was utilized for the configurations of the OCS and the packet networks. The authors addressed the hardware and software challenges and emphasized on the need for near real-time analysis of application requirements to optimally obtain hybrid switch scheduling decisions. The work in [225] addressed the challenges associated with handling long-lived elephant flows of background applications while running Hadoop jobs in Helios hybrid data centers with SDN control. Although SDN control for electronic switches can provide alternative routes to reduce congestion and allow prioritizing packets in the queues, such techniques largely increase the switches CPU and RAM requirements with elephant flows. Alternatively, [224] proposed detecting elephant flows and redirecting them via the high bandwidth OCS, to improve the performance of Hadoop. To reduce the latency of multi-hop connections between servers in 2D Torus data centers, the work in [225] proposed the use of SDN-controlled MEMS to bypass electronic links and directly connect the servers. Results based on an emulation testbed and all-to-all traffic pattern indicated that optical bypassing can reduce the end-to-end latency for 11 of the 15 hosts by 11%. To improve the efficiency of multicasting and incasting in workloads such as HDFS read, join, VM provisioning, and in-cluster software updates, a hybrid architecture based on Optical Space Switches (OSS) was proposed in [225] to establish point-to-point links on-demand to connect passive splitters and combiners. The splitters transparently duplicate the data optically at the line rate and the combiners aggregate incast traffic under orchestration system control with TDM. Compared with small scale electronic non-blocking switches, similar performance was obtained indicating potential gains with the optical accelerators at larger scale, where non-blocking performance is not attained by electronic switches. Unlike the above work which fully offloads multicast traffic to the optical layer, HERO in [227] was proposed to integrate optical passive splitters and FSO modules with electronic switching to handle multicast traffic. During the optical switches configuration, HERO multicasts through the electronic switches, then migrates to optical multicasting. HERO exhibited linear increase in the completion time with the increase in the flow sizes and significantly outperformed Binomial, and ring Message Passing Interface (MPI) broadcasting algorithms with electronic switching only for messages less than or equal to, and greater than 12 kBytes in size, respectively.

A software-defined and controlled hybrid OPS and OCS data center is proposed and examined in [228] for multi-tenants dynamic Virtual Data Center (VDC) provisioning. A topology manager was utlilized to build the VDCs with different provisioning granularities (i.e. macro and micro) in an Architecture-on-Demand (AoD) node with OPS, and OCS modules in addition to ToRs and servers. For VMs placement, a network-aware heuristic that jointly considers computing and networking resources and takes the wavelengths continuity, optical devices heterogeneity, and the VM dynamics into account was considered. Improvements by 30.1% and 14.6% were achieved by the hybrid topology with 8, and 18 wavelengths per ToR switch, respectively, compared to using OCS only. The work in [229] proposed a 3D MEMS crossbar to connect server blades for scalable stream processing systems. Software-based control, routing, and scheduling mechanisms were utilized to adapt the topology to graph computational needs while accounting for MEMS reconfiguration and protocol stack delays. To overcome the high blocking ratio and scalability limitations of single hop MEMS-based OCS, the authors in [230] proposed a distributed multi-hop OCS that utilizes WDM and SDM technologies integrated with multi-rooted tree based data centers. A multi wavelengths optical switch based on Microring Resonators (MR) is designed to ensure fast switching. A modification to Fat-tree by replacing half of the electronic core, aggregation, and access switches with the OCS was proposed and distributed control was utilized at each optical switch with low bandwidth copper links to interconnect and combine control with EPS. Compared with Hedera and DARD, much faster optical path setup (i.e. 126.144 μs) was achieved with much lower control messaging overhead.

*B. Data Center Routing:*

In [231], a "reduce tasks" placement problem was analyzed in multi-rack environments to decrease cross-rack traffic based on two greedy approaches. Compared to original Hadoop with random reduce task placements, up to 32% speedup in completion time among different workloads was achieved. A scalable DCN-aware load balancing technique for key distribution and routing in the shuffling phase of MapReduce was proposed in [232] while considering DCN bandwidth constraints and addressing data skewness. A centralized Heuristic with two subproblems; network flow and load balancing, was examined and compared to three state-of-the-art techniques, load balancing-based; LPT, fairness-based; LEEN, and the default routing algorithm with hash-based key distribution in MapReduce. For synthetic and realistic traffic, the network-aware load balancing algorithm outperformed the others by 40% in terms of completion time and achieved maximum load per reduce comparable to that of LPT. To improve shuffling and reduce its routing costs under varying data sizes and data reduction ratios, a joint intermediate data partitioning and aggregation scheme was proposed in [233]. A decomposition-based distributed online algorithm was proposed to dynamically adjust data partitioning by assigning keys with larger data sizes to reduce tasks closer to map tasks while optimizing the placement and migration of aggregators that merge the same key traffic from multiple map tasks before sending them to remote reduce tasks. For large scale computations (i.e. 100 keys), and compared to random hash-based partitioning with no aggregation and with random 6 aggregators placement, the scheme resulted in 50%, and 26% reduction in the completion time, respectively. DCP was proposed in [234] as an efficient and distributed cache sharing protocol to reduce the intra data center traffic in Fat-tree data centers by eliminating the need to retransmit redundant packets. It utilized a packets cache in each switch for the eliminations and a Bloom filter header field to store and share cached packets information among switches. Simulation results for 8-ary fat-tree data center showed that DCP eliminated the retransmission by 40-60%. To effectively use the bandwidth in BCube data centers, the work in [235] proposed and optimized two schemes for in-network aggregation at the servers and switches. The first for incast traffic was modeled as minimal incast aggregation tree problem, and the second for shuffling traffic was modeled as minimal shuffle aggregation subgraph problem. 2-approximation efficient algorithms named IRS-based, and SRS-based were suggested for incast and shuffling traffic, respectively. Moreover, an effective forwarding scheme based on in-switch and in-packet Bloom filters was utilized at a cost of 10 Bytes per packet to ease related flows identification. The results for SRS-based revealed traffic reduction by 32.87% for a small BCube, and 53.33 % for a large-scale BCube with 262,144 servers.

The optimization of data transfers throughput in scientific applications was addressed in [236] while considering the impact of combining different levels of TCP flows pipelining, parallelism, and concurrency and the heterogeneity in the files sizes. Recommendations were presented such as the use of pipelining only for file sizes less than a threshold related to the bandwidth latency product and with different levels related to file size ranges. To improve the throughput, routing scalability, and upgrade flexibility in electronic switching data centers with random topologies, Space Shuffle (S2) was proposed in [237] as a greedy key-based multi-path routing mechanism that operates in multiple ring spaces. Results based on fine-grained packet-level simulations that considered the finite sizes of shared buffers and forwarding tables and the acknowledgment (ACK) packets indicated improved scalability compared to Jellyfish, and higher throughput and shorter paths than SWDC and Fat-tree. However, the overheads associated with packets reordering were not considered. An oblivious distributed adaptive routing scheme in Clos-based data centers was proposed in [238] and was proven to converge to non-blocking assignments with minimal out-of-the-order packet delivery via approximate Markov models and simulations when the flows are sent at half the available bandwidth. While transmitting at full bandwidth with strictly and rearrangeable non-blocking routing resulted in exponential convergence time, the proposed approach converged in less than 80 μs in a 1152 nodes cluster and showed weak dependency on the network size at the cost of minimal delay due to retransmitting first packets of redirected flows. The work in [239] utilized stochastic permutations to generate bursty traffic flows and statistically evaluated the expected throughput of several layer 2 single path and multi-path routing protocols in oversubscribed spine-leaf data centers. Those included deterministic single path selections based on hashing source or destination, worst-case optimal single path, ECMP-like flow-level multipathing, and a stateful Round Robin-based packet-level multipathing (packets spraying). Simulation results indicated that the throughput of the ECMP-like multipath routing is less than the deterministic single path due to flow collisions as 40% of the flows experienced 2 fold slowdown and the packet spraying outperformed all examined protocols. The authors in [240] proposed DCMPTCP to improve MPTCP through three mechanisms; Fallback for rACk-local traffic (FACT) to reduce unnecessary sub-flows creation for rack-local traffic, ADAptive packet scheduler (ADA) to estimate flow lengths and enhance their division, and Signal sHARE control (SHARE) to enhance the congestion control for the short flows with many-to-one patterns. Compared with two congestion control mechanisms typically used with MPTCP; LIA and XMP, 40% reduction in inter rack flows transmission time was achieved.

The energy efficiency of routing big data applications traffic in data centers was considered in [241]-[243]. In [241], preemptive flow scheduling and energy efficient routing were combined to improve the utilization in Fat-tree data centers by maximizing switches sharing while exclusively assigning each flow to needed links during its schedule. Compared to links bandwidth sharing with ECMP, and flow preemption with ECMP, additional energy savings by 40% and 30% were achieved, respectively at the cost of increased average completion time. To improve the energy efficiency of MapReduce-like systems, the work in [242] examined combining VM assignments with TE. A heuristic; GEERA was designed to first cluster the VMs via minimum *k*-cut, and then assign them via local search, while accounting for the traffic patterns. Compared with other load-balancing techniques (i.e. Randomized Shortest Path (RSP), and integral optimal and fractional solutions), an average additional 20% energy saving was achieved, and total average savings by 60% in Fat-tree, and 30% in BCube data centers were achieved. A GreenDCN framework was proposed in [243] to green switch-centric data center networks (Fat-tree as focus) by optimizing VM placements and TE while considering the network features such as the regularity and role of switches, and the applications traffic patterns. The energy-efficient VM assignment algorithm (OptEEA), transforms VMs into super VMs with heavy traffic, assigns jobs to different pods through *k*-means clustering, and finally assigns super VMs to racks through minimum *k*-cut. Then, the energy-aware routing algorithm (EER) utilizes the first fit decreasing algorithm and MPTCP to balance the traffic across a minimized number of switches. Compared with greedy VM assignments and shortest path routing, GreenDCN achieved 50% reduction in the energy consumption.

The optimization of routing traffic between VMs and in multi tenant data center was considered in [244]-[246]. VirtualKnotter in [244] utilized a two-step heuristic to optimize VM placement in virtualized data centers while accounting for the congestion due to core switch over-subscription and unbalanced workload placements. While other TE schemes operate for fixed source and destination pairs, VirtualKnotter considered reallocating them through optimizing VM migration to further reduce the congestion. Compared to a baseline clustering-based VM placement, a reduction by 53% in congestion was achieved for production data center traffic. To allow effective multiplexing of applications with different routing requirements, the authors in [245] proposed an online Topology Switching (TS) abstraction that defines a different logical typology and routing mechanism per application according to its goals (e.g. bisection bandwidth, isolation, and resilience). Results based on simulations for an 8-ary Fat-tree indicated that the tasks achieved their goals with TS while a unified routing via ECMP with shortest path and isolation based on VLAN failed to guarantee the different goals. The work in [246] addressed the fairness of bandwidth allocation and link sharing between multiple applications in private cloud data centers. A distributed algorithm based on dual-based decomposition was utilized to assign link bandwidths for flows based on maximizing the social welfare across the applications while maintaining performance-centric fairness with controlled relaxation. The authors assumed that the bottlenecks are in the access link of the PM and considered workloads where half of the tasks communicate with the other half and no data skew. To evaluate the algorithm, two different scenarios for applications allocation and communication requirements were used and the results indicated the flexibility and the fast convergence of the proposed algorithm.

SDN-based solutions to optimize the routing of big data applications traffic in various data center topologies were discussed in [247]-[257]. MILPFlow was developed in [247] as a routing optimization tool set that utilizes MILP modeling based on an SDN topology description, that define the characteristics of the data center, to deploy data path rules in OpenFlow-based controllers. To improve the routing of shuffling traffic in Fat-tree data centers, an application-aware SDN routing scheme was proposed in [248]. The scheme included a Fat-tree manager, MapReduce manager, links load monitor, and a routing component. The Fat-tree manager maintains information about the properties of different connections to prioritize the assignment of flows with less paths flexibility. Emulation results indicated a reduction in the shuffling time by 20% and 10% compared to Round Robin-based ECMP under no skew, and with skew, respectively. Compared to Spanning Tree and Floodlight forwarding module (shortest path-based routing), a reduction around 60% was achieved. To enhance the shuffling between map and reduce VMs under background traffic in OpenFlow-controlled clusters, the work in [249] suggested dynamic flows assignment to queues with different rates in OVS and LINC software switches. Results showed that prioritizing Hadoop traffic and providing more bandwidth to straggler reduce tasks can reduce the completion time by 42% compared to solely using a 50 Mbps queue. In [250], a dynamic algorithm for workload balancing between different racks in a Hadoop cluster was proposed. The algorithm estimates the processing capabilities of each rack and accordingly modifies the allocation of unfinished tasks to racks with least completion time and higher computing capacity. The proposed algorithm was found to decrease the completion time of the slowest jobs by 50%. A network Overlay Framework (NoF) was proposed in [251] to gurantee the bandwidth requirements of Hadoop traffic at run time. NoF achieves this by defining networks topologies, setting communication paths, and prioritizing traffic. A fully virtualized spine-leaf cluster was utilized to examine the impact on job execution time when redirecting Hadoop flows through the overlay network controlled by NoF and a reduction in the completion time by 18-66% was achieved. An Application-Aware Network (ANN) platform with SDN-based adaptive TE was examined in [252] to control the traffic in Hadoop clusters at fine-grain level to achieve better performance and resources allocation. An Address Resolution Protocol (ARP) resolver algorithm for flooding avoidance was proposed instead of STP to update the routing tables. Compared with MapReduce in oversubscribed links, SDN-based TE resulted in completion time improvement between 16-337× for different workloads.

To dynamically reduce the volume of shuffling traffic and green data exchange, the work in [253] utilized spate coding-based middleboxes under SDN control in 2-tier data centers. The scheme uses a sampler to obtain side information (i.e. traffic from map to reduce tasks residing in the same node) in addition to the coder/decoder at the middleboxes. Then, OpenFlow is used to multicast the coded packets according to an optimized grouping of multicast trees. Compared to native Vanilla Hadoop, reduction in the exchanged volume by 43%, and 59% were achieved when allocating the middlebox in the aggregation and a ToR switch, respectively. Compared to Camdoop-like in-network aggregation, the reduction percentages were 13%, and 38%. As an improvement of MPTCP, the authors in [254] proposed a responsive centralized scheme that adds subflows dynamically and selects best route according to current traffic conditions in SDN-controlled Fat-tree data centers. A controller that employs Hedera's demand estimation to perform subflow route calculation and path selection, and a monitor per server to adjust the number of subflows dynamically were utilized. Compared with ECMP and Hedera under shuffling with background traffic, the responsive scheme achieved 12% improvement in completion time while utilizing a lower number of subflows. To overcome the scalability issues of centralized OpenFlow-based controllers, the work in [255] proposed a distributed algorithm at each switch; LocalFlow that exploits the knowledge of its active flows and defines the forwarding rules at flowlets, individual flows and sub-flows resolutions. LocalFlow targets data centers with symmetric topologies and can tolerate asymmetries caused by links and node failures. It allows but reduces flow spatial splitting, while aggregating flows per destination, only if the load imbalance exceeded a slack threshold. To avoid the pitfalls of packets reordering, the duplicate acknowledgment (dup-ACK) duration at end-hosts is slightly increased. LocalFlow improved the throughput by up to 171%, 19%, and 23% compared to ECMP, MPTCP, and Hedera, respectively. XPath was proposed in [256] to allow different applications to explicitly route their flows without the overheads of establishing paths and dynamically adding them to routing tables of commodity switches. A 2-step compression algorithm was utilized to enable pre-installing very huge number of desired paths into commodity switches IP tables in large-scale data centers with limited table sizes. First, to reduce the number of unique IDs for paths, non-conflicting paths such as converging and disjoint ones are clustered into path sets and each set is assigned a single ID. Second, the sets are mapped to IP addresses based on Longest Prefix Matching (LPM) to reduce the number of IP tables entries. Then, a logically centralized SDN-based controller can be utilized to dynamically update the IDs of the paths instead of the routing tables, and to handle failures. For MapReduce shuffling traffic, XPath enabled choosing non-conflicting parallel paths and achieved 3× completion time reduction compared to ECMP. The recent work in [257] discussed the need for applying Machine Learning (ML) techniques to bridge the gap between large-scale DCN topological information and various networking applications such as traffic prediction, monitoring, routing, and workloads placement. A methodology named Topology2Vec was proposed to generate topology representation results in the form of low dimensional vectors from the actual topologies while accounting for the dynamics of links and nodes availability due to failures or congestion. To demonstrate the usability of Topology2Vec, the problem of placing SDN controllers to reduce end-to-end delays was addressed. A summary of the data center focused optimization studies is given in Table V.

TABLE V SUMMARY OF DATA CENTER -FOCUSED OPTIMIZATION STUDIES - I

| Ref | Objective | Application/ platform | Tools | Benchmarks/workloads | Experimental Setup/Simulation environment |
|---|---|---|---|---|---|
| [204]* | Hadoop performance model for multi-rack clusters | Hadoop 0.21.0 | Modelling for processing time | MapReduce program to read, encrypt, then sort | 160 servers in 5 racks (4 CPU cores, 8GB RAM, 2 250GB disk), Gigabit Ethernet Cisco Catalyst 3750G and 2960-S, Bonnie++, Netperf |
| [205]* | Simulation approach to evaluate data center topology effects on Hadoop | Hadoop 1.x | ns-2 network simulator, DiskSim, C++, Tcl, Python | TeraSort, Search, Index | MRPerf-based simulations for (single rack, double rack, tree-based and DCell) topologies with 2, 4, 8, 16 nodes each with (2 Xeon Quad 2.5 GHz core, 4 750GB SATA), 1 Gbps Ethernet |
| [206]* | Evaluating job finish time for MapReduce workloads in different data centers topologies | Hadoop 1.x | CloudSimExMapReduce simulator, CloudDynTop | Scientific workloads | Simulations for hierachical, fat-tree, CamCube, DCell, and hypothetically optimal topology for MapReduce, 6-9 servers, 1 Gbps and 10 Gbps links |
| [207]* | Optimizing shuffling operations in electronic, hybrid, and optical data center topologies | Google's MapReduce | MILP | Sort | Spine-leaf, Fat-tree, BCube, DCell, c-Through, Helios, Proteus topologies with 16 nodes, 10 Gbps links |
| [208]* | Evaluation of data centers Network Power Effectiveness (NPE) Flyways: wireless or wired additional links on-demand between congested ToR switches | - | Regular and energy-aware routing algorithms | One-to-one and all-to-all TCP traffic flows | Simulations for Fat-tree, Vl2, FBFLY, BCube, DCell, FiConn with similar network diameter, ∼8000 or ∼100 servers, and a mix of 1, 10 GbE links |
| [209]* | Data path performance evaluation in spine-leaf architectures | - | Network simulation cradle | Realistic workload with a mix of small and large jobs, and bursty traffic | OMNET++-based Simulations for spine-leaf data center (100 servers in 4 racks) with oversubscription ratios (1:1, 2.5:1, 5:1) and 10, 40, 100 Gigabit Ethernet |
| [210]* | FFTree: Adjustment to pods in Fat-tree for higher bandwidth and flexibility | Google's MapReduce | Flexibility in pods design defined by edge offset, BulkSendApplication | WordCount | Ns3-based simulations for 16 Data Nodes in Linux containers connected by TapBridge NetDevice |
| [211]*, [212]* | Flyways: wireless or wired additional links on-demand between congested ToR switches | MapReduce-like data mining | Central controller, MPLS label switching, optimization problem | Production MapReduce workloads | Simulation for 1500 servers in 75 racks with 20 server per rack) and additional Flyways with (0.1, 0.6, 1, and 2) Gbps capacity per Flyway |
| [213]* | Failure-tolerant and spectrum efficient wireless data center for big data applications | - | Bimodal degree distribution, space and time division multiple access scheme | 125 Mbytes input data, 20 Mbytes intermediate data | Simulations for a spherical rack with 200 servers 10 of them are hub servers, |
| [214]* | Camdoop: MapReduce-like system in CamCube data centers | Hadoop 0.20.2, Dryad, Camdoop | key-based routing, in-network prcessing, spanning tree | WordCount, Sort | 27 servers CamCube (Quad Core 2.27 GHz CPU, 12GB RAM, 32GB SSD), 1 Gbps Quadport NIC, 2 1 Gbps dual NIC, packet level simulator for 512 server CamCube |
| [215]* | In-network distributed query processing system to increase throughput | Apache Solr | NaaS box attached to ToR for in-network processing | Queries from 75 clients | Solr cluster (1 master, 6 workers), 1 Gbps Ethernet for servers, 10 Gbps for NaaS box |
| [216]* | RAMCube : BCube-oriented design for resilient key-value store | Key-value store | RPC through Ethernet, one hop allocation to recovery server | Key-value workloads with set, get, delete operations | BCube(4,1) with 16 servers (2.27GHz Quad core CPU, 16GB RAM, 7200 RPM 1TB disk) and 8 8-port DLink GbE, ServerSwitch 1 Gbps NIC in each server |
| [217]* | Traffic-Aware VM Placement Problem (TVMPP) in data centers | - | Two-tier approximate algorithm | VM-to-VM global and partitioned and traffic, production traces | Tree, VL2, Fat-tree, and BCube data centers with 1024 VMs |
| [218]* | Oktopus: Intra data center network virtualization for predictable performance | - | Greedy allocation algorithms, rate-limiting at end hosts | Symmetric and asymmetric VM-to- VM static and dynamic traffic | 25 nodes with 5 ToRs and core switch (2.27GHz CPU, 4GB RAM, 1 Gbps), Simulations for multi-tenant data center with 16000 servers in 4 racks and 4 VMs per server |
| [219]* | VM to PM mapping based on architectural and resources constraints | - | Traffic Amount First (TAF) heuristic | Uniformly random VM-to-VM traffic | Simulations for Fat-tree, VL2, and Tree-based data centers in large-scale (60 VMs) and small-scale (10 VMs) settings |
| [220]* | NetDEO: VM placement in data centers data centers and efficient system upgrade | Multi-tier web applications | Meta-heuristic based on Simulated annealing | Synthesized traces | Simulations for data center networks with heterogeneous servers and different topologies (non-homogeneous Tree, FatTree, BCube) |
| [221]* | Emulator for SDN-based data centers with reconfigurable topologies | Hadoop 1.x | OVS | Simulated Hadoop shuffle traffic generator | Testbed with 1 primary server (2.4 GHz, 4GB RAM), 12 client workstations (dual-core CPU, 2GB RAM), 2 Pica8 Pronto 3290 SDN-enabled Gigabit Ethernet switches) |
| [222]* | SDN-approach in Application-Aware Networks (AAN) to improve Hadoop | Hadoop 2.x | Ryu, OVS, lightweight REST-API, ARP resolver | HiBench benchmark (sort, word count, scan, join, PageRank) | 1 master and 8 slaves in GENI testbed (2.67GHz CPU, 8GB RAM) 100 Mbps per link. |
| [223]* | Application-aware run-time SDN-based control for data centers with Hadoop | - | 2D Torus topology configuration algorithms | - | Hybrid data center with OpenFlow-enabled ToR electrical switches connected to an electrical core switch and MEMS-based core optical switch |
| [224]* | Experimental evaluation for MapReduce in c-Through and Helios | Hadoop 1.x | Topology manager, OpenFlow, circuit switch manager | TrintonSort (900 GB) and TCP long-lived traffic | 24 servers in 4 racks, 5 Gbps packet link and 10 Gbps optical link Monaco switches, Glimmerglass MEMS |
| [225]* | Measuring latency in Torus-based hybrid optical/electrical switching data centers | - | Network Orchestrator Module (NOM) | Real-time network traffic via Iperf | 2D Torus network testbed constructed with 4 MEMS sliced from a 96×96 MEMS- based CrossFiber LiteSwitcM and two Pica8 OpenFlow switches to emulate 16 NIC |
| [226]* | Acceleration of incast and multicast traffic with on-demand passive optical modules | - | Floodlight, Redis pub/sub messages, integer program to configure OSS | Two sets of 50 multicast jobs (500MB-5GB) with 4 and 7 groups | 8 nodes hybrid cluster testbed with optical splitters and combiners connected with controlled Optical Space Switch (OSS) |
| [227]* | HERO: Hybrid accelerated delivery for multicast traffic | MPICH for Message Passing Interface (MPI) | SDN controller, greedy optical links assignment algorithm | 50 multicast flows with uniform flows (100MB-1GB) and groups (10-100) | Small free-space optics mutlicast testbed with passive optical 1:9 splitters for throughput and delay evaluation, flow-level simulations for 100 racks in spine and leaf hybrid topology |
| [228]* | Multi-tenant virtual optical data center with network-aware VM placement | Virtual Data Center (VDC) composition | OpenDayLight, network-aware VM placement based on variance | 2 VDC, randomly generated 500 VDCs with Poisson requests arrival | FPGA optoelectronics 12×10GE ToRs, SOA-based OPS switch, Polatis 192×192 MEMS- based switch as optical back-plane, simulations for 20/40 servers per rack, 8 racks |
| [229]* | Reconfigurable 3D MEMS based data center for optimized stream computing systems | System S | Scheduling Optimizer for Distributed Applications (SODA) | Multiple jobs for streaming applications | 3 IBM Bladecenter-H chassis, 4 HS21 blade servers per chassis, Calient 320×320-port 3D MEMS |
| [230]* | Hybrid switching architecture for cloud applications in Fat-tree data centers | - | Multi wavelength optical switch, distributed control for scheduling | 3 synthetic traffic patterns with mice and elephant flows | OPNET simulations for the proposed hybrid OCS and EPS switching in Fat-tree with 128 servers to evaluate the average end-to-end delay and the network throughput. |
| [231]° | Reduce tasks placements to reduce cross-rack traffic | Hadoop 1.x | Linear-time greedy algorithm, binary search | WordCount, Grep, PageRank, *k*-means, Frequency Pattern Match | Cluster with 4 racks; A, B , C, and D with 7, 5, 4, and 4 servers (Intel Xeon E5504, E5520, E5620, and E5620 CPUs, 8GB RAM), 1 GbE links |
| [232]° | Data center network-aware load balancing to optimize shuffling in MapReduce with skewed data | - | Greedy algorithm | Synthetic and Wikipedia page-to-page link datasets | Simulations for 12-ary Fat-tree with 1 Gbps links |
| [233]° | Joint intermediate data partitioning and aggregation to improve the shuffling phase of MapReduce | - | Profiling, decomposition-based distributed algorithm | Dump files in Wikimedia, WordCount, random shuffling | Simulations for three-tier data center with 20 nodes |
| [234]° | DCP: Distributed cache protocol to reduce redundant packets transmission in Fat-tree | - | In-memory store, bloom filter header | Randomly generated packets with Zipf distribution | Simulations for Fat-tree with k=16 (1024 servers, 64 core switches, 28 aggregation, 128 edge), simulations |
| [235]° | In-network aggregation in BCube for scalable and efficient data shuffling | Hadoop 1.x | IRS-based incast, SRS-based shuffle, Bloom filter | WordCount with combiner | 61 VMs in 6 nodes emulated BCube (2 8-cores CPU, 24GB RAM, 1TB disk), large-scale simulations for BCube(8,5) with |
| [236]° | Application-level TCP tuning for data transfers through pipelining, parallelism, and concurrency | - | Recursive Chunk Division, Parallelism-Concurrency Pipelining | Bulk data transfers (512KB-2GB) per file | high-speed networking testbeds and cloud networks Emulab-based emulations, AWS EC2 instances |
| [237]° | Space Shuffle (S2): greedy routing on multiple rings to improve throughput, scalability, and flexibility | - | Greediest routing, MILP | Random permutation traffic | Fine-grained packet-level event-based simulations for the proposed data center, Jellyfish, SWDC, and Fat-tree |
| [238]° | Non-blocking distributed oblivious adaptive routing in Clos-based data centers for big data applications | - | Approx. Markov chain models to predict convergence time | Random permutation traffic of 245KB flows | OMNet++-based simulations for InfiniBand network (three-level Clos DCN with 24 input, 24 output, and 48 intermediate switches and 1152 nodes), 40 Gbps links |
| [239]° | Evaluation for different routing protocols on the performance of spine and leaf data centers | - | Stochastic permutations for traffic generation | Bursty traffic, delay-sensitive workloads | Flow-level simulations for spine and leaf data center with 8 spine switches, 16 leaf switches, and 128 end nodes |
| [240]° | DCMPTCP: improved MPTCP for load balancing in data centers with rack local and inter racks traffic | - | Fallback for rACk-local Traffic, ADAptive packet scheduler | Many-to-one traffic, data mining and web search traffic | Ns3-based simulations for Spine and leaf data center with 8 spine, 8 leaf switches, and 64 nodes per leaf switch, 10 and 40 Gbps links |
| [241]° | Greening data centers by Flow Preemption (FP) and Energy-Aware Routing (EAR) | - | Algorithm for the FP and EAR scheme | 10k flows wit exponential distribution with mean of 64MB | Simulations for 24-ary Fat-tree |
| [242]° | Improving the energy efficiency of routing in data centers by joint VM assignments and TE | - | Approximate-algorithm (GEERA) | Uniform random traffic and number of VMs | Fat-tree (4-ary and 8-ary), BCube(2,2) and BCube(8,2) |
| [243]° | GreenDCN: scalable framework to green data center network through optimized VM placement and TE | - | Time-slotted algorithms; optEEA, EER | Synthetic jobs with normal distribution for no. of servers | Simulations for 8-ary and 12-ary Fat-tree data centers, 2 VMs per server, identical switches with 300W max power and max processing speed of 1 Tbps |
| [244]° | VirtualKnotter: efficient online VM placement and migration to reduce congestion in data centers | - | Multiway θ-Kernighan-Lin and simulated annealing | Synthetic and realistic traffic | Dynamic simulations |
| [245]° | Topology switching to allow multiple applications to use different routing schemes | - | Allocator, centralized topology server | Randomly generated routing tasks | Simulations for 16-ary Fat-tree with 1 Gbps links |
| [246]° | Performance-centric fairness in links bandwidth allocation in private clouds data centers | - | Gradient projection-based distributed algorithm | Two scenarios for applications traffic | Simulations for a private data center with homogeneous nodes |
| [247]° | MILPFlow: a Tool set for modeling and data paths deployment in OpenFlow-based data centers | - | MILP, OVS | Video streaming, Iperf traffic | Mininet-based emulation for 4-ary Fat-tree in VirtualBox-based VMs, 1 GbE NIC |
| [248]° | Application-aware SDN routing in data centers | Hadoop 1.x | Floodlight controller, managers, links monitor, routing component | WordCount | EstiNet-based emulation for 20 OpenFlow switches in 4-ary Fat-tree with 16 nodes |
| [249]° | Hadoop acceleration in an OpenFlow-based cluster | Cloudera distribution of Hadoop | Floodlight, OVS and LINC switches | Sort (0.4 MB - 4GB), Iperf for background traffic | 10 VMs in 3 nodes in Cloudera connected by a physical switch |
| [250]° | SDN-controlled dynamic workload balancing to improve completion time of Hadoop jobs | - | Balancing algorithm based on estimation and prediction | WordCount | Mumak-based simulations for a data center with three racks |
| [251]° | SDN-based Network Overlay Framework (NoF) to define networks based on applications requirements | Hadoop 2.3.0 | Configuration engine, OVS, POX controller | TeraGen, TeraSort, iperf for background traffic | Virtualized testbed with spine-leaf virtual switches, VirtualBox for VMs |
| [252]° | SDN-approach in Application-Aware Networks (AAN) to improve Hadoop | Hadoop 1.x | Ryu, OVS, lightweight REST-API, ARP resolver | HiBench benchmark (sort, word count, scan, join, PageRank) | 1 master and 8 slaves in GENI testbed (2.67GHz CPU, 8GB RAM), 100 Mbps per link |
| [253]° | Dynamic control for data volumes through SDN control and spade coding in cloud data centers | Vanilla Hadoop | Sampler, spate coding-based middleboxes for coding/decoding | TeraSort, Grep | Prototype: 2-tier data center with 8 nodes (12 cores CPU, 128GB RAM, 1TB disk), Testbed: 8 VMs controlled by Citrix XenServer and connected with OVS |
| [254]° | Responsive multipath TCP for optimized shuffling in SDN-based data centers | - | Demand estimation and route calculation algorithms | Random, permutation, and shuffling traffic | NS3-based simulations for 8-ary Fat-tree with 1 Gbps links and customized SDN controller |
| [255]° | LocalFlow: local link balancing for scalable and optimal flow routing in data centers | - | Switch-local algorithm | pcap packet traces, MapReduce-style flows | Packet-level network simulations (stand-alone and htsim-based) for 8-ary and 16-ary Fat-tree, VL2, oversubscribed variations |
| [256]° | XPath: explicit flow-level path control in commodity switches in data centers | - | 2 step compression algorithm | Random TCP connections, sequential read/write, shuffling | Testbed: 6-ary Fat-tree testbed with Pronto Broadcom 48-port Ethernet switches, Algorithm evaluation for BCube, Fat-tree, HyperX, and VL2 with different scales |
| [257]° | Topology2Vec: DCN representation learning for networking applications | - | Biased random walk sampling ML-based network partitioning | Real-world Internet topologies from the Topology Zoo | Simulations |

* Data center topology, ° Data center routing.

### *C. Scheduling of Flows, Coflows, and Jobs in Data Centers:*

Scheduling big data traffic at the flow level was addressed in [258]-[261]. *Orchestra* [258], a task-aware centralized cluster manager, aimed to reduce the average completion time of various data transfer patterns for batch, iterative, and interactive workloads. Within each transfer, a Transfer Controller (TC) was used for monitoring and updating sources associated with destination sets. For broadcast, a BitTorrent-like protocol, *Cornet*, was utilized and for shuffle, an algorithm named Weighted Shuffle Scheduling (WSS) was used. Orchestra allows scheduling at the transfer level between applications stages where concurrent transfers belonging to the same job can be optimized through an Inter-Transfer Controller (ITC) that can utilize FIFO, fair, and priority-based scheduling. Broadcast was improved by 4.5 times compared to native Hadoop with status quo and high priority transfers were improved by 1.7 times. *Seawall* in [259] is an edge scheduler that used a hypervisor-based mechanism to ensure fair sharing of DCN links between tenants. *FlowComb* was proposed in [260] as a centralized and transparent network management framework to improve the utilization of clusters and reduce the processing time of big data applications. FlowComb utilized software agents installed in nodes to detect transfer requests and report to the centralized engine and OpenFlow to enforce forwarding rules and paths. An improvement of 35% in completion time of sort workloads was achieved with 60% path enforcement. In [261], distributed flow scheduling was proposed to achieve adaptive routing and load balancing through DiFS. For several traffic patterns, DiFS achieved better aggregate throughput compared to ECMP, and comparable or higher throughput compared to Hedera and DARD.

Optimizing coflows scheduling, which is more applications-aware than flows scheduling, to minimize the completion time of workloads was the focus in [262]-[265]. *Varys* was proposed in [262] as a coordinated inter-coflow scheduler in data centers targeting predictable performance and reduced completion time for big data applications. A greedy co-flow scheduler and a per-flow rate allocator were utilized to identify the slowest flow in a co-flow and adjust the rate of accompanying flows to the slowest one so networking resources can be used in other co-flows. Trace-driven simulations indicated that Varys achieved 3.66×, 5.53×, and 5.65× improvements compared to fair sharing, per-flow scheduling, and FIFO, respectively. *Baraat* in [263] suggested decentralized task-aware scheduling for co-flows to reduce their tail completion times. FIFO with limited multiplexing (FIFO-LM) is used to schedule the tasks while avoiding head-of-line blocking for small flows by changing the multiplexing level when heavy tasks arrive. Compared to pFabric [622] and *Orchestra*, the completion time of 95% of MapReduce workloads was reduced by 43% and 93%, respectively. A decentralized coflow-aware scheduling system (D-CAS) that dynamically and explicitly set the priorities of flows according to the maximum load in the senders was proposed in [264] to minimize coflows completion time. D-CAS achieved a performance close to Varys with up to 15% difference and outperformed Baraat by 1.4 and 4 times for homogeneous and heterogeneous workloads, respectively. *Rapier* in [265] integrated routing and scheduling at the coflow level to optimize the performance of big data applications in DCNs with commodity switches. Compared to Varys with ECMP and optimized routing only, about 80% and 60% improvement in coflow completion time was achieved.

Scheduling traffic in DCNs with SDN environments was addressed in [266]-[269]. Online scheduling of multicast flows in Fat-tree DCNs was considered in [266] to ensure bounds on congestion and improve the throughput and load balancing. A centralized Bounded Congestion Multicast Scheduling (BCMS) algorithm was developed for use with OpenFlow and improvements compared to VLB and random scheduling were reported. *Pythia* in [267] focused on improving the network performance under skewed workloads. A run-time intermediate data size prediction and centralized OpenFlow-based control were utilized and up to 46% reduction in completion time was obtained compared to ECMP. SDN Openflow network with Combined Input and Crosspoint Queued (CICQ) switches was proposed to dynamically schedule packets of different big data applications in [268]. An application-aware computing and networking resource allocation scheme based on neural network predictions was developed to allocate adequate resources so that the SLA is minimally violated. For five different DCN applications, the scheme achieved less than 4% SLA violation rate at the cost of 9.21% increase in the energy consumption. The authors in [269], proposed and experimentally demonstrated a heuristic for Bandwidth-Aware Scheduling with SDN (BASS) to reduce the minimum job completion time in Hadoop clusters. The heuristic prioritizes scheduling the tasks locally but considers remote assignment while accounting for links assignment for data transfers if the total completion time is reduced. BASS was found to reduce the completion time by 10%.

The work reported in [270]-[273] focused on scheduling traffic in optical and hybrid DCNs. In [270], the gaps between OCS interconnects performance and latency-sensitive applications requirements were addressed. Two scheduling algorithms: centralized Static Circuit Flexible Topology (SCFT) and distributed Flexible Circuit, Flexible Topology (FCFT) were proposed and up to 2.44× improvement over Mordia [283] was achieved. Resource allocation in NEPHELE was addressed in [271] while accounting for the SDN controller delay and random allocation of iterative MapReduce tasks. Compared to Mordia, NEPHELE uses multiple WDM rings instead of one to enable efficient use of resources, and introduces an application-aware and a feedback-based synchronous slotted scheduling algorithms. In [272], different end-to-end continuous-time scheduling algorithms were designed for a proposed hybrid packet optical DCNs by the same authors, and were compared using random traffic. Effective traffic scheduling for a proposed Packet-Switched Optical Network (PSON) DCN with space switches and layers of AWGRs was examined in [273]. To treat traffic flows differently and optimally based on their type, three machine-learning flow detection algorithms were used and were compared in terms of accuracy and classification speed. Scheduling algorithms that consider priority of flows, and occupancy of buffers were proposed and improvements in terms of packet loss ratio and average delay compared to Round Robin were reported.

The authors in [274] maximized the throughput of data retrieval for single and multiple applications in tree-based DCNs while accounting for links bandwidths and allocation of data replicas. The proposed approximation algorithm achieved near optimal performance and improved retrieval time compared to random scheduling. A topology-aware heuristic for data placement that minimizes remote data access costs in geo-distributed MapReduce clusters was proposed in [275] based on optimized replica balanced distribution tree. *CoMan* was proposed in [276] to improve bandwidth utilization and reduce completion time of several big data frameworks in Multiplexed Datacenters. Virtual Link Group (VLG) abstraction was utilized to define shared bandwidth resources pool and an approximation algorithm was developed. Compared to ECMP with ElasticSwitch [133], the bandwidth utilization and completion time were improved by 2.83× and 6.68×, respectively. The authors in [177] proposed probabilistic map and reduce tasks scheduling algorithms that consider the data center topology and the links statuses in computing the cost and latency of data transmission. Based on the calculations, the algorithm incorporates randomness in the assignments where the tasks with the least transmission time get scheduled on available slots with higher probability. Thus, the locality and completion time are balanced without enforcement. Compared to a coupling scheduling method and to fair scheduling, reductions by 17%, and 46% were achieved. A Network-Aware Scheduler (NAS) was examined in [278] for improving the shuffling phase in clusters with multi racks. NAS utilizes three algorithms: Map Task Scheduling (MTS) to balance cross node traffic caused by skewness, Congestion-Avoidance Reduce Task Scheduling (CA-RTS) to reduce cross-rack traffic, and Congestion-Reduction RTS (CR-RTS) to prioritize light-shuffle traffic jobs. Compared to state-of-the-art schedulers like fair and delay, improvements in the throughput by up to 62%, and reductions in the average completion time by up to 44% and cross rack traffic by up to 40% were reported. A fine-grained framework; Time-Interleaved Virtual Cluster (TIVC) that accounts for the dynamics of big data applications bandwidth requirements was proposed and tested in a system named PROTEUS in [279]. Based on observing, via profiling four workloads, that the traffic peaked only during 30-60% of the execution time, TIVC targeted reducing the over-utilization of bandwidth and allowing overlapped scheduling of jobs. First, the user's application was profiled and a cost model based on bandwidth cap and performance trade-offs was generated, then a dynamic programming-based spatio-temporal jobs allocation algorithm was used to maximize the utilization and revenue. Compared with Oktopus in [218] for mixed batch jobs, a reduction by 34.5% in the completion time was achieved. For dynamically arriving jobs at 80% load, PROTEUS reduced rejection ratio to 3.4% from 9.5% reported for Oktopus.

The authors in [280] explored the benefits of ahead-planning with MapReduce and DAG-based production workloads and developed *Corral* that jointly optimizes data placement and tasks scheduling at the rack-level. Corral modified HDFS and AM to enable rack selection with data and tasks placement and hence, increase the locality and reduce the overall cross-rack traffic. Compared to native YARN, with capacity scheduler, the makespan was reduced by up to 33% while reducing cross-rack traffic by up to 90%. Simulation results for integrating Corral with Varys [262] showed clear improvement over using capacity scheduler with Varys and over using Corral with TCP indicating the importance of integrating network and data/tasks scheduling solutions. The authors in [281] proposed *Mercury* as an extension of YARN that enables hybrid resource management in large clusters by allowing AM of applications to request guaranteed containers via a centralized scheduling or queueable containers via distributed scheduling according to their needs. Compared to default YARN, an average improvement by 35% in throughput was obtained. The work in [282], optimized the allocation of container resources in terms of reduced congestion and improved data locality by considering the networking resources and the data locations. Compared to default placement in YARN, 67% reduction in completion time for network-extensive workloads was obtained.

The energy efficiency of data centers through workloads and traffic scheduling was considered in [283]-286]. The work in [283] examined the performance-energy tradeoffs when using the Low Power Idle (LPI) link sleep mode of the EEE standard [639] with MapReduce workloads. The timing parameters of entering and leaving the LPI mode in 10GbE links were optimized while utilizing packet coalescing (i.e delaying outgoing packets during the quite mode and aggregating them for transmission in the following active mode). Depending on the superscription ratio and workloads, EEE achieved power saving between 5 and 8 times compared to legacy Ethernet. A detailed study of the optimum packet coalescing setting under different TCP settings and with shallow and deep buffers (i.e. 128kB and 10MB per port) was provided and an additional improvement by a factor of two was achieved. Applications scheduling in fewer machines for energy saving in networking devices of large scale data centers was considered in [284] through a Hierarchical Scheduling Algorithm (HSA). The workloads were allocated at the nodes according to their subsets (i.e. connectivity with ToR switches), then according to the levels of higher layer switches and associate data transmission costs. HSA was tested with a Dynamic Max Node Sorting Method and two classical bin packing problem solutions and the proposed method resulted in improved performance and stability. *Willow* in [285] aimed to reduce the energy consumption of switches in Fat-tree through SDN-based dynamic scheduling while accounting for the performance of network-limited elastic flows. The number of activated switches as well as usage duration was considered in computing the routing path per flow. Compared to ECMP and classical heuristics (i.e. simulated annealing and particle swarm optimization), up to 60% savings were achieved. The authors in [286] proposed *JouleMR* as a green-aware and cost-effective tasks and jobs scheduling framework for MapReduce workloads that maximized renewable energy usage while accounting for brown energy dynamic pricing and renewable energy storage. To obtain a plan for the jobs/tasks scheduling, performance-energy consumption models were utilized to guide a two-phase heuristic. The first phase optimizes the start time of tasks/jobs to reduce the brown energy usage and the second phase optimizes the assigned resources per task/job to further reduce it while satisfying soft deadlines. Compared to Hadoop with YARN, a reduction by 35% in brown energy usage and by 21% in overall energy consumption was obtained.

*D. Performance Improvements based on Advances in Computing, Storage, and Networking Technologies:*

Comparisons of scale-up and scale-out systems for Hadoop were discussed in [287]-[290]. In [287], the authors showed that running Hadoop workloads with sub Tera-scale on a single scaled-up server can be more efficient than scale-out clusters. For the scale-up system, several optimizations were suggested to optimize storage, number of concurrent tasks, heartbeats, JVMs heap size, and the shuffling operation. A hybrid Hadoop cluster with scale-out and scale-up servers was proposed in [288]. Based on profiling the performance of workloads on scale-out and scale-up clusters, a scheduler that assigns each job to the best choice was designed. A study to measure the impact on HDFS of improved networking such as InfiniBand and 10Gigabit Ethernet, selected protocol, and SSD storage systems was presented in [289]. With HDD storage, enhanced networking achieved up to 100% performance improvement, while with SSD, the improvement was by up to 219%. The authors in [290] discussed the need to optimize the tasks scheduling and data placement for data-centric workloads in compute-centric (i.e. HPC) clusters. A comprehensive evaluation was conducted to study the impact of storing intermediate RDDs on RAM, local SSD, and Lustre clusters, which requires write locks, on Spark workloads with different compute and I/O requirements. An Enhanced Load Balancer scheduler was suggested and a Congestion-Aware Task Dispatching to SSD mechanism was also proposed. Enhancements by 25%, and 41.2% were achieved. It was also found that increasing the chunk size to 128 MB from 32 MB reduced the job execution time by 15.9% as scheduling overheads were reduced.

The performance of big data applications under SSD storage was addressed in [291]-[294]. The work in [291] examined the performance improvements and economics (i.e. cost-per-performance) of partial and full use of SSDs in MapReduce clusters. A comparison between SATA Hard Disk Drives (HDDs) and Peripheral Component Interconnect express (PCIe) SSDs with the same aggregate I/O bandwidth showed that SSDs achieved up to 70% better performance at 2.4× the cost of HDDs. When upgrading HDDs systems with SSDs, proper configurations with multiple HDFSs should be considered. The authors in [292], discussed the need to modify Hadoop frameworks when using SSDs and proposed a direct map output collection, pre-read model for HDFS, a reduce scheduler to address skewness, and a modified data blocks placement policy to extend the lifetime of SSDs. The first two methods achieved improvements by 30%, and 18%, while the scheduler produced improvements by 12% and the overall improvement was by 21%. To accelerate I/O intensive and memory-intensive MapReduce jobs, *mpCache*, that dynamically caches input data from HDD into SATA SSDs, was proposed and examined in [293] and average speedups by 2× were gained. The aggregate performance of Non-Volatile Memory Express (NVMe) SSDs with high number of Docker containers for database applications was evaluated in [294]. Under optimized selection of instances count, up to 6× improved write throughput can be obtained compared to one instance. With multiple applications, the performance can degrade by 50%.

Optimizing data transfers from networked storage systems were addressed in [295]-[297]. Google traces were used in [295] to examine the performance of Hadoop with Network Attached Storage (NAS). MRPerf [111] was utilized to evaluate the slow down factor of using a NAS, that requires rack-remote transfers, instead of conventional Direct Attached Storage (DAS) for different number of racks and different workloads. The results showed that CPU-intensive workloads had the least slow down factor and that TeraSort workloads can benefit the most from an assumed enhanced NAS. *FlexDAS* in [296] was proposed as a switch based on SAS expanders to form a Disk Area Network (DAN) to provide flexibility in connecting and disconnecting computing nodes to HDDs while maintaining the I/O performance of DAS. Results based on a prototype showed that FlexDAS can detach and attach a HDD in 1.16 seconds and for three I/O intensive applications, it achieved the same I/O performance of DAS. The work in [297] addressed optimizing and balancing the I/O operation required for Terabits data transfers from centralized Parallel Storage Systems (PFS) over WAN for competing scientific applications. An end-to-end layout-aware optimization based on a scheduling algorithm for sources (i.e. disks) and for sinks (i.e. requesting servers) was utilized to coordinate the use of network and storage and to reduce I/O imbalances caused by congestion in some disks.

Remote Direct Memory Access (RDMA) that enables zero-copy (i.e. accessing the RAM without the intervention of the operating system), was utilized in [298], [299] to improve the performance of big data applications. Fast Remote Memory (FaRM) was proposed in [298] as a system for key-value and graph stores that regards RAMs of the machines in a cluster as a shared address space. FaRM utilizes one-sided RDMA lock-free serializable reads, RDMA writes for fast message passing, and single machine transactions with optimized locality. A kernel driver, PhyCo, was developed to enable addressing 2GB memory regions to overcome the issue of limited entries in NIC page tables. Compared to TCP/IP, FaRM achieved improved rate by 11×, and 9×, for request sizes of 16 and 512 bytes, respectively. Hadoop-A in [299] reduced the overheads of materializing intermediate data in MapReduce by proposing a virtual shuffling mechanism that utilizes RDMA in InfiniBand and optimizes the use of memory buffers. Improvements by up to 27% compared to default shuffling and 12% power consumption reduction were achieved.

The studies in [300]-[309] focused on utilizing improved CPUs, GPUs, and Field Programmable Gate Arrays (FPGAs) to accelerate big data applications. An investigation of the inefficiencies of scale-out clusters for cloud applications based on micro-architectural characterization of their workloads was conducted in [300]. It was found that scale-out workloads exhibit high instruction-cache miss rate, low instruction and memory level parallelism, and have low CPU core to core bandwidth requirements, and recommendations for specialized servers were provided. The author in [301] proposed a system for exascale computation that includes low-power processors, byte-addressable NAND flash as main memory, and Dynamic RAM (DRAM) as cache for the flash. A total of 2550 nodes in 51 boards were designed to have their main memories connected as Hoffman-Singleton graph, which provides 2.6 PB capacity, 9.1 TB/s bisection bandwidth and reduced latency to at most four hops. Mars in [302] was proposed as a MapReduce runtime system that utilized NVIDIA GPUs, AMD GPUs, and multi-core CPUs to accelerate the computations. To ease the programming of MapReduce on GPUs, MarsCUDA programming framework was introduced to automate the task partitioning, data allocation, and threads allocation to key-value pairs. The results showed that Mars outperformed a CPU-only based MapReduce system for multi-core CPUs, Phoenix, by 22 times. The authors in [303] proposed coupling the use of CPUs and GPUs for accelerating map tasks with MPTCP to reduce shuffling time for improving the performance and reliability of Hadoop.

FPMR was proposed in [304] as a framework to help developers boost MapReduce for machine learning and data mining in FPGAs while masking the details of task scheduling and data management and communications between FPGAs and CPU. A processor scheduler was implemented in FPGA to control the assignments of map and reduce tasks, and a Common Data Path (CDP) was built to perform data transfers without redundancy to several tasks. As an example, compute-intensive rank algorithm; RankBoost was mapped into FPMR. Compared to CPU-only implementation, speedups by 16.74×, and 31.8× were achieved without and with CDP, respectively. SODA in [305] introduced a framework for software defined accelerators based on FPGA. To target heterogeneous architectures, SODA utilizes a component based programming model for the accelerators, and Dataflow execution phases to parallelize the tasks with an out-of-order scheduling scheme. In a Constrained Shortest Path Finding (CSPF) algorithm for SDN applications in 128 nodes, acceleration by 43.75× was achieved. To accelerate MapReduce with high energy efficiency in hosting data centers, the authors in [306] extended the current High Level Synthesis (HLS) toolflow to include the MapReduce framework, which allows customizing the accelerations and scaling to data center level. To allow easy transition from C/C++ to Register Transfer Level (RTL) design, the tasks and data paths in MapReduce were decoupled, the accessible memory for each task was isolated, and a dataflow computational approach was forced. The throughput was improved by up to 4.3× while having a two order of magnitude improvement in the energy efficiency compared to a multi-core processor. A Multilevel NoSQL cache implemented in FPGA-based in-NIC and software-based in-kernel caches was proposed in [307] to accelerate NoSQL. To efficiently process entries with different sizes, multiple heterogeneous Processing Elements (PEs) (i.e. string PE, hash PE, list PE, and set PE) were utilized instead of pipelining. FPGA-based NICs were also utilized in [308] to perform one-at-a-time processing for data streaming instead of micro-batch processing of Apache Spark Streaming and suggested filtering the data in the NIC to reduce processed data.

Experimental results demonstrated that WordCount throughput was improved by 22× and change point detection latency was reduced by 94.12%. The authors in [309] examined the challenges associated with partitioning large graphs for the Graphlet counting algorithm from bioinformatics and proposed a framework for reconfigurable acceleration in FPGAs that achieved an order of magnitude improvement compared to processing in quad-core CPUs and better scalability.

The challenges for big data applications in disaggregated data centers were discussed in [310], [311]. In [310], the network requirements for supporting big data applications in disaggregated data centers were discussed. The key findings were that existing networking hardware that provides between 40 Gbps and 100 Gbps data rates is sufficient for rack-scale and even data center-scale disaggregation with some workloads, if the network software is improved and the protocols used were deadline-aware. In [311], the authors demonstrated the feasibility of composable rack-scale disaggregation for heterogeneous big data workloads and negligible latency impact was recorded for some of the workloads such as Memchached. Table VI provides the second part of the summary of data center focused optimization studies reported in this section.

TABLE VI SUMMARY OF DATA CENTER -FOCUSED OPTIMIZATION STUDIES - II

| Ref | Objective | Application/ platform | Tools | Benchmarks/workloads | Experimental Setup/Simulation environment |
|---|---|---|---|---|---|
| [258]* | Orchestra: Task-aware cluster management to reduce the completion time of data transfers | Spark | BitTorrent-like protocol, Weighted Shuffle Scheduling | Facebook traces | EC2 instances, DETERlab cluster |
| [259]* | Seawall: Centralized network bandwidth allocation scheme in multitenant environments | - | Shim layer, Strawman bandwidth allocator | Web workloads | Cosmos; 60 nodes in 3 racks (Xeon 2.27 GHz CPU, 4GB RAM), 1 Gbps NIC ports |
| [260]* | FlowComb: Transparent network-management framework for high utilization and fast processing | Hadoop 1.x | Software agents, OpenFlow | Sort 10GB | 14 nodes |
| [261]* | Distributed Flow Scheduling (DiFS) for adaptive routing in hirarchical data centers | - | Imbalance Detection and, Explicit Adaption algorithm | Deterministic, random, and shuffling flows (120 GB) | Packet-level stand-alone simulator for Fat-tree with 16 hosts and 1024 hosts |
| [262]* | Varys: Centralized co-flow scheduling to reduce completion time of data-intensive applications | - | Smallest-Effective-Bottleneck First (SEBF) heuristic | Facebook traces | Extra large high-memory instances in a 100-machine EC2 cluster, task-level trace-driven simulations |
| [263]* | Baraat: Decentralized task-aware scheduler for co-flows in DCNs to reduce tail completion time | Memchached | FIFO limited multiplexing, Smart Priority Class (SPC) | Bing, Cosmos, and Facebook traces | 20 nodes in 5 racks testbed (Quad core, 4GB RAM), 1 GbE flow-based simulations, ns-2 large-scale simulations |
| [264]* | D-CAS: A decentralized coflow-aware scheduling in large-scale DCNs | - | Online 2-approximation scheduling algorithm | Facebook traces | Python-based trace-driven simulations |
| [265]* | Rapier: coflow-aware joint routing and scheduling in DCNs | - | Online heuristic, OpenFlow | Many-to-many communication patterns | Spine-leaf DCN with 9 nodes (4 CPU cores,8GB RAM, 500GB disk) large-scale event-based flow-level simulations for 512 nodes Fat-tree and VL2 |
| [266]* | Low-complexity online multicast flows scheduling in Fat-tree DCNs | - | BCMS algorithm | Unicast and mixed synthetic traffic | Event-driven flow simulations for 32-ary Fat-tree |
| [267]* | Pythia: real-time prediction of data communication volume to accelerate MapReduce | Hadoop 1.1.2 | Application-specific instrumentation, OpenFlow | HiBench benckmarks: Sort and Nutch index | 10 nodes (12 CPU cores, 128 GB RAM, 1 SATA disk) in two racks, OpenFlow-enabled ToRs |
| [268]* | Application-aware Resources Allocation scheme for SDN-based data centers | - | Neural networks | Different applications traffic | Cloudsim-based simulations for 20 servers with 80 VMs |
| [269]* | BASS: Bandwidth-Aware Scheduling with SDN in Hadoop clusters | Hadoop 1.2.1 | OVS, time-slotted bandwidth allocation algorithm | WordCount, Sort | 5-nodes cluster |
| [270]* | Scheduling in dynamic OCS-switching data centers for delay-sensitive applications | - | SCFT and FCFT scheduling algorithms | NAS Parallel Benchmark (NPB) suite | OCSEMU emulator with 20 nodes (E5520 CPU, 24 GB RAM, 500 GB disk) |
| [271]* | Slotted resource allocation in an SDN control-based hybrid DCN | - | Online and incremental scheduling heuristics | Iterative MapReduce workloads | OMNET++ 4.3.1-based packet-level simulations |
| [272]* | Scheduling algorithms in hybrid packet optical data centers | - | MILP, heuristics | 10k-20k random requests | 1024 nodes in two-tier network (4 core MEMS switches, and 64 ToRs) |
| [273]* | Scheduling in packet-switched optical DCNs | - | Mahout, C4.5, and Naïve Bayes Discretization (NBD) | Random traffic | Simulations for 80 ToRs connected by two 40×40 AWGRs and two 1 × 2 space switches |
| [274]* | Max-throughput data transfer scheduling to minimize data retrieval time | - | MILP, heuristic | 1-1, many-1 Bulk transfers | Simulations for three-tier and Fat-tree DCNs with 128 nodes and VL2 with 160 nodes. |
| [275]* | Topology-aware data placement in geo-distributed data centers | Hadoop 2.2.0 | Heuristic scheduler | WordCount, TeraSort, k-means | 18 nodes, TopoSim MapReduce simulator |
| [276]* | CoMan: Global in-network management in multiplexed data centres | - | 3/2 approximation algorithm | 55 flows from 10 applications | Emulation for a Fat-tree DCN with 10 switches and 8 servers (using Pica8 3297), trace-driven simulations |
| [277]* | Network-aware MapReduce tasks placement to reduce transmission costs in DCNs | Hadoop 1.2.1 | Probabilistic tasks scheduling algorithm, | Wordcount, Grep from BigDataBench, Terasort | Palmetto HPC platform with 1978 slave nodes (16 cores CPU, 16GB RAM, 3GB disk), 10 GbE |
| [278]* | Network-aware Scheduler (NAS) for high performance shuffling | Hadoop 2.x | MTS, CA-RTS, and CR-RTS scheduling algorithms | Facebook traces | 40 nodes in 8 racks cluster with 1 GbE for cross-rack links Trace-driven event-based simulations for 600 nodes in 30 racks with 200 users |
| [279]* | PROTEUS: system for Temporally Interleaved Virtual Cluster (TIVC) abstraction for multi-tenant data centers | Hive, Hadoop 0.20.0 | Profiling, dynamic programming | Sort, WordCount, Join, aggregation | Profiling: 33 nodes (4 cores 2.4GHz CPU, 4GB RAM), Gigabit switch prototype: 18-nodes in three-tier data center with NetFPGA switches, simulations |
| [280]* | Corral: offline scheduling framework to reduce cross-rack traffic for production workloads | Hadoop 2.4 | Offline planner, cluster scheduler | SWIM and Microsoft traces, TPC-H | 210 nodes (32 cores) in 7 racks, 10GbE large-scale simulations for 2000 nodes in 50 racks |
| [281]* | Mercury: Fine-grained hybrid resources management and scheduling | Hadoop 2.4.1 | Daemon in each node, extensions to YARN | Gridmix and Microsoft's production traces | 256 nodes cluster (2 8-core CPU, 128GB RAM, 10 3TB disks), 10 Gbps in rack, 6 Gbps between racks |
| [282]* | Optimizing containers allocation to reduce congestion and improve data locality | Hadoop 2.7, Apache Flink 0.9 | Simulated annealing, modification to AM | *k*-means, connected components | 8 nodes (8-core CPU, 32GB RAM, SATA disk) in a Fat-tree topology 1 GbE, OpenFlow 1.1 |
| [283]* | Examining the impact of the Energy Efficient Ethernet on MapReduce workloads | Hadoop 1.x | EEE in switches, packets coalescing | TeraSort, Search, Index | MRperf-based packet-level simulations for two racks cluster with up to 80 nodes |
| [284]* | Energy-aware hierarchical applications scheduling in large DCNs | - | k th max node sorting, dynamic max node sorting algorithms | Uniform / normal random applications demands | C++ based simulations for up to 4096 nodes with 32, 64, 128, and 256 ports switches |
| [285]* | Willow: Energy-efficient SDN-based flows scheduling in data centers with network-limited workloads | Hadoop 1.x | Online greedy approximation algorithm | Locally generated MapReduce traces | 16 nodes (AMD CPU, 2GB RAM), 1 GbE, Simulations for Fat-tree and Fat-tree with disabled core switches, 1 GbE |
| [286]* | JouleMR: Cost-effective and energy-aware MapReduce jobs and tasks scheduling | Hadoop 2.6.0 | Two-phase heuristic | TeraSort, GridMix Facebook traces | 10 nodes cluster (2 6-cores CPUs, 24GB RAM, 500GB disk), 10 GbE simulations for 600 nodes in tree-structured DCN, with1GbE links |
| [287]* | Evaluation of scale-up vs scale-out systems for Hadoop workloads | Vanilla Hadoop | Parameters optimization to Hadoop in scale-up servers | Log processing, select, aggregate, join, TeraSort, k-means, indexing | 16 nodes (Quad-core CPU, 12GB RAM, 160GB HDD, 32GB SSD), 1 GbE, Dell PowerEdge910 (4 8-core CPU, 512GB RAM, 2 600GB HDD, 8 240GB SSD) |
| [288]° | Hybrid scale-out and scale-up Hadoop cluster | Hadoop 1.2.1 | Automatic Hadoop parameters optimization, scheduler to select cluster | WordCount, Grep, Terasort, TestDFSIO, Facebook traces | 12 nodes (2 4-core CPU, 16GB RAM, 193GB HDD), 10 Gbps Mytinet, 2 nodes (4 6-core CPU, 505GB RAM, 91GB HDD), 10 Gbps Mytinet, OrangeFS |
| [289]° | Measuring the impact of InfiniBand and 10Gigabit on HDFS | Hadoop 0.20.0 | Testing different network interfaces and technologies | Sort, random write, sequential write | 10 nodes cluster (Quad-core CPU, 6GB RAM, 250GB HDD or (up to 4), 64GB SSD), InfiniBand DDR 16 Gbps links and 10GbE |
| [290]° | Optimizing HPC clusters for dual compute-centric and data-centric computations | Spark 0.7.0 | Enhanced Load Balancer, Congestion-Aware Task Dispatching | GroupBy, Grep, Logistic Regression (LR) | Hyperion cluster with 101 nodes ( 2 8-core CPUs, 64GB RAM, 128GB SSD) in 2 racks, InfiniBand QDR 32 Gbps links |
| [291]° | Cost-per-performance evaluation of MapReduce clusters with SSDs and HDDs | Hadoop 2.x | Standalone evaluation with different storage and application configurations | Teragen, Terasort, Teravalidate, WordCount, shuffle, HDFS | (8 core CPU, 48GB RAM ) 10 GbE, single rack, SSDs 1.3TB, HHDs 2TB setups 6 HDDs, 11 HDDs, 1 SSD, (6 HHDs +1 SSD) |
| [292]° | Improving Hadoop framework for better utilization of SSDs | Hadoop 2.6.0, Spark 1.3 | Modified map data handling, pre-read in HDFS, reduce scheduler, placement policy | Terasort, DFSIO | 16 nodes (8 core CPU, 256GB RAM, 600GB HDD, 512GB SATA SSD, 10GbE) 8 nodes (6 core CPU, 256GB RAM, 250GB HDD, 800GB NVMe SSD, 10GbE) Ceres-II 39 nodes (6 core CPU, 64GB RAM, 960GB NVMe SSD, 40GbE) |
| [293]° | mpCache: SSD-based acceleration for MapReduce workloads in commodity clusters | Hadoop 2.2.0 | Modified HDFS, admission control policy, main cache replacement scheme, | Grep, WC - Wikipedia classification - Netflix, PUMA | 7 nodes (2 8-core CPUs, 20MB cache, 32GB RAM, 2TB SATA HDD, 2 160GB SATA SSD) |
| [294]° | Evaluation of IO-intensive applications with Docker containers | Cassandra 3.0, MySQL 5.7, FIO 2.2 | Docker's data volume | Cassandra-stress, TPC-C benchmark | (32 core hyper-threaded CPU, 20480KB cache, 128GB RAM, and 3 960GB NVMe SSD) |
| [295]° | Examining the performance of different Hadoop schedulers and impact of NAS on Hadoop | - | Modification to MRPerf to implement Fair share scheduler and Quincy | Synthesized Google traces for 9174 jobs | MRperf-based simulations for 2 and 4 racks clusters |
| [296]° | Improving the flexibility of direct attached storage via a Disk Area Network (DAN) | Hadoop 2.0.5, Cassandra Swift 1.10.0 | FlexDAS switch, SAS expanders, Host Bus Adapters (HBAs) | TeraSort for 1TB, YCSB COSBench 0.3.0.0 | 12 nodes (4 core CPU, 48GB RAM), 10GbE and 48 external HDDs |
| [297]° | Optimizing bulk data transfers from parallel file systems via layout-aware I/O operations | Scientific applications | Layout-aware source algorithm, layout-aware sink algorithm | Offline Storage Tables (OSTs) with 20MB and 256MB files | Spider storage system at the Oak Ridge Leadership computing Facility with 96 DataDirect Network (DDN) S2A9900 RAID controllers for 13400 1TB SATA HDDs |
| [298]° | FaRM: Fast Remote Memory system using RDMA to improve in-memory key-value store | Key-value and graph stores | Circular buffer for RDMA messaging, PhyCo kernel driver, ZooKeeper | YCSB | 20 nodes (2 8-core CPUs, 128GB RAM, 240GB SSD), 40Gbps RDMA over Converged Ethernet (RoCE) |
| [299]° | HadoopA: Virtual shuffling for efficient data movement in MapReduce | Hadoop 0.20.0 | virtual shuffling through three-level segment near-demand merging, and merging sub-trees | TeraSort, Grep, WordCount, and Hive workloads | 21-nodes cluster (dual socket quad-core 2.13 GHz Intel Xeon, 8 GB RAM, 500 GB disk, 8 PCI-Express 2.0 bus), InfiniBand QDR switch, 48-port 10GigE |
| [300]° | Micro-architectural characterization of scale-out workloads | - | Intel VTune for characterization | Caching, NoSQL, MapReduce, media PARSEC, SPEC2006, TPC-E | PowerEdge M1000e (2 Intel 6-core CPUs, 32KB L1 cache, 256KB L2 cache, 12MB L3 cache, 24GB RAM) |
| [301]° | A 2 PetaFLOP, 3 Petabyte, 9 TB/s and 90 kW cabinet for exascale computations | - | Hoffman-Singleton topology | - | 2,550 nodes (64-bit Boston chip, 64GB DDR3 SDRAM, 1TB NAND flash) |
| [302]° | Mars: Accelerating MapReduce with GPUs | Hadoop 1.x | CUDA, Hadoop streaming, GPU Prefix Sum routine, GPU Bitonic Sort routine | String match, matrix multiplication, MC, Black-scholes, similarity score, PCA | 240-core NVIDIA GTX280+ 4-core CPU), (128-core NVIDIA 8800GTX + 4-core CPU), (320 core ATI Radeon HD 3870 + 2-core CPU) |
| [303]° | Using GPUs and MPTCP to improve Hadoop performance | Hadoop 1.x | CUDA, Hadoop pipes, MPTCP | Terasort, WordCount and PiEstimate DataGen | Node1 (Intel i7 920, 24GB RAM, 4 1TB HDD), node2 (Intel Quad Q9400, NVIDIA GT530 GPU, 4GB RAM, 500GB HDD), heterogeneous 5 nodes |
| [304]° | FPMR: a MapReduce framework on FPGA to accelerate RankBoost | - | On-chip processor scheduler, Common Data Path (CDP) | RankBoost | 1 node (Intel Pentium 4, 4GB RAM), Altera Stratix II EP2S180F1508 FPGA, Quartus II 8.1 and ModelSim 6.1-based simulations |
| [305]° | SODA: software defined acceleration for big data with heterogeneous reconfigurable multicore FPGA resources | - | Vivado high level synthesis tools, out-of-order scheduling algorithm | Constrained Shortest Path Finding (CSPF) for SDN | Xilinx Zynq FPGA, ARM Cortex processor |
| [306]° | HLSMapReduceFlow: Synthesizable MapReduce Accelerator for FPGA-coupled Data Centers | - | High-Level Synthesis-based MapReduce dataflow | WordCount, histogram, Matrix multiplication, linear regression, PCA, k-means | Virtex-7 FPGA |
| [307]° | FPGA-based in-NIC and software-based of NetFPGA, in-Kernel caches for NoSQL | NoSQL | DRAM and multi processing elements in NetFPGA, Netfilter framework for kernel cache | USR, SYS, APP, ETC, VAR | NetFPGA-10G ( Xilinx Virtex-5 XC5VTX240T) |
| [308]° | FPGA-based processing in NIC for Spark streaming | Spark 1.6.0, Scala 2.10.5 | one-at-a-time methodology operations | WordCount, change point detection | NetFPGA-10G (Xilinx Virtex-5 XC5VTX240TFFG1759-2) as NIC in server node (Intel core i5 CPU, 8GB DRAM) and a client node |
| [309]° | FPGA-acceleration for large graph processing | - | Graph Processing Elements (GPEs), memory interconnect network, run-time management unit | Graphlet counting algorithm | Convey HC-1 server with Virtex-5 LX330s FPGA |
| [310]° | Network requirements for big data application in disaggregated data centers | Hadoop, Spark, GraphLab, Timely dataflow, Memchached, HERD, SparkSQL | Page-level memory access, block-level distributed data placement RDMA and integrated NICs | WordCount, sort, collaborative filtering, YCSB | EC2 instances (m3.2xlarge, c3.4xlarge), with virtual private network (VPC), simulations and emulations |
| [311]° | A composable architecture for rack-scale disaggregation for big data computing | Memchached, Giraph, Cassandra | Empirical approaches for resources provisioning, PCIe switches | 100k Memcached operations, TopKPagerank, 10k Cassandra operations | H3 RAID array with 12 SATA drives and single PCIe Gen3 × 4 port, PCIe switch with 16 Gen2 × 4 port, host bus adapter connected to IBM 3650M4 |

* Scheduling in data centers, °Performance improvements based on advances in technologies.

## VIII. Summary of Optimization Studies and Future Research Directions:

Tremendous efforts were devoted in the last decade to address various challenges in optimizing the deployment of big data applications and their infrastructures. The ever increasing data volumes and the heterogeneous requirements of available and proposed frameworks in dedicated and multi-tenant clusters are still calling for further improvements of different aspects of these infrastructures. We categorized the optimization studies of big data applications into three broad categories which are at the applications-level, cloud networking-level, and at the data center-level. Application-level studies focused on the efforts to improve the configuration or structure of the frameworks. These studies were further categorized into optimizing jobs and data placements, jobs scheduling, and completion time, in addition to benchmarking, production traces, modeling, profiling and simulators for big data applications. Studies at the cloud networking level addressed the optimization beyond single cluster deployments, for example for transporting big data and/or for in-network processing of big data. These studies were further categorized into cloud resource management, virtual assignments and container assignments, bulk transfers and inter data center networking, and SDN, EON, and NFV optimization studies. Finally, the data center-level studies focused on optimizing the design of the clusters to improve the performance of the applications. These studies were further categorized into topology, routing, scheduling of flows, coflows, and jobs, and advances in computing, storage, and networking technologies studies. In what follows, we highlight key challenges for big data application deployments and provide some insights into research gaps and future directions.

**Big Data Volumes:** There is a huge gap in most of the studies between the actual volumes of big data and the tested traffic or workloads volumes. The main reason is the relatively high costs of experimenting in large clusters or renting IaaS. This calls for improving existing simulators or performance models to enable accurate testing of systems at large scale. Data volumes are continuing to grow beyond the capabilities of existing systems due to many bottlenecks in computing and networking. This will require continuing the investment in scale-up systems to incorporate different technologies such as SDN and advanced optical networks for intra and inter data center networking. Another key challenge with big data is related to the veracity and value of big data which calls for cleansing techniques prior to processing to eliminate unnecessary computations.

**Workload characteristics and their modeling:** Big data workload characteristics and available frameworks will keep changing and evolving. Most of the studies of big data address only the MapReduce framework while few considered other variations like key-value store, streaming and graph processing applications or a mixture of applications. Also, several studies have utilized scaled and outdated production traces where only high level statistics are available or a subset of workloads in micro benchmarks is available for the evaluations which might not be very representative. Thus, there is a need for more comprehensive, enhanced, and updated benchmarking suites and production traces to reflect more recent frameworks and workload characteristics.

**Resources allocation:** Different workloads can have uncorrelated CPU, I/O, memory, and networking resources requirements and placing those together can improve the overall infrastructure utilization. However, isolating resources such as cache, network, and I/O via recent management platforms at large scale is still challenging. Improving some of the resources can change the optimum configurations for applications. For example, improving the networking can reduce the need to maintain data locality, and hence, the stress on tasks scheduling is reduced. Also, improving the CPU can change CPU-bound workloads into I/O or memory-bound workloads. Another challenge is that there is still a gap between the users' knowledge about their requirements and the resources they lease which leads to non-optimal configurations and hence, waste of resources, delayed response, and higher energy consumption. More tools to aid users in understanding their requirements or their workloads are required for better resource utilization.

**Performance in proposed clusters:** Most of the research that enhances big data frameworks, reported in in Section III, was carried in scale-out clusters while overlooking the physical layer characteristics and focusing on the framework details. Alternatively, most of the studies in Sections V, and VII have considered custom-build simulators, SDN switches-based emulations, or small scale-up/out clusters while focusing on the hardware impact, considering only a subset of configurations, and oversimplifying the frameworks characteristics. Although it might sound more economical to scale out infrastructures as the computational requirements increase, this might not be sufficient for applications with strict QoS requirements as scaling-out depends on high level of parallelism which is constrained by the network between the nodes. Hence, improving the networking and scaling up the data centers are required and are gaining research and industrial considerations. With improving DCNs, many tradeoffs should be considered including the scalability, agility, end-to-end delays, in addition to the complexity of the routing and scheduling mechanisms required to fully exploit the improved bandwidth links. This will mostly be satisfied by application-centric DCN architectures that utilize SDN to dynamically vary the topologies at run-time to match the requirements of deployed applications.

**Clusters heterogeneity:** Potential existence of hardware with different specifications for example due to replacements in very large clusters can lead to completion time imbalance among tasks. This requires more attention in big data studies and accurate profiling of the performance in all nodes to improve task scheduling to reduce the imbalance.

**Multi-tenancy environments:** Multi-tenancy is a key requirement enabled by virtualization of cloud data centers where workloads of different users are multiplexed in the same infrastructure to improve the utilization. In current cloud services, static cluster configurations and manual adjustments are still followed but are not optimal. There is still lack of dynamic job allocation and scheduling for multi-users. In such environments, the performance isolation between users due to sharing resources such as network and I/O is not yet widely addressed. Also, fairness between users and optimal pricing while maintaining acceptable QoS for all users is still a challenging research topic requiring more comprehensive multi-objective studies.

**Geo-distributed frameworks:** The challenges faced in geo-distributed frameworks were summarized in Subsection IV-C. Those include the need for modifying the frameworks which were originally designed for single cluster deployments. There is a need for new frameworks for offers and pricing, optimal resources allocation in heterogeneous clusters, QoS guarantees, resilience, and energy consumption minimization. A key challenges with workloads in geo-distributed frameworks is that not all workloads can be divided as the whole data set may need to reside in a single cluster, also transporting some data sets from remote locations can have high data-access latency. Extended SDN control between transport networks and within data centers is a promising research area to jointly optimize path computations, provision distributed resources, and reduce jobs completion time. It is also a promising research area to improve big data applications and frameworks in geo-distributed environments.

**Power consumption:** Current infrastructures have a trade-off between energy efficiency and the applications performance. Most cloud providers still favor over-provisioning to meet SLA over reducing the power consumption. Reducing the power consumption while maintaining the performance is an area that should be explored further in designing future systems. For example, it is attractive to consider more energy-efficient components, more efficient geo-distributed frameworks to reduce the need for transporting big data. It is also attractive to perform progressive computations in the network as the data transitions, while considering agile technologies such SDN, VNE, and NFV, which can improve the energy efficiency of big data applications, however, the impact of those strategies on the applications performance should be comprehensively addressed.

## IX. CONCLUSIONS

Improving the performance and efficiency of big data frameworks and applications is an ongoing critical research area as they are becoming the mainstream for implementing various services including data analytics and machine learning at large scale. These are services that continue to grow in importance. Support should also be developed for other services deployed fully or partially in cloud and fog computing environments with ever increasing volumes of generated data. Big data interacts with systems at different levels starting from the acquisition of data through wireless and wired access networks from users and IoTs, to transmission through WAN networks, into different types of data centers for storage and processing via different frameworks. This survey paper surveyed big data applications and the technologies and network infrastructure needed to implement them. It identified a number of key challenges and research gaps in optimizing big data applications and infrastructures. It has also comprehensively summarized early and recent efforts towards improving the performance and/or energy efficiency of such big data applications at different layers. For the convenience of readers with different backgrounds, brief descriptions of big data applications and frameworks, cloud computing and related emerging technologies, and data centers are provided in Sections II, IV, and VI, respectively. The optimization studies, that appear in Section III focus on the frameworks, those that appear in Section V focus on cloud networking, with Section VII focusing on data centers, and finally comprehensive summaries are given in Tables I-VI. The survey paid attention to a range of existing and proposed technologies and focused on different frameworks and applications including MapReduce, data streaming and graph processing. The survey considered different optimization metrics (e.g. completion time, fairness, cost, profit, and energy consumption), reported studies that considered different representative workloads, optimization tools and mathematical techniques, and considered simulation-based and experimental evaluations in clouds and prototypes. We provided some future research directions in Section VIII to aid researchers in identifying the limitations of current solutions and hence determine the area where future technologies should be developed in order to improve big data applications, their infrastructures and performance.

ACKNOWLEDGEMENTS

Sanaa Hamid Mohamed would like to acknowledge Doctoral Training Award (DTA) funding from the UK Engineering and physical Sciences Research Council (EPSRC). This work was supported by the Engineering and Physical Sciences Research Council, INTERNET (EP/H040536/1), STAR (EP/K016873/1) and TOWS (EP/S016570/1) projects. All data are provided in full in the results section of this paper.

**Sanaa Hamid Mohamed**

received the B.Sc. degree (honors) in electrical and electronic engineering from the University of Khartoum, Sudan, in 2009 and the M.Sc. degree in electrical engineering from the American University of Sharjah (AUS), United Arab Emirates, in 2013. She received a full-time graduate teaching assistantship, MSELE program, AUS in 2011. She received a Doctoral Training Award (DTA) from the UK Engineering and Physical Sciences Research Council (EPSRC) to fund her PhD studies at the University of Leeds in 2015. Currently, is a PhD student at the Institute of Communication and Power Networks, University of Leeds, UK. She has held research and teaching positions at the University of Khartoum, AUS, Sudan University of Science and Technology, and Khalifa University between 2009 and 2013. She has been an IEEE member since 2008 and a member of the IEEE communication, photonics, computer, cloud computing, software defined networks, and sustainable ICT societies. Her research interests include wireless communications, optical communications, optical networking, software-defined networking, data center networking, big data analytics, and cloud and fog computing.

**Taisir E. H. El-Gorashi**

received the B.S. degree (first-class Hons.) in electrical and electronic engineering from the University of Khartoum, Khartoum, Sudan, in 2004, the M.Sc. degree (with distinction) in photonic and communication systems from the University of Wales, Swansea, UK, in 2005, and the PhD degree in optical networking from the University of Leeds, Leeds, UK, in 2010. She is currently a Lecturer in optical networks in the School of Electrical and Electronic Engineering, University of Leeds. Previously, she held a Postdoctoral Research post at the

University of Leeds (2010– 2014), where she focused on the energy efficiency of optical networks investigating the use of renewable energy in core networks, green IP over WDM networks with datacenters, energy efficient physical topology design, energy efficiency of content distribution networks, distributed cloud computing, network virtualization and Big Data. In 2012, she was a BT Research Fellow, where she developed energy efficient hybrid wireless-optical broadband access networks and explored the dynamics of TV viewing behavior and program popularity. The energy efficiency techniques developed during her postdoctoral research contributed 3 out of the 8 carefully chosen core network energy efficiency improvement measures recommended by the GreenTouch consortium for every operator network worldwide. Her work led to several invited talks at GreenTouch, Bell Labs, Optical Network Design and Modelling conference, Optical Fiber Communications conference, International Conference on Computer Communications, EU Future Internet Assembly, IEEE Sustainable ICT Summit and IEEE 5G World Forum and collaboration with Nokia and Huawei.

**Jaafar M. H. Elmirghani (M'92–SM'99)**
is the Director of the Institute of Communication and Power Networks within the School of Electronic and Electrical Engineering, University of Leeds, UK. He joined Leeds in 2007 and prior to that (2000–2007) as chair in optical communications at the University of Wales Swansea he founded, developed and directed the Institute of Advanced Telecommunications and the Technium Digital (TD), a technology incubator/spin-off hub. He has provided outstanding leadership in a number of large research projects at the IAT and TD. He received the Ph.D. in the synchronization of optical systems and optical receiver design from the University of Huddersfield UK in 1994 and the DSc in Communication Systems and Networks from University of Leeds, UK, in 2014. He has co-authored Photonic Switching Technology: Systems and Networks, (Wiley) and has published over 500 papers. He has research interests in optical systems and networks. Prof. Elmirghani is Fellow of the IET, Fellow of the Institute of Physics and Senior Member of IEEE. He was Chairman of IEEE Comsoc Transmission Access and Optical Systems technical committee and was Chairman of IEEE Comsoc Signal Processing and Communications Electronics technical committee, and an editor of IEEE Communications Magazine. He was founding Chair of the Advanced Signal Processing for Communication Symposium which started at IEEE GLOBECOM'99 and has continued since at every ICC and GLOBECOM. Prof. Elmirghani was also founding Chair of the first IEEE ICC/GLOBECOM optical symposium at GLOBECOM'00, the Future Photonic Network Technologies, Architectures and Protocols Symposium. He chaired this Symposium, which continues to date under different names. He was the founding chair of the first Green Track at ICC/GLOBECOM at GLOBECOM 2011, and is Chair of the IEEE Sustainable ICT Initiative within the IEEE Technical Activities Board (TAB) Future Directions Committee (FDC) and within the IEEE Communications Society, a pan IEEE Societies Initiative responsible for Green and Sustainable ICT activities across IEEE, 2012-present. He is and has been on the technical program committee of 38 IEEE ICC/GLOBECOM conferences between 1995 and 2019 including 18 times as Symposium Chair. He received the IEEE Communications Society Hal Sobol award, the IEEE Comsoc Chapter Achievement award for excellence in chapter activities (both in 2005), the University of Wales Swansea Outstanding Research Achievement Award, 2006, the IEEE Communications Society Signal Processing and Communication Electronics outstanding service award, 2009, a best paper award at IEEE ICC'2013, the IEEE Comsoc Transmission Access and Optical Systems outstanding Service award 2015 in recognition of "Leadership and Contributions to the Area of Green Communications", received the GreenTouch 1000x award in 2015 for "pioneering research contributions to the field of energy efficiency in telecommunications", the 2016 IET Optoelectronics Premium Award and shared with 6 GreenTouch innovators the 2016 Edison Award in the "Collective Disruption" Category for their work on the GreenMeter, an international competition, clear evidence of his seminal contributions to Green Communications which have a lasting impact on the environment (green) and society. He is currently an editor of: IET Optoelectronics, Journal of Optical Communications, IEEE Communications Surveys and Tutorials and IEEE Journal on Selected Areas in Communications series on Green Communications and Networking. He was Co-Chair of the GreenTouch Wired, Core and Access Networks Working Group, an adviser to the Commonwealth Scholarship Commission, member of the Royal Society International Joint Projects Panel and member of the Engineering and Physical Sciences Research Council (EPSRC) College. He was Principal Investigator (PI) of the

£6m EPSRC INTelligent Energy awaRe NETworks (INTERNET) Programme Grant, 2010-2016 and is currently PI of the £6.6m EPSRC Terabit Bidirectional Multi-user Optical Wireless System (TOWS) for 6G LiFi Programme Grant, 2019-2024. He has been awarded in excess of £30 million in grants to date from EPSRC, the EU and industry and has held prestigious fellowships funded by the Royal Society and by BT. He was an IEEE Comsoc Distinguished Lecturer 2013-2016.